\documentclass[a4paper,11pt]{article}
\pdfoutput=1
\PassOptionsToPackage{x11names,dvipsnames}{xcolor}
\usepackage{jheppub}
\usepackage[T1]{fontenc}
\usepackage[compat=1.1.0]{tikz-feynman}
\usepackage{tikzsymbols}
\usepackage{upgreek}
\usepackage{natbib}
\usepackage{appendix}
\usepackage{mathtools, slashed, multirow}
\usepackage{subcaption}
\usepackage{cleveref}
\usepackage{tikz}
\usetikzlibrary{tikzmark,calc}
\usepackage{color, colortbl}
\usepackage{bigfoot}
\usepackage{mathrsfs}
\usepackage{soul}
\usepackage[normalem]{ulem}
\usepackage{orcidlink}
\usepackage{adjustbox}
\usepackage{tcolorbox}
\usepackage[x11names,dvipsnames]{xcolor}

\newcommand{\eq}{Eq\,.~}
\newcommand{\eqs}{eqs\,.~}
\newcommand{\Eq}{Eq\,.~}
\newcommand{\Eqs}{Eqs\,.~}

\newcommand{\keV}{\text{keV}}

\newcommand{\GeV}{\text{GeV}}
\newcommand{\TeV}{\text{TeV}}
\newcommand{\SM}{\text{SM}}

\newcommand{\lphiS}{\lambda_{\Phi S}}
\newcommand{\lphiH}{\lambda_{\Phi H}}
\newcommand{\lphi}{\lambda_{\Phi}}

\newcommand{\ULmuLtau}{{{\rm U}(1)_{\rm L_\mu-L_\tau}^{}}}

\newcommand{\Qmuta}{{\text{Q}_{\mu\tau}^{}}}
\newcommand{\zmuta}{{\text{Z}_{\mu\tau}^{}}}
\newcommand{\mzmuta}{{m^{}_{\text{Z}_{\mu\tau}^{}}}}

\newcommand{\gmuta}{{g_{\mu\tau}^{}}}

\newcommand{\vmuta}{{v_{\mu\tau}^{}}}

\newcommand{\mone}{m_{\phi_1^{}}}
\newcommand{\mtwo}{m_{\phi_2^{}}}
\newcommand{\one}{\phi_1^{}}
\newcommand{\two}{\phi_2^{}}

\newcommand{\deltam}{\delta m}

\newcommand{\mhtwo}{m_{\rm h_2^{}}}
\newcommand{\Ztwo}{\mathcal{Z}_2}
\newcommand{\salpha}{\sin\alpha}

\newcommand{\B}{{\text{B}}}
\newcommand{\epsA}{{\epsilon^{}_{\gamma\zmuta}}}
\newcommand{\epsZ}{{\epsilon^{}_{Z\zmuta}}}
\definecolor{MyDarkBlue}{rgb}{0.1, 0.1, 0.8}
\definecolor{SBlue}{rgb}{0.2, 0.4, 0.7} 
\definecolor{MyLightBlue}{rgb}{0.22,0.51,0.9}
\definecolor{MyGreen}{rgb}{0.0, 0.5, 0.0}
\definecolor{BrickRed}{rgb}{0.8, 0.25, 0.33}
\definecolor{deepmagenta}{rgb}{0.8, 0.0, 0.8}
\definecolor{mediumtealblue}{rgb}{0.0, 0.33, 0.71}
\definecolor{warmblack}{rgb}{0.0, 0.26, 0.26}
\definecolor{bostonuniversityred}{rgb}{0.8, 0.0, 0.0}
\definecolor{junglegreen}{rgb}{0.16, 0.67, 0.53}
\definecolor{lightcornflowerblue}{rgb}{0.6, 0.81, 0.93}
\definecolor{mypink1}{rgb}{0.858, 0.188, 0.478}
\definecolor{mypink2}{RGB}{219, 48, 122}
\definecolor{mypink3}{cmyk}{0, 0.7808, 0.4429, 0.1412}
\definecolor{mygray}{gray}{0.2}
\definecolor{ForestGreen}{RGB}{34,139,34}
\arxivnumber{2610.12443}
\title{A scalar-extended $\boldsymbol{\rm U(1)_{ L_{\mu} - L_{\tau}}}$ explanation of the LUX-ZEPLIN 248 keV excess}

\author[a,b]{Dipankar Pradhan\orcidlink{0000-0002-2450-6677},}
\emailAdd{dipankar.pradhan@iopb.res.in}
\affiliation[a]{Institute of Physics, Sachivalaya Marg, Bhubaneswar, Odisha 751005, India}
\affiliation[b]{Homi Bhabha National Institute, BARC Training School Complex, Anushakti Nagar, Mumbai 400094, India}
\author[c]{Abhik Sarkar\orcidlink{0000-0003-1449-2934}}
\emailAdd{asarkar@lpsc.in2p3.fr}
\affiliation[c]{Laboratoire de Physique Subatomique et de Cosmologie, Université Grenoble-Alpes, CNRS/IN2P3, 53 Avenue des Martyrs, 38026 Grenoble, France}
\abstract{The recently observed dark matter (DM) nuclear recoil event at an energy of $248 \pm 23\,(\text{stat.}) \pm 23\,(\text{syst.})\,\text{keV}$ at the LUX-ZEPLIN (LZ) experiment, favoring an inelastic scattering interpretation, has motivated various beyond the Standard Model (BSM) scenarios aimed at explaining the observed excess. In particular, this provides an opportunity to probe models that naturally accommodate inelastic DM scattering. In this work, we explore one such possibility by extending the $\rm U(1)_{L_{\mu}-L_{\tau}}$ model with two complex scalar fields, $\Phi$ and $S$, carrying charges $-1$ and $+2$, respectively, under the $\rm U(1)_{L_{\mu}-L_{\tau}}$ gauge symmetry. These charge assignments allow a trilinear interaction between the two scalar fields, which is crucial for generating the DM mass splitting. Following the spontaneous breaking of the $\rm U(1)_{L_{\mu}-L_{\tau}}$ symmetry, with $S$ acquiring a vacuum expectation value and generating a mass for the BSM gauge boson $Z_{\mu\tau}$, the real and imaginary components of $\Phi$ acquire a mass splitting. This naturally gives rise to an inelastic DM scattering scenario that can accommodate the LZ excess. We perform a likelihood-based assessment of the sensitivity of the model parameters to the observed event while consistently accounting for other relevant phenomenological constraints. Finally, we discuss the prospects for probing the viable parameter space at future experiments.}
\makeatletter
\gdef\@fpheader{}
\makeatother
\begin{document}
\begin{flushright}
IOP/BBSR/2026-13
\end{flushright}
\maketitle
\flushbottom
\section{Introduction}
The existence of dark matter (DM) \cite{Cirelli:2024ssz} is one of the most compelling pieces of evidence for physics beyond the Standard Model (BSM). However, its particle nature remains elusive, as direct and indirect detection experiments, as well as collider-based searches, have so far yielded no conclusive evidence for DM. The recent observation of a high-energy nuclear recoil event with an energy of $248 \pm 23\,(\text{stat.}) \pm 23\,(\text{syst.})\,\text{keV}$ at the LUX-ZEPLIN (LZ) experiment \cite{LZ:2026axp}, distinguishable from the expected backgrounds, has opened up a new avenue for DM searches. Although the event corresponds to a global significance of $2.6\sigma$ (and a local significance of $3.4\sigma$) with respect to the background-only expectation, the observation of additional events with similar characteristics in future data could provide evidence for GeV-scale DM undergoing inelastic scattering off a nucleus \cite{Tucker-Smith:2001myb}. Following this observation, several BSM scenarios have been proposed to explain the observed excess \cite{Sheikh:2026lxx, Das:2026buc, Zhu:2026dag, Choudhury:2026qjz, DiMauro:2026ldr, Dent:2026bji, Girmohanta:2026qzo, Freese:2026sga, Barman:2026omh, Anwar:2026ffj, Chattopadhyay:2026ryw, Bisal:2026khf, Lee:2026jxl, Kumar:2026lgi, Cabo-Almeida:2026uqw, Okada:2026upm, Mahapatra:2026glu, He:2026hqz, Palmisano:2026kuj, Baer:2026yrt, Baer:2026fpy, Borah:2026ris, Arcadi:2026kev, Nagata:2026pbj, Okada:2026fef, Paul:2026okh, Bose:2026szs, Maier:2026qpr, Gemmell:2026yaw, Jho:2026zzj, Lou:2026idn, Das:2026uyy, Bose:2026ndd,Jeesun:2026vzo,He:2026idw,Bandyopadhyay:2026gjw,Yin:2026jnn,Kalita:2026hvb,Visinelli:2026kgt,Lian:2026hpm,Lian:2026bze,De:2026win,Kotlarski:2026pep,CentellesChulia:2026ats,Nguyen:2026lui,Unwin:2026qep,Unwin:2026rdp,Chakraborti:2026gmu,Das:2026jsq,Capolupo:2026uyi}.

In this work, we explore an extension of the $\ULmuLtau$-gauged model with two additional complex scalar fields \cite{He:1990pn, Baek:2017sew, Altmannshofer:2014cfa, Nomura:2026qyq, Wang:2026ytg, Qi:2021rhh, Baek:2008nz}. The spontaneous breaking of the $\ULmuLtau$ symmetry, triggered by one of the scalars acquiring a non-zero vacuum expectation value (VEV), renders the other scalar stable and hence a viable DM candidate. With an appropriate choice of charge assignments, a trilinear interaction between the two scalars is allowed. Following $\ULmuLtau$ symmetry breaking, this interaction generates a mass splitting between the real and imaginary components of the DM candidate, thereby enabling inelastic scattering off a nucleus that can potentially account for the observed high-energy nuclear recoil event at LZ. The minimal $\ULmuLtau$ model, as well as various extensions of its scalar sector, has been studied extensively in the literature \cite{Okada:2026fef, Okada:2026eol, Kumar:2026lgi, Ahmed:2026qjg, Du:2026lpa,De:2025hay}. However, the possibility of generating a mass splitting within the scalar DM sector and its connection to the high-energy nuclear recoil event observed at LZ provide the primary novel motivation for this work.

In addition to its dark sector realization, the $\ULmuLtau$ model is connected to a variety of phenomenological signatures. The spontaneous breaking of the $\ULmuLtau$ symmetry renders the corresponding gauge boson, $\zmuta$, massive through the Higgs mechanism and induces kinetic and mass mixing with the Standard Model (SM) gauge bosons, $\gamma$ and $Z$. The characteristic charge assignment of the model leads to tree-level couplings of $\zmuta$ to muons ($\mu$), taus ($\tau$), and their corresponding neutrinos, while couplings to the remaining SM fermions arise only at the one-loop level and are consequently suppressed. This distinctive interaction structure imposes stringent constraints from observables in the $\mu$-$\tau$ sector, including contributions to the anomalous magnetic moments of the muon and tau. The tree-level couplings to muon and tau neutrinos can also be probed through neutrino trident production. Conversely, the loop-suppressed couplings to other SM fermions significantly weaken constraints from indirect detection and existing collider searches. Nevertheless, future facilities, particularly muon colliders, can provide a highly sensitive probe of this framework, as the model predicts measurable deviations in di-muon and di-tau production processes and related observables.

The work is organized as follows. In~\autoref{sec:model}, we discuss the model, including the relevant charge assignments and symmetry-breaking pattern, along with the associated theoretical and low-energy constraints. In~\autoref{sec:relic}, we investigate the parameter space consistent with the observed DM relic abundance. In~\autoref{sec:DD}, we study the prospects for direct detection, with particular emphasis on inelastic scattering at LZ, and perform a dedicated likelihood-based statistical analysis to determine the corresponding allowed parameter space. In~\autoref{sec:collider}, we discuss constraints from indirect detection and existing LHC searches. We also explore the prospects for probing the model at future muon colliders. Finally, we summarize the results and present our conclusions in~\autoref{sec:summary}.
\section{Extended $\boldsymbol{\ULmuLtau}$ Model}
\label{sec:model}
The $\ULmuLtau$ model is a well-motivated BSM scenario based on the gauging of the anomaly-free $\rm L_\mu - L_\tau$ symmetry. Under this symmetry, the muon ($\mu$) and muon neutrino ($\nu_\mu$) carry charge $+a$, while the tau ($\tau$) and tau neutrino ($\nu_\tau$) carry charge $-a$, where $a$ is a rational number. All other SM particles are neutral. The symmetry is vector-like, with the left and right-handed components of the charged leptons carrying identical $\rm L_\mu - L_\tau$ charges. The corresponding gauge symmetry is associated with a gauge boson, $\zmuta$, which mediates the new $\ULmuLtau$ interaction and acquires a mass via spontaneous symmetry breaking.

In this work, we extend the $\ULmuLtau$-gauged model by introducing two complex scalar fields, $S$ and $\Phi$, that transform non-trivially under the $\ULmuLtau$ gauge symmetry. We assign $\ULmuLtau$ charges $\Qmuta=\pm2$ to the muon and tau leptons, respectively, and $Q_{\mu\tau}=+2$ and $-1$ to the complex scalar fields $S$ and $\Phi$. The spontaneous symmetry breaking induced by $\langle S\rangle=(\vmuta/\sqrt{2})$ leaves an exact residual $\mathcal{Z}_2$ symmetry \cite{Babu:2002tx,Wang:2004mg,Petersen:2009ip,Luhn:2006hla,Baek:2022ozm}, under which all SM fields are $\mathcal{Z}_2$-even, while the scalar $\Phi$, carrying $Q_{\mu\tau}^{\Phi}=-1$, is rendered $\mathcal{Z}_2$-odd, transforming as $\Phi\to-\Phi$ under $\mathcal{Z}_2$. This residual symmetry ensures the stability of $\Phi$.
\begin{table}[htb!]
\centering
\begin{adjustbox}{width=0.5\textwidth, keepaspectratio}
\begin{tabular}{|c|c|cc|}\hline
\rowcolor{gray!20}Fields  & 
$\rm \mathcal{G}_{\SM}$
&\tikzmarknode{X}{$\quad\ULmuLtau$}
&\tikzmarknode{Y}{$\quad\mathcal{Z}_2\quad$}\\
\hline
\rowcolor{yellow!10}$\mu_{\mathtt{L}}\,,\nu_{\mu}$ &$1,~2,~-1/2$&$+2$& $+1$ \\
\rowcolor{yellow!10}$\mu_{\mathtt{R}}$ &$1,~1,~-1$&$+2$& $+1$\\
\rowcolor{red!10}$\tau_{\mathtt{L}}\,, \nu_{\tau}$ &$1,~2,~-1/2$&$-2$& $+1$ \\
\rowcolor{red!10}$\tau_{\mathtt{R}}$ &$1,~1,~-1$&$-2$& $+1$\\\hline
\rowcolor{green!10}$S$     & $1,~1,~0$&$+2$& $+1$   \\
\rowcolor{cyan!10}$\Phi$ & $1,~1,~0$&$-1$& $-1$   \\
\rowcolor{brown!10}$\zmuta$ & $1,~1,~0$&$~~0$& $+1$   \\\hline
\end{tabular}
\end{adjustbox}
\caption{Relevant fields and their charges under the SM gauge symmetry $\mathcal{G}_{\rm SM}$, $\rm U(1)_{L_\mu - L_\tau}$, and $\mathcal{Z}_2$ symmetries, respectively. Here, $\rm \mathcal{G}_{\rm SM} \equiv \text{SU}(3)_c \otimes \text{SU}(2)_L \otimes \text{U}(1)_Y$ denotes the SM gauge group.}
\label{tab:tab1}
\end{table}
The relevant fields and their corresponding charges are defined in~\autoref{tab:tab1}. The extended Lagrangian is given by:
\begin{gather}
\begin{split}
\mathcal{L}\supset\mathcal{L}^{\rm kin}(S,\Phi,\mathtt{Z}_{\mu\tau})+\mathcal{L}_{\mu\tau}-V(\Phi,S,H)-V(H,S)\,,
\end{split}
\label{eq:lag1}
\end{gather}
where, the complex scalar field $\Phi$ is composed of two real scalar fields, $\phi_1$ and $\phi_2$, defined as
\begin{align}
\Phi=\frac{\phi_1+i\phi_2}{\sqrt{2}}\,,
\end{align}
with the assumption that $\langle\Phi\rangle=0$, a residual symmetry is preserved, ensuring the stability of the lighter real scalar state, $\phi_1$ for $\mone<\mtwo$ or $\phi_2$ for $\mtwo<\mone$. In addition to this criterion, the stability of $\phi_2$ as a DM candidate requires $\tau_{\phi_1}<\tau_{\rm univ}$. However, if $\tau_{\phi_1}>\tau_{\rm univ}$\footnote{In this regime, both endothermic and exothermic inelastic scattering processes will be active.}, both DM states contribute to the total DM relic abundance~\cite{Gemmell:2026yaw}, detailed in \autoref{app:stability}. The kinetic terms can be expressed as,
\begin{align}
\mathcal{L}^{\rm kin}(S,\Phi,\mathtt{Z}_{\mu\tau})\,=\,-\frac{1}{4}(\mathtt{Z}_{\mu\tau})^{\alpha\beta}(\mathtt{Z}_{\mu\tau})_{\alpha\beta}+\left(\mathcal{D}_{\alpha}^{}S\right)^\ast\left(\mathcal{D}^{\alpha}S\right)+\left(\mathcal{D}_{\alpha}^{}\Phi\right)^\ast\left(\mathcal{D}^{\alpha}\Phi\right)\,,
\end{align}
where the $\mathtt{Z}_{\mu\tau}^{\alpha\beta}$ and $\mathcal{D}^{\alpha}_{}$, corresponds to $S$ and $\Phi$ respectively, are given by:
\begin{subequations}\begin{align}
\mathtt{Z}_{\mu\tau}^{\alpha\beta}&=\partial^{\alpha} \mathtt{Z}_{\mu\tau}^{\beta}-\partial^{\beta} \mathtt{Z}_{\mu\tau}^{\alpha}\,, \\
\mathcal{D}^{\alpha}_{}\Phi&=\partial^{\alpha}\Phi-iQ_{\mu\tau}^\Phi\gmuta \mathtt{Z}_{\mu\tau}^{\alpha}\Phi\,,\\
\mathcal{D}^{\alpha}_{}S&=\partial^{\alpha}S-iQ_{\mu\tau}^S\gmuta \mathtt{Z}_{\mu\tau}^{\alpha} S\,.
\end{align}\end{subequations}
The modified kinetic term for the second and third generation leptons in the presence of the $\ULmuLtau$ symmetry is given by:
\begin{align}\begin{split}
\mathcal{L}_{\mu\tau}=\gmuta\sum_{j=\mu,\tau}\left[Q_{\mu\tau}^j\mathtt{Z}_{\mu\tau}^\alpha\bar{\ell}^j_{\mathtt{L}}\gamma_{\alpha}^{}\ell^j_{\mathtt{L}}+Q_{\mu\tau}^j\mathtt{Z}_{\mu\tau}^\alpha\bar{\ell}^j_\mathtt{R}\gamma_{\alpha}^{}\ell^j_\mathtt{R}\right]\,.
\label{eq:kinetic_term}
\end{split}\end{align}
\noindent
The scalar potentials can be written as follows:
\begin{align}
V(H,S) = 
-\mu_{H}^2(H^{\dagger}_{}H)
-\mu_{S}^{2}(S^\ast S)
+\lambda_{H}(H^{\dagger}H)^{2}  
+\lambda_S\left(S^\ast S\right)^2
+\lambda_{SH}(S^\ast S)(H^{\dagger}H)\,,
\label{eq:VHS}
\end{align}
and
\begin{gather}
\begin{split}
V(\Phi,S,H)=&\mu_\Phi^2 (\Phi^\ast\Phi)+\lphi (\Phi^\ast\Phi)^2+\lphiS (\Phi^\ast\Phi)(S^\ast S)\\&+\lphiH (\Phi^\ast\Phi)(H^\dagger H)+\mu(\Phi^2S + {\rm h.c.})\,.
\end{split}
\label{eq:lag2}
\end{gather}
Here, we assume the associated couplings are real to preserve CP invariance, while the final term arises from the specific charge assignments of the BSM scalars.
\subsection*{Spontaneous Symmetry Breaking}
In this section, we discuss the spontaneous breaking of $\ULmuLtau$ and the SM electroweak $\rm \text{SU}(2)_L \otimes U(1)_Y$ gauge symmetries, with $\langle S \rangle \neq 0$, $\langle H \rangle \neq 0$, and $\langle \Phi \rangle = 0$. For the potential $V(H,S)$ given in Eq.~\eqref{eq:VHS}, we choose $\lambda_H, \lambda_S > 0$ and $\mu_H^2, \mu_S^2 < 0$, which ensure that the potential has a minimum corresponding to the following vacuum:
\begin{subequations}\begin{align}
H=
\begin{pmatrix}
H^+\\ \dfrac{v_{\rm EW}^{}+h+ i H^0}{\sqrt{2}}
\end{pmatrix}
\to 
\langle H\rangle=
\begin{pmatrix}
0\\ \dfrac{v_{\rm EW}^{}}{\sqrt{2}}
\end{pmatrix}\,;\\
S=\dfrac{1}{\sqrt{2}}\left(\vmuta+\mathcal{S}+i \mathcal{A}\right)
\to 
\langle S\rangle=\dfrac{1}{\sqrt{2}}\vmuta\,.
\end{align}\end{subequations}
Therefore, two scalar fields acquire non-zero VEVs, which break $\rm \mathcal{G}_{\rm SM} \otimes U(1)_{L_\mu - L_\tau} \to \text{SU}(3)_c \otimes \text{U}(1)_{\rm EM} \otimes \mathcal{Z}_2$. In the above, $H^{\pm,0}$ and $\mathcal{A}$ denote the Nambu-Goldstone bosons that disappear in the unitary gauge after the spontaneous breaking of the electroweak and $\ULmuLtau$ symmetries. Barring strong interactions, the remnant symmetry is $\rm \text{U}(1)_{\rm EM} \otimes \mathcal{Z}_2$. The $\ULmuLtau$ gauge boson $\zmuta$ acquires a mass through the spontaneous breaking of this symmetry, which is given by $\mzmuta = Q_{\mu\tau}^S \gmuta \vmuta$. Simultaneously, the scalar potential $V(\Phi,S,H)$ in \eq\eqref{eq:lag2}, after the spontaneous breaking of the $\ULmuLtau$ symmetry, becomes
\begin{eqnarray}
\label{eq:lag3}
\begin{split}
V(\phi_1,\phi_2,\mathcal{S},H)=
&\dfrac{1}{2}\left(\mu_\Phi^2+\dfrac{1}{2}\lphiS v_{\mu\tau}^2+\sqrt{2}\mu \vmuta\right) \phi_1^2
+\dfrac{1}{2}\left(\mu_\Phi^2+\dfrac{1}{2}\lphiS v_{\mu\tau}^2-\sqrt{2}\mu \vmuta\right) \phi_2^2\\
+&\dfrac{1}{4}\lphi (\phi_1^2+\phi_2^2)^2
+\dfrac{1}{4}\lphiS{\mathcal{S}}^2 (\phi_1^2+\phi_2^2)
+\dfrac{1}{2}\lphiS\vmuta \mathcal{S}(\phi_1^2+\phi_2^2)\\
+&\dfrac{1}{2}\lphiH (\phi_1^2+\phi_2^2)(H^\dagger H)+\dfrac{1}{2}\mu  \mathcal{S}(\phi_1^2-\phi_2^2)\,.
\end{split}
\end{eqnarray}
Consequently, the mass terms of $\phi_1$ and $\phi_2$ after electroweak symmetry breaking (EWSB) can be written as,
\begin{subequations}
\begin{gather}
\label{eq:mphi1}
\mone^2=\mu_\Phi^2+\dfrac{1}{2}\lphiS v_{\mu\tau}^2+\dfrac{1}{2}\lphiH v^2_{\rm EW}+\sqrt{2}\mu \vmuta\,,\\
\mtwo^2=\mu_\Phi^2+\dfrac{1}{2}\lphiS v_{\mu\tau}^2+\dfrac{1}{2}\lphiH v^2_{\rm EW}-\sqrt{2}\mu \vmuta\,.
\label{eq:mphi2}
\end{gather}
\end{subequations}
Following \eqs\eqref{eq:mphi1} and \eqref{eq:mphi2}, we can easily find out,
\begin{subequations}
\begin{gather}
\mu=\dfrac{m_{\phi_1}^2-m^2_{\phi_2}}{2\sqrt{2}\vmuta}\,,\\
2\mu^2_\Phi=(m_{\phi_1}^2+m^2_{\phi_2})-(\lphiS\ v_{\mu\tau}^2+\lphiH v^2_{\rm EW})\,.
\label{eq:muphi}
\end{gather}
\end{subequations}
Therefore, the DM mass splitting between $\phi_1$ and $\phi_2$ can be written as $\deltam=\mone-\mtwo$, which in the small splitting approximation is given by
\begin{gather}
\deltam=\dfrac{2\sqrt{2}\mu\vmuta}{\mone+\mtwo}\simeq \dfrac{\sqrt{2}\mu\vmuta}{\mtwo}\,.
\label{eq:deltam}
\end{gather}
The symmetry breaking associated with the extended scalar sector also provides an avenue for a first-order phase transition in the BSM framework, which has been explored in~\cite{Guo:2025qes,Hooper:2025fda}.
\subsection*{Scalar Mass Mixing}
Simultaneously, the potential term $V(H, S)$ (\eq\eqref{eq:VHS}), after EWSB, induces mixing between the scalar field $\mathcal{S}$ and the CP-even component of the SM Higgs field $h$, generating a non-diagonal mass term proportional to $h\mathcal{S}$. By imposing an orthogonal transformation with mixing angle $\alpha$, we obtain the physical basis $(h_1,~h_2)$, where one of the states corresponds to the observed SM Higgs boson of mass, $125.2$ GeV, while the other is a neutral BSM scalar. The original basis can be expressed in terms of the physical basis as follows:
\begin{gather}
h=h_2\sin\alpha+h_1\cos\alpha\,,\\
\mathcal{S}=h_2\cos\alpha-h_1\sin\alpha\,,
\end{gather}
where the mixing angle is determined by the relation
\begin{align}
\alpha= \frac{1}{2}\tan^{-1}{\left(\frac{2\lambda_{SH} v_{\rm EW} \vmuta}{\mu_S^2-\mu_H^2-2\lambda_H v^2_{\rm EW}+\dfrac{1}{2}\lambda_{SH} v^2_{\rm EW}}\right)}\,.
\end{align}
Analogously, the internal parameters $(\lambda_H,~\lambda_S,~\lambda_{SH}, \mu_S,~\mu_H)$ are readily expressed in terms of the external model parameters $(\sin\alpha,~m_{h_1},~m_{h_2})$ as follows:
\begin{subequations}\begin{gather}
\lambda_H=\dfrac{1}{2v^2_{\rm EW}}(m_{h_1}^2\cos^2\alpha+m_{h_2}^2\sin^2\alpha)\,,\\
\lambda_{S}=\dfrac{1}{2v_{\mu\tau}^2}(m_{h_2}^2\cos^2\alpha+m_{h_1}^2\sin^2\alpha)\,,\\
\lambda_{SH}=\dfrac{\sin2\alpha}{2\vmuta v_{\rm EW}}(m_{h_2}^2-m_{h_1}^2)\,,\\
\mu_H^2=\lambda_H v^2_{\rm EW}+\dfrac{1}{2}\lambda_{SH} v_{\mu\tau}^2\,,\\
\mu_S^2=\lambda_{S} v_{\mu\tau}^2+\dfrac{1}{2}\lambda_{SH} v^2_{\rm EW}\,.
\end{gather}\end{subequations}
\subsection*{Gauge Kinetic Mixing}
\begin{figure}[htb!]
\centering
\begin{tikzpicture}
\begin{feynman}
\vertex(b){\(\color{black}{\gamma/Z }\)};
\vertex[right=2cm of b] (c);
\vertex[right=2cm of c] (d);
\vertex[right=2cm of d] (e){\(\color{black}{\zmuta}\)};
\diagram*{
(b) -- [line width=0.25mm, boson, arrow size=1.2pt, style=black,ultra thick] (c),
(c) -- [line width=0.25mm, fermion, half left, arrow size=1.5pt, style=black,ultra thick, edge label'={\(\color{black}{\ell}\)}] (d),
(d) -- [line width=0.25mm, fermion, half left,arrow size=1.5pt, style=black,ultra thick, edge label'={\(\color{black}{\ell}\)}] (c),
(d) -- [line width=0.25mm, boson, arrow size=1.2pt, style=gray, ultra thick] (e)};
\node at (c)[circle,fill,style=black,inner sep=1pt]{};
\node at (d)[circle,fill,style=black,inner sep=1pt]{};
\end{feynman}
\end{tikzpicture}
\caption{One-loop Feynman diagrams associated with radiatively generated kinetic mixing between SM neutral gauge bosons ($\gamma/Z$) and $\zmuta$. Here, $\ell=\mu,$ and $\tau$.}
\label{feyn:zmutau}
\end{figure}
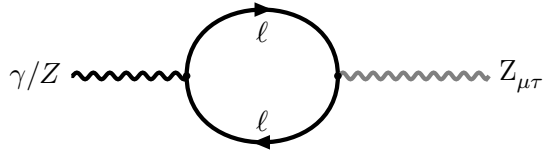

In this framework, we consider a minimal $\ULmuLtau$ scenario in which tree-level kinetic mixing of the $\zmuta$ gauge boson with the SM gauge bosons is absent. However, a non-zero kinetic mixing is radiatively generated at the one-loop level~\cite{Hapitas:2021ilr}, mediated by the $\mu$ and $\tau$ leptons carrying the $\ULmuLtau$ charge, as illustrated in~\autoref{feyn:zmutau}.

An effective one-loop Lagrangian describing the non-zero kinetic mixing among the field strength tensors of the Standard Model photon ($F_{\alpha\beta}$), the $\mathtt{Z}$ boson ($\mathtt{Z}_{\alpha\beta}$), and the $\mathtt{Z}_{\mu\tau}$ gauge boson ($Z^{\alpha\beta}_{\mu\tau}$) can be expressed as \cite{Bauer:2022nwt, Babu:1997st},
\begin{align}
\mathcal{L}_{\epsilon}=-\dfrac{1}{2}\epsilon_{\gamma\zmuta}^{}F_{\alpha\beta}{Z}_{\mu\tau}^{\alpha\beta}-\dfrac{1}{2}\epsilon_{Z\zmuta}^{}{\mathtt{Z}}_{\alpha\beta}{Z}_{\mu\tau}^{\alpha\beta}\,,
\label{eq:kin-mix}
\end{align}
where $F_{\alpha\beta}=\partial_\alpha A_\beta-\partial_\beta A_\alpha\,,\text{ and }\mathtt{Z}_{\alpha\beta}=\partial_\alpha \mathtt{Z}_\beta-\partial_\beta \mathtt{Z}_\alpha$. Following \cite{Amaral:2020tga,Foldenauer:2024cdp}, we obtain the mixing parameters $\epsilon_{\gamma\zmuta}^{}$ and $\epsilon_{Z\zmuta}^{}$,
\begin{align}
\epsilon_{\gamma\zmuta}^{}\simeq \dfrac{e g_{\mu\tau}}{12\pi^2}\ln\left(\dfrac{m_{\tau}^2}{m_{\mu}^2}\right)\,,
\quad\text{and}\quad
\epsilon_{Z\zmuta}^{}\simeq \dfrac{g_V^{\ell} g_{\mu\tau}}{12\pi^2}\ln\left(\dfrac{m_{\tau}^2}{m_{\mu}^2}\right)\,.
\label{eq:loop-mixing}
\end{align}
Here, $g_{V}^{\ell}=\dfrac{g}{4c_W}(4s_W^2-1)=\left(3m_\mathtt{Z}^2-4m_W^2\right)/\left(2v_{\rm EW}m_\mathtt{Z}\right)$, where $g=e/s_W$ denotes the $\mathrm{SU(2)_L}$ gauge coupling constant and $s_W$ is the sine of the Weinberg angle. The masses of the SM $Z$ and $W$ bosons are denoted by $m_\mathtt{Z}$ and $m_W$, respectively. The radiatively generated kinetic-mixing terms in \eq\eqref{eq:kin-mix} can then be diagonalized by a non-unitary transformation \cite{Hamaguchi:2020djq}:
\begin{align}
\begin{pmatrix}
A_{}^{\alpha}\\\mathtt{Z}_{}^{\alpha}\\\mathtt{Z}^{\alpha}_{\mu\tau}
\end{pmatrix}
\to
\begin{pmatrix}
\;1\qquad &0\qquad&-\epsilon_{\gamma \zmuta}^{}\delta_{\B\zmuta}\;\\
\;0\qquad&1\qquad&-{\epsilon_{Z \zmuta}^{}}{\delta_{\B\zmuta}}\;\\
\;0\qquad&0\qquad&\delta_{\B\zmuta}\;\\
\end{pmatrix}
\begin{pmatrix}
A_{}^{\alpha}\\\mathtt{Z}_{}^{\alpha}\\\mathtt{Z}^{\alpha}_{\mu\tau}
\end{pmatrix}\,.
\end{align}
For notation simplification, we have redefined the parameter as:
$\delta_{\B\zmuta}^2=\left(1-\epsilon_{\gamma\zmuta}^2-\epsilon_{Z\zmuta}^2\right)^{-1}$. Moreover, the gauge kinetic mixing induces a non-diagonal mass term in the Lagrangian,
\begin{equation}
\delta m^{}_{Z\zmuta}\mathtt{Z}_\alpha \mathtt{Z}_{\mu\tau}^\alpha=-\delta_{\B\zmuta}\epsZ \dfrac{e^2 v^2_{\rm EW}}{4c_w^2s_w^2}\mathtt{Z}_\alpha \mathtt{Z}_{\mu\tau}^\alpha=-\delta_{\B\zmuta}\epsZ m_Z^2 \mathtt{Z}_\alpha \mathtt{Z}_{\mu\tau}^\alpha\,.
\end{equation}
The corresponding mass matrix is,
\begin{equation}
\rm M^2=
\begin{pmatrix}
m_\mathtt{Z}^2&\delta m^{}_{Z\zmuta}\\
\delta m^{}_{Z\zmuta}& \quad (m_{\zmuta}^2+m_\mathtt{Z}^2\epsilon_{Z\zmuta}^2)\delta_{\B\zmuta}^2
\end{pmatrix},
\end{equation}
which can subsequently be diagonalized using an orthogonal matrix transformation,
\begin{equation}
R(\vartheta)=\begin{pmatrix}
\cos\vartheta&\sin\vartheta\\
-\sin\vartheta&\cos\vartheta
\end{pmatrix}\,,
\end{equation}
and 
$\begin{pmatrix}
\mathtt{Z}^{\alpha}\\
\mathtt{Z}_{\mu\tau}^{\alpha}
\end{pmatrix} 
= R(\vartheta)\begin{pmatrix}
Z^{\alpha}\\
Z_{\mu\tau}^{\alpha}
\end{pmatrix}\,,$
where
\begin{align}
\vartheta= \frac{1}{2} \tan^{-1}{\left(\frac{2\ \epsZ m_\mathtt{Z}^2\delta_{\B\zmuta}}{m_\mathtt{Z}^2-m^2_{\zmuta}\delta_{\B\zmuta}^2}\right)}\,.
\label{eq:tan2theta}
\end{align}

\noindent Following this, we rewrite the fields in their mass basis as follows:
\begin{subequations}
\begin{gather}
\mathtt{Z}^\alpha_{}= \cos\vartheta\ Z^{\alpha}+\sin\vartheta\ Z^{\alpha}_{\mu\tau}\,,\\ 
\mathtt{Z}^\alpha_{\mu\tau}= \cos\vartheta\ Z^{\alpha}_{\mu\tau} -\sin\vartheta\ Z^{\alpha}\,.
\end{gather}
\end{subequations}
\subsection*{Theoretical Constraints}
In order to ensure a stable ground state, the quartic couplings appearing in the Lagrangian must satisfy the following conditions \cite{Biswas:2016yan}:
\begin{gather}
\lambda_i>0\,,\quad\lambda_{HS}+2\sqrt{\lambda_H\lambda_S}>0\,,\quad\lambda_{H\Phi}+2\sqrt{\lambda_H\lambda_\Phi}>0\,,\quad\lambda_{S\Phi}+2\sqrt{\lambda_S\lambda_\Phi}>0\,,\\
\begin{split}
\sqrt{\lambda_{SH}+2\sqrt{\lambda_S\lambda_H}}
\sqrt{\lambda_{\Phi H}+2\sqrt{\lambda_\Phi\lambda_H}}
\sqrt{\lambda_{S\Phi}+2\sqrt{\lambda_S\lambda_\Phi}}\\
+2\sqrt{\lambda_S\lambda_\Phi\lambda_H}+\lambda_S\sqrt{\lambda_H}
+\lambda_S\sqrt{\lambda_\Phi}+\lambda_H\sqrt{\lambda_\Phi}>0\,,
\end{split}
\end{gather}
where $i=\{S,~H,~\Phi\}$. Additionally, upper bounds on the quartic and gauge couplings can be derived by requiring the theory to remain perturbative, which is given by,
\begin{align}
\lambda_i<4\pi\text{ and } \gmuta<\sqrt{4\pi}\,.
\end{align}
Further theoretical constraints on these couplings arise from imposing perturbative unitarity on the scattering amplitudes; see Ref\,.~\cite{Bhattacharya:2024nla} for details. For our phenomenological analysis, we take the following parameters as external inputs:
\begin{align}
\{\mtwo\,,~\deltam\,,~\mzmuta\,,~\gmuta\,,~\mhtwo\,,~\lphiH\,,~\lphiS\,,~\salpha\}\,.
\end{align}
Before delving into the detailed analysis, we first restrict the parameter space to ensure that the model satisfies the relevant theoretical constraints. For a given choice of the external parameters $(\mtwo,~\mu,~\vmuta,~\lphiH,~\lphiS)$, there may exist regions of the parameter space where $\mu_\Phi^2$ becomes negative, which follows from \eq\eqref{eq:deltam}.
\begin{figure}[htb!]
\centering
\includegraphics[width=0.7\linewidth]{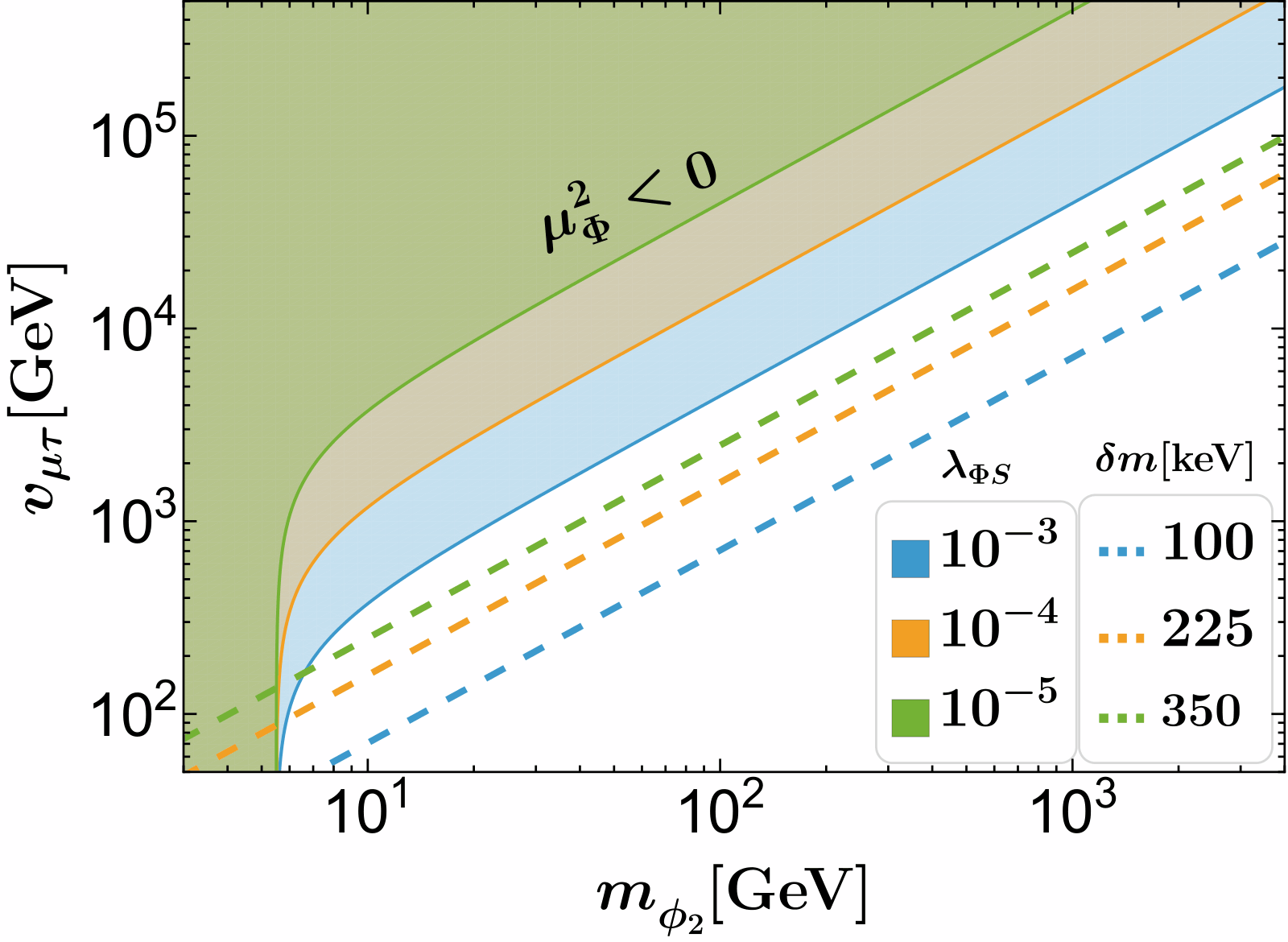}
\caption{The thick dashed lines indicate the points satisfying the required splitting in the $\mtwo-\vmuta$ plane for three different values of $\deltam$ and $\mu=10~\keV$. The shaded regions are excluded by $\mu_\Phi^2<0$ for three different choices of $\lphiS$, while the tiny splitting $\deltam$ and $\lphiH$ leave the shaded region nearly invariant.} 
\label{fig:mass-splitting}
\end{figure}
Although such parameter choices may remain within the theoretical bounds, we impose the additional requirement $\mu_\Phi^2>0$, as shown in \autoref{fig:mass-splitting}. Guided by this consideration, we choose $\lambda_{\Phi S}\approx 10^{-3}-10^{-4}$ throughout the analysis. The resulting reduced set of free parameters is then given by
\begin{align}
\{\mtwo\,,~\deltam\,,~\mzmuta\,,~\gmuta\,,~\mhtwo\,,~\salpha\,,~\lphiH\}\,,
\label{eq:freepara}
\end{align}
which we vary in the subsequent phenomenological study.
\subsection*{Low-Energy Constraints}
\label{sec:constraint}
The tree-level couplings of the $\zmuta$ boson to the muon, tau, and their corresponding neutrinos make the $\ULmuLtau$ model subject to constraints from precision measurements involving these leptons. One such constraint arises from the anomalous magnetic moments, $(g-2)$, of the muon and tau. Since the experimental precision of $(g-2)_\tau$ is approximately $\mathcal{O}(10^5)$ less precise than that of $(g-2)_\mu$, the latter provides a significantly stronger constraint on $\mzmuta$ and the associated coupling $\gmuta$. Other important constraints arise from neutrino trident production, $B_s-\bar{B}_s$ oscillations and solar capture of DM.

\paragraph{Muon $\boldsymbol{(g-2)}$} In addition to the SM contribution, the anomalous magnetic moment $(g-2)_{\ell}$, with $\ell=\mu\text{ and }\tau$, receives an additional one-loop radiative contribution from $\zmuta$ exchange. This contribution can be expressed as~\cite{Baek:2001kca, Pospelov:2008zw, Altmannshofer:2016oaq, Heeck:2011wj}:
\begin{align}
\Delta a_\ell=\dfrac{(Q^\ell_{\mu\tau}\gmuta)^2}{4\pi^2}\int_0^1 \dfrac{x^2(1-x)}{x^2+(1-x)(m^2_\zmuta/m^2_\ell)}\,.
\end{align}
The current limit on $\Delta a_{\mu}$ is $38(63) \times 10^{-11}$, which is a $0.6\sigma$ discrepancy \cite{Muong-2:2025xyk,Muong-2:2026qnz,Aliberti:2025beg}. The corresponding limits on $\gmuta$ are shown in \autoref{fig:scan1}.

\paragraph{Neutrino Trident} The neutrino-nucleon scattering process $\nu_\mu N \to \nu_\mu \mu^+ \mu^- N$ is commonly referred to as neutrino trident production, owing to the final-state configuration consisting of a neutrino and a charged lepton-antilepton pair. Such processes can be mediated by a new gauge boson $\zmuta$ in models with $\ULmuLtau$ flavor symmetry. The measurement of trident cross-sections at experiments like CCFR (DUNE) provides a robust lower bound on the ratio $\mzmuta/(Q^{\ell}_{\mu\tau}\gmuta)$ \cite{Altmannshofer:2014pba}  (\cite{Altmannshofer:2019zhy}), structured as follows: 
\begin{gather}
4.9\,\TeV \gtrsim \dfrac{\mzmuta}{Q^{\ell}_{\mu\tau}\gmuta} \gtrsim 540~(500)\,\GeV\,,
\qquad \text{with } \mzmuta \gtrsim 10\,\GeV\,.
\end{gather}
The upper bound corresponds to the $B_s^0 - \bar{B}_s^0$ mixing constraints if the model is required to account for the $B_s$-meson anomalies. The resulting CCFR bounds and complementary exclusion regions are illustrated in \autoref{fig:scan1}.

\paragraph{Solar Capture} DM can be captured by the Sun via elastic and inelastic scattering with solar nuclei and subsequently annihilate in the core to produce neutrinos detectable by IceCube. Owing to their distinct energy spectra relative to the atmospheric background, these neutrinos provide a probe of DM annihilation, though null results constrain the parameter space. In our framework, while the solar capture rate is substantial, the resulting neutrino event rate is insufficient to probe the parameter space of interest. Consequently, the IceCube constraints are omitted. The corresponding calculations are detailed in~\autoref{sec:IceCube}.
\section{Relic Density}
\label{sec:relic}
\begin{figure}[htb!]
\centering
\begin{adjustbox}{width=1.0\textwidth, keepaspectratio}
\subfloat[]{
\begin{tikzpicture}
\begin{feynman}
\vertex (a);
\vertex[above left=0.45cm and 0.9cm of a] (a1){\(\phi_i\)};
\vertex[above right=0.45cm and 0.9cm  of a] (a2){\(h_{1,2}\)}; 
\vertex[below = 0.9cm of a] (b); 
\vertex[below left=0.45cm and 0.9cm of b] (c1){\(\phi_i\)};
\vertex[below right=0.45cm and 0.9cm of b] (c2){\(h_{1,2}\)};
\diagram*{
(a1) -- [line width=0.35mm, scalar, arrow size=0.8pt,style=black] (a),
(a2) -- [line width=0.35mm, scalar, arrow size=0.8pt,style=black] (a),
(a) -- [line width=0.35mm,  scalar, arrow size=0.8pt, edge label={\(\color{black}{\phi_i}\)}, style=black!50] (b),
(b) -- [line width=0.35mm, scalar, arrow size=0.8pt] (c1),
(b) -- [line width=0.35mm,scalar, arrow size=0.8pt] (c2)};
\node at (a)[circle,fill,style=black,inner sep=1pt]{};
\node at (b)[circle,fill,style=black,inner sep=1pt]{};
\end{feynman}
\end{tikzpicture}\label{fig:feyn01}}
\subfloat[]{
\begin{tikzpicture}
\begin{feynman}
\vertex (a);
\vertex[above left=0.45cm and 0.9cm of a] (a1){\(\phi_i\)};
\vertex[above right=0.45cm and 0.9cm  of a] (a2){\(V\)}; 
\vertex[below = 0.9cm of a] (b); 
\vertex[below left=0.45cm and 0.9cm of b] (c1){\(\phi_i\)};
\vertex[below right=0.45cm and 0.9cm of b] (c2){\(V\)};
\diagram*{
(a1) -- [line width=0.35mm, scalar, arrow size=0.8pt,style=black] (a),
(a2) -- [line width=0.35mm, boson, arrow size=0.8pt,style=black] (a),
(a) -- [line width=0.35mm,  scalar, arrow size=0.8pt, edge label={\(\color{black}{\phi_j}\)}, style=black!50] (b),
(b) -- [line width=0.35mm, scalar, arrow size=0.8pt] (c1),
(b) -- [line width=0.35mm,boson, arrow size=0.8pt] (c2)};
\node at (a)[circle,fill,style=black,inner sep=1pt]{};
\node at (b)[circle,fill,style=black,inner sep=1pt]{};
\end{feynman}
\end{tikzpicture}\label{fig:feyn02}}
\subfloat[]{
\begin{tikzpicture}
\begin{feynman}
\vertex (a);
\vertex[above left=0.9cm and 0.9cm of a] (a1){\(\phi_i\)};
\vertex[above right=0.9cm and 0.9cm  of a] (a2){\(h_{1,2}\)}; 
\vertex[below left=0.9cm and 0.9cm of a] (c1){\(\phi_i\)};
\vertex[below right=0.9cm and 0.9cm of a] (c2){\(h_{1,2}\)};
\diagram*{
(a1) -- [line width=0.35mm, scalar, arrow size=0.8pt,style=black] (a),
(a2) -- [line width=0.35mm, scalar, arrow size=0.8pt,style=black] (a),
(a) -- [line width=0.35mm, scalar, arrow size=0.8pt] (c1),
(a) -- [line width=0.35mm, scalar, arrow size=0.8pt] (c2)};
\node at (a)[circle,fill,style=black,inner sep=1pt]{};
\end{feynman}
\end{tikzpicture}\label{fig:feyn03}}
\subfloat[]{
\begin{tikzpicture}
\begin{feynman}
\vertex (a);
\vertex[above left=0.9cm and 0.9cm of a] (a1){\(\phi_i\)};
\vertex[above right=0.9cm and 0.9cm  of a] (a2){\(V\)}; 
\vertex[below left=0.9cm and 0.9cm of a] (c1){\(\phi_i\)};
\vertex[below right=0.9cm and 0.9cm of a] (c2){\(V\)};
\diagram*{
(a1) -- [line width=0.35mm, scalar, arrow size=0.8pt,style=black] (a),
(a2) -- [line width=0.35mm, boson, arrow size=0.8pt,style=black] (a),
(a) -- [line width=0.35mm, scalar, arrow size=0.8pt] (c1),
(a) -- [line width=0.35mm, boson, arrow size=0.8pt] (c2)};
\node at (a)[circle,fill,style=black,inner sep=1pt]{};
\end{feynman}
\end{tikzpicture}\label{fig:feyn04}}
\end{adjustbox}

\begin{adjustbox}{width=1.0\textwidth, keepaspectratio}
\subfloat[]{\begin{tikzpicture}
\begin{feynman}
\vertex (a);
\vertex[above left=0.9cm and 0.9cm  of a] (a1){\(\phi_i\)};
\vertex[below left=0.9cm and 0.9cm  of a] (a2){\(\phi_i\)}; 
\vertex[right=0.9cm of a] (b); 
\vertex[above right=0.9cm and 0.9cm of b] (c1){\(V/W^+\)};
\vertex[below right=0.9cm and 0.9cm of b] (c2){\(V/W^-\)};
\diagram*{
(a1) -- [line width=0.35mm,  scalar, arrow size=0.8pt,style=black] (a),
(a) -- [line width=0.35mm, scalar, arrow size=0.8pt,style=black] (a2),
(a) -- [line width=0.35mm, scalar,arrow size=0.8pt, edge label={\(\color{black}{h_{1,2}}\)}, style=black!50] (b),
(b) -- [line width=0.35mm, boson, arrow size=0.8pt] (c1),
(b) -- [line width=0.35mm, boson, arrow size=0.8pt] (c2)};
\node at (a)[circle,fill,style=black,inner sep=1pt]{};
\node at (b)[circle,fill,style=black,inner sep=1pt]{};
\end{feynman}
\end{tikzpicture}\label{fig:feyn05}}
\subfloat[]{\begin{tikzpicture}
\begin{feynman}
\vertex (a);
\vertex[above left=0.9cm and 0.9cm  of a] (a1){\(\phi_i\)};
\vertex[below left=0.9cm and 0.9cm  of a] (a2){\(\phi_i\)}; 
\vertex[right=0.9cm of a] (b); 
\vertex[above right=0.9cm and 0.9cm of b] (c1){\(\ell/q\)};
\vertex[below right=0.9cm and 0.9cm of b] (c2){\(\bar\ell/\bar{q}\)};
\diagram*{
(a1) -- [line width=0.35mm,  scalar, arrow size=0.8pt,style=black] (a),
(a) -- [line width=0.35mm, scalar, arrow size=0.8pt,style=black] (a2),
(a) -- [line width=0.35mm, scalar,arrow size=0.8pt, edge label={\(\color{black}{h_{1,2}}\)}, style=black!50] (b),
(b) -- [line width=0.35mm, fermion, arrow size=0.8pt] (c1),
(c2) -- [line width=0.35mm, fermion, arrow size=0.8pt] (b)};
\node at (a)[circle,fill,style=black,inner sep=1pt]{};
\node at (b)[circle,fill,style=black,inner sep=1pt]{};
\end{feynman}
\end{tikzpicture}\label{fig:feyn06}}
\subfloat[]{\begin{tikzpicture}
\begin{feynman}
\vertex (a);
\vertex[above left=0.9cm and 0.9cm  of a] (a1){\(\phi_i\)};
\vertex[below left=0.9cm and 0.9cm  of a] (a2){\(\phi_i\)}; 
\vertex[right=0.9cm of a] (b); 
\vertex[above right=0.9cm and 0.9cm of b] (c1){\(h_{1,2}\)};
\vertex[below right=0.9cm and 0.9cm of b] (c2){\(h_{1,2}\)};
\diagram*{
(a1) -- [line width=0.35mm,  scalar, arrow size=0.8pt,style=black] (a),
(a) -- [line width=0.35mm, scalar, arrow size=0.8pt,style=black] (a2),
(a) -- [line width=0.35mm, scalar,arrow size=0.8pt, edge label={\(\color{black}{h_{1,2}}\)}, style=black!50] (b),
(b) -- [line width=0.35mm, scalar, arrow size=0.8pt] (c1),
(c2) -- [line width=0.35mm, scalar, arrow size=0.8pt] (b)};
\node at (a)[circle,fill,style=black,inner sep=1pt]{};
\node at (b)[circle,fill,style=black,inner sep=1pt]{};
\end{feynman}
\end{tikzpicture}\label{fig:feyn07}}
\end{adjustbox}

\begin{adjustbox}{width=1.0\textwidth, keepaspectratio}
\subfloat[]{
\begin{tikzpicture}
\begin{feynman}
\vertex (a);
\vertex[above left=0.45cm and 0.9cm of a] (a1){\(\phi_i\)};
\vertex[above right=0.45cm and 0.9cm  of a] (a2){\(h_{1,2}\)}; 
\vertex[below = 0.9cm of a] (b); 
\vertex[below left=0.45cm and 0.9cm of b] (c1){\(\phi_j\)};
\vertex[below right=0.45cm and 0.9cm of b] (c2){\(V\)};
\diagram*{
(a1) -- [line width=0.35mm, scalar, arrow size=0.8pt,style=black] (a),
(a2) -- [line width=0.35mm, scalar, arrow size=0.8pt,style=black] (a),
(a) -- [line width=0.35mm,  scalar, arrow size=0.8pt, edge label={\(\color{black}{\phi_i}\)}, style=black!50] (b),
(b) -- [line width=0.35mm, scalar, arrow size=0.8pt] (c1),
(b) -- [line width=0.35mm,boson, arrow size=0.8pt] (c2)};
\node at (a)[circle,fill,style=black,inner sep=1pt]{};
\node at (b)[circle,fill,style=black,inner sep=1pt]{};
\end{feynman}
\end{tikzpicture}\label{fig:feyn08}}
\subfloat[]{
\begin{tikzpicture}
\begin{feynman}
\vertex (a);
\vertex[above left=0.45cm and 0.9cm of a] (a1){\(\phi_i\)};
\vertex[above right=0.45cm and 0.9cm  of a] (a2){\(V\)}; 
\vertex[below = 0.9cm of a] (b); 
\vertex[below left=0.45cm and 0.9cm of b] (c1){\(\phi_j\)};
\vertex[below right=0.45cm and 0.9cm of b] (c2){\(h_{1,2}\)};
\diagram*{
(a1) -- [line width=0.35mm, scalar, arrow size=0.8pt,style=black] (a),
(a2) -- [line width=0.35mm, boson, arrow size=0.8pt,style=black] (a),
(a) -- [line width=0.35mm,  scalar, arrow size=0.8pt, edge label={\(\color{black}{\phi_j}\)}, style=black!50] (b),
(b) -- [line width=0.35mm, scalar, arrow size=0.8pt] (c1),
(b) -- [line width=0.35mm,scalar, arrow size=0.8pt] (c2)};
\node at (a)[circle,fill,style=black,inner sep=1pt]{};
\node at (b)[circle,fill,style=black,inner sep=1pt]{};
\end{feynman}
\end{tikzpicture}\label{fig:feyn09}}
\subfloat[]{\begin{tikzpicture}
\begin{feynman}
\vertex (a);
\vertex[above left=0.9cm and 0.9cm  of a] (a1){\(\phi_i\)};
\vertex[below left=0.9cm and 0.9cm  of a] (a2){\(\phi_j\)}; 
\vertex[right=0.9cm of a] (b); 
\vertex[above right=0.9cm and 0.9cm of b] (c1){\(h_{1,2}\)};
\vertex[below right=0.9cm and 0.9cm of b] (c2){\(V\)};
\diagram*{
(a1) -- [line width=0.35mm,  scalar, arrow size=0.8pt,style=black] (a),
(a) -- [line width=0.35mm, scalar, arrow size=0.8pt,style=black] (a2),
(a) -- [line width=0.35mm, boson,arrow size=0.8pt, edge label={\(\color{black}{V}\)}, style=black!50] (b),
(b) -- [line width=0.35mm, scalar, arrow size=0.8pt] (c1),
(c2) -- [line width=0.35mm, boson, arrow size=0.8pt] (b)};
\node at (a)[circle,fill,style=black,inner sep=1pt]{};
\node at (b)[circle,fill,style=black,inner sep=1pt]{};
\end{feynman}
\end{tikzpicture}\label{fig:feyn010}}
\subfloat[]{\begin{tikzpicture}
\begin{feynman}
\vertex (a);
\vertex[above left=0.9cm and 0.9cm  of a] (a1){\(\phi_i\)};
\vertex[below left=0.9cm and 0.9cm  of a] (a2){\(\phi_j\)}; 
\vertex[right=0.9cm of a] (b); 
\vertex[above right=0.9cm and 0.9cm of b] (c1){\(f\)};
\vertex[below right=0.9cm and 0.9cm of b] (c2){\(\bar f\)};
\diagram*{
(a1) -- [line width=0.35mm,  scalar, arrow size=0.8pt,style = black] (a),
(a) -- [line width=0.35mm, scalar, arrow size=0.8pt,style = black] (a2),
(a) -- [line width=0.35mm, boson,arrow size=0.8pt, edge label = {\(\color{black}{X}\)}, style=black!50] (b),
(b) -- [line width=0.35mm, fermion, arrow size=0.8pt] (c1),
(c2) -- [line width=0.35mm, fermion, arrow size=0.8pt] (b)};
\node at (a)[circle,fill,style=black,inner sep=1pt]{};
\node at (b)[circle,fill,style=black,inner sep=1pt]{};
\end{feynman}
\end{tikzpicture}
\label{fig:feyn011}}
\end{adjustbox}
\caption{Feynman diagrams relevant to DM thermal freeze-out, where $f=\{q,~\ell,~\nu_\ell\}$, $V=\{Z,~\zmuta\}$, and $i,j=\{1,2\}$ ($i\neq j$), with $q$, $\ell$, and $\nu_\ell$ denoting the SM quarks, leptons, and neutrinos, respectively.}
\label{fig:feynman-relic1}
\end{figure}
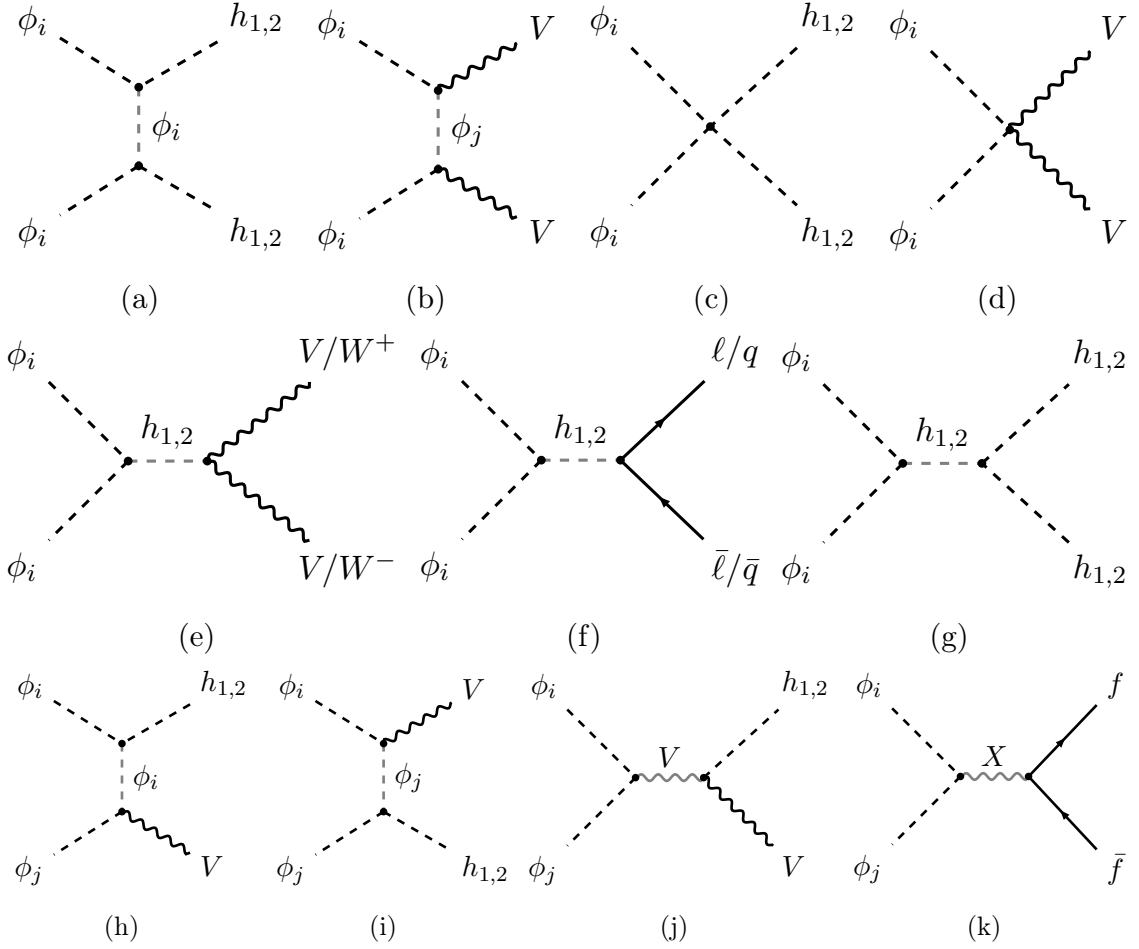

The observed relic abundance of DM in the present Universe is precisely measured by cosmological observations, with the \texttt{Planck} data giving $\Omega_{\rm DM}h^2 \simeq 0.12 \pm 0.0012$ \cite{Planck:2018vyg}. One of the most popular thermal production mechanisms for explaining this abundance of DM is the weakly interacting massive particle (WIMP) paradigm, in which DM was initially in thermal equilibrium with the Standard Model thermal bath and subsequently froze out as the Universe expanded and interactions could no longer maintain thermal equilibrium. In the standard WIMP scenario, the relic density is determined primarily by the thermally averaged DM annihilation cross section, $\langle\sigma v\rangle$. In our model, we therefore investigate the parameter space in which the DM candidate achieves the observed relic density via thermal freeze-out. The Feynman diagrams for DM annihilation, semi-annihilation, and co-annihilation channels are shown in~\autoref{fig:feynman-relic1}.

Assuming that the heavier dark-sector particle has a lifetime shorter than the age of the Universe, only the lighter state remains stable and accounts for the total present-day DM relic density. The present-day DM number density can therefore be determined by solving the coupled Boltzmann equations (BEQs),
\begin{align}
\frac{dn_{\phi_i}}{dt}+3\mathcal{H}n_{\phi_i}
={}&-\langle\sigma v\rangle_{\phi_i\phi_i}
\left[n_{\phi_i}^2-(n_{\phi_i}^{\rm eq})^2\right]\nonumber\\
&-\langle\sigma v\rangle_{\phi_1\phi_2}
\left[n_{\phi_1}n_{\phi_2}
-n_{\phi_1}^{\rm eq}n_{\phi_2}^{\rm eq}\right]
+\mathcal{C}_i^{\rm dec}
+\mathcal{C}_i^{\rm conv},
\label{eq:relic1}
\end{align}
where $i=1,2$; $\mathcal{H}$ is the Hubble expansion rate, $n_{\phi_i}$ and $n_{\phi_i}^{\rm eq}$ denote the number density and equilibrium number density of the $i^{\rm th}$ dark-sector particle, respectively, and $\langle\sigma v\rangle_{ij}$ is the thermally averaged annihilation or coannihilation cross section. The terms $\mathcal{C}_i^{\rm dec}$ and $\mathcal{C}_i^{\rm conv}$ account for the decay and DM-conversion processes, respectively. The total DM number density is given by
\begin{equation}
n_\phi\equiv n_{\phi_1}+n_{\phi_2}.
\end{equation}

Alternatively, when the two dark-sector states remain in chemical equilibrium, and the effects of explicit DM-conversion processes can be incorporated into the coannihilation treatment \cite{Edsjo:1997bg}, the coupled BEQs can be reduced to a single BEQ for the total DM number density. In this case, the relevant annihilation and coannihilation channels (see \autoref{fig:feynman-relic1}) are described by an effective thermally averaged cross section ($\langle\sigma v\rangle^{\rm eff}$),
\begin{equation}
\frac{dn_{\phi}}{dt}+3\mathcal{H}n_{\phi}
=-\langle\sigma v\rangle^{\rm eff}
\left[n_{\phi}^2-(n_{\phi}^{\rm eq})^2\right],
\label{eq:relic2}
\end{equation}
where
\begin{equation}
\langle\sigma v\rangle^{\rm eff}
=\sum_{i,j=1,2}
\langle\sigma v\rangle_{ij}
\frac{n_{\phi_i}^{\rm eq}n_{\phi_j}^{\rm eq}}
{(n_\phi^{\rm eq})^2}\,.
\label{eq:coannihillation}
\end{equation}
For approximately degenerate states with $\deltam\ll T_{\rm FO}$ and $g_{\phi_1}=g_{\phi_2}$, the DM equilibrium number density becomes
$n_{\phi_1}^{\rm eq}\simeq n_{\phi_2}^{\rm eq}\simeq n_\phi^{\rm eq}/2$, and consequently the \Eq\eqref{eq:coannihillation} reduces to
\begin{equation}
\langle\sigma v\rangle^{\rm eff}
\simeq\frac{1}{4}
\left[
\langle\sigma v\rangle_{\phi_1\phi_1}
+\langle\sigma v\rangle_{\phi_2\phi_2}
+2\langle\sigma v\rangle_{\phi_1\phi_2}
\right].
\end{equation}
However, here we do not calculate it analytically; instead, we implement the model using \texttt{FeynRules} \cite{Alloul:2013bka}, and the resulting model files are fed to \texttt{micrOMEGAs6} \cite{Alguero:2023zol} to determine the parameter space consistent with the observed DM relic density. Finally, the DM relic density is governed by
\begin{align}
\Omega_{\rm DM}h^2 = 2.744 \times 10^8 m_{\two}(Y_{\phi}/s)_{T=T_0}\,,
\end{align}
where $Y_\phi=n_\phi/s$ and $T_0=2.35\times 10^{-13}~\GeV$ is the present-day temperature of the Universe. The results, representing the relic-density-allowed points, are shown in \autoref{fig:scan1}, \ref{fig:scan2}, and \ref{fig:scan3} for different benchmark choices in subsequent sections. However, the analysis in this section varies only $\mtwo,~\mzmuta,~\deltam,$ and $\sin\alpha$, while keeping the remaining parameters fixed to the values indicated in the figure caption.
\section{Direct Detection}
\label{sec:DD}
Currently, there are three main approaches to probing DM: direct detection (DD), indirect detection (ID), and collider searches. Among these, DD is a leading method for exploring specific DM mass regimes, as it relies on elastic or inelastic interactions between DM particles and the visible sector to produce observable recoil signals. Depending on the detector technology and sensitivity, a wide range of DM masses can be investigated. Based on the target material and the nature of the observable signal, DM-nucleus scattering is most sensitive to GeV-scale DM, whereas DM-electron scattering is relevant for lower masses in the MeV range. Since this work focuses on a DM mass range from a few GeV to the TeV scale, we therefore concentrate exclusively on DM-nucleus scattering, employing the liquid-xenon detector framework characteristic of the LUX-ZEPLIN (LZ) experiment.
\begin{figure}[htb!]
\centering
\begin{adjustbox}{width=0.5\textwidth}
\begin{tcolorbox}[colback=gray!5, colframe=black!10, boxrule=0.8pt, arc=4mm, boxsep=0pt, left=0pt, right=0pt, top=0pt, bottom=0pt, width=0.55\textwidth, halign=center]
\begin{tikzpicture}
\begin{feynman}
\vertex(a);
\vertex[above left=0.5 cm and 1.25cm of a] (a1){\(\rm \color{black}{\phi_2}\)};
\vertex[above right=0.5 cm and 1.25cm of a] (a2){\(\rm \color{black}{\phi_2}\)};
\vertex[below =1.5cm of a] (d);
\vertex[below left=0.5 cm and 1.25cm of d] (d1){\(\rm \color{black}{\mathbb{N}}\)};
\vertex[below right=0.5 cm and 1.25cm of d] (d2){\(\rm \color{black}\mathbb{N}\)};
\diagram*{
(a1) -- [line width=0.25mm, scalar, arrow size=1.2pt, style=black,ultra thick] (a),
(a2) -- [line width=0.25mm, scalar, arrow size=1.2pt, style=black,ultra thick] (a),
(a) -- [line width=0.25mm, scalar, arrow size=1.2pt, style=gray, ultra thick, edge label'={\(\color{black}{ h_1/h_2}\)}] (d),
(d1) -- [line width=0.25mm, fermion, arrow size=1.2pt, style=mediumtealblue,ultra thick] (d),
(d) -- [line width=0.25mm, fermion, arrow size=1.2pt, style=mediumtealblue,ultra thick] (d2)};
\node at (a)[circle,fill,style=black,inner sep=1pt]{};
\node at (d)[circle,fill,style=mediumtealblue,inner sep=1pt]{};
\end{feynman}
\end{tikzpicture}
\begin{tikzpicture}
\begin{feynman}
\vertex(a);
\vertex[above left=0.5 cm and 1.25cm of a] (a1){\(\rm \color{black}{\phi_2}\)};
\vertex[above right=0.5 cm and 1.25cm of a] (a2){\(\rm \color{black}{\phi_1}\)};
\vertex[below =0.75cm of a] (c);
\vertex[below =0.75cm of c] (d);
\vertex[below left=0.5 cm and 1.25cm of d] (d1){\(\rm \color{black}{\mathbb{N}}\)};
\vertex[below right=0.5 cm and 1.25cm of d] (d2){\(\rm \color{black}\mathbb{N}\)};
\diagram*{
(a1) -- [line width=0.25mm, scalar, arrow size=1.2pt, style=black,ultra thick] (a),
(a2) -- [line width=0.25mm, scalar, arrow size=1.2pt, style=black,ultra thick] (a),
(a) -- [line width=0.25mm, boson, arrow size=1.2pt, style=black,ultra thick, edge label'={\(\color{black}{\zmuta }\)},insertion=1] (c),
(c) -- [line width=0.25mm, boson, arrow size=1.2pt, style=mediumtealblue, ultra thick, edge label={\(\color{black}{ \gamma/Z}\)}] (d),
(d1) -- [line width=0.25mm, fermion, arrow size=1.2pt, style=mediumtealblue,ultra thick] (d),
(d) -- [line width=0.25mm, fermion, arrow size=1.2pt, style=mediumtealblue,ultra thick] (d2)};
\node at (a)[circle,fill,style=black,inner sep=1pt]{};
\node at (c)[circle,fill,style=black,inner sep=1pt]{};
\node at (d)[circle,fill,style=mediumtealblue,inner sep=1pt]{};
\end{feynman}
\end{tikzpicture}
\end{tcolorbox}
\end{adjustbox}
\caption{The relevant Feynman diagrams corresponding to elastic (left) and inelastic (right) DM-nuclear scattering processes.}
\label{feyn:phi2-n}
\end{figure}
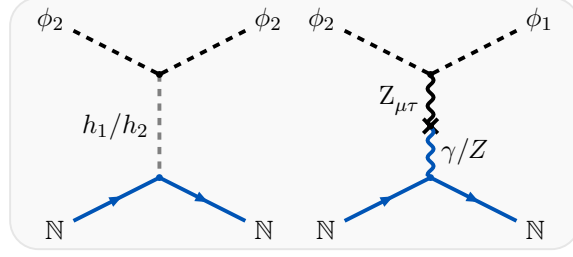

A major focus of this work is to investigate the possibility of explaining the observed inelastic nuclear recoil signal within the extended $\ULmuLtau$ framework. Before discussing the inelastic nuclear scattering regime, we first examine the constraints from elastic nuclear recoils provided by LZ~\cite{LZ:2025zpw}, which are also relevant to our model. \autoref{feyn:phi2-n} illustrates the Feynman diagrams for the elastic (left) and inelastic (right) nuclear recoil processes. The detailed implications of the bounds on the elastic scattering cross sections and the excess arising from inelastic scattering are discussed below.
\subsection{Elastic Nuclear Recoil}
The differential nuclear recoil rate for elastic DM–nucleus scattering can be expressed as \cite{Lin:2019uvt},
\begin{align}
\frac{dR^{\rm Elastic}}{dE_R} = N_T\frac{\rho_{0}}{m_{\phi_2}} \int_{v > v_{\min}}  v ~d^3{\bf v} ~ f_E(\mathbf{v}, t) \frac{d\sigma_A^{\rm Elastic}}{dE_R}\,,
\end{align}
where $N_T$ is the number of target nuclei per unit detector mass, $\rho_{0}=0.4 \text{ GeV/cm}^3$ is the local DM density, $m_{\phi_2}$ is the DM mass, $m_A$ is the target nucleus mass, and $f_E(\mathbf{v}, t)$ is the normalised DM velocity distribution in the lab frame~\cite{Asadi:2026iot, Langhoff:2026ujr}. In elastic scattering, the minimum velocity required to produce a recoil energy $E_R$ is given by
\begin{align}\label{eq:elas}
v^{\rm el}_{\min}(E_R) = \sqrt{\frac{m_A E_R}{2 \mu_{\phi_2 A}^2}},\quad \text{with} \quad\mu_{\phi_2 A} = \frac{m_{\phi_2} m_A}{m_{\phi_2} + m_A}\,,
\end{align}where $\mu_{\phi_2 A}$ denotes the reduced mass of the DM-nucleus system. The differential DM-nucleus elastic scattering rate, assuming spin-independent interactions, can be generalized as,
\begin{align}
\frac{d\sigma_A^{\rm Elastic}}{dE_R}=\dfrac{\sigma_{\phi_2\rm N}^{\rm Elastic}}{\mu_{\phi_2 \rm N}^2}\dfrac{m_AA^2}{2v^2}F^2_A(q)\Theta\left(v-v^{\rm el}_{\rm min}\right)\,,
\end{align}
where, $\Theta$ function ensures that no scattering takes place below the $v^{\rm el}_{\rm min}$ threshold. The elastic spin-independent DM-nucleon scattering cross-section $(\sigma_{\phi_2\rm N}^{\rm Elastic})$ is given by \cite{Bhattacharya:2021rwh},
\begin{align}
\sigma_{\rm \phi_2N}^{\rm Elastic}=\dfrac{f_N^2\mu_{\phi_2N}^2m_N^2}{4\pi v_{\rm EW}^2m^2_{\phi_2}}\left(\cos\alpha\dfrac{\lambda_{h_1\phi_2\phi_2}}{m_{h_1}^2}+\sin\alpha\dfrac{\lambda_{h_2\phi_2\phi_2}}{m^2_{h_2}}\right)^2\,,
\label{eq:phi2N}
\end{align}
where $f_N\simeq0.308$ is the nucleon form factor and $\mu_{\phi_2\rm N}=\dfrac{m_Nm_{\phi_2}}{m_N+m_{\phi_2}}$ stands for the reduced mass of DM-nucleon system and 
\begin{align}
\begin{split}
\lambda_{h_1\phi_2\phi_2}=(\mu\sqrt{2}-\lphiS\vmuta)\cos\alpha-(\lphiH v)\sin\alpha\,,\\
\lambda_{h_2\phi_2\phi_2}=(\lphiS\vmuta-\mu\sqrt{2})\sin\alpha-(\lphiH v)\cos\alpha\,.
\end{split}
\end{align}

\begin{figure}[htb!]
\centering
\includegraphics[width=0.75\linewidth]{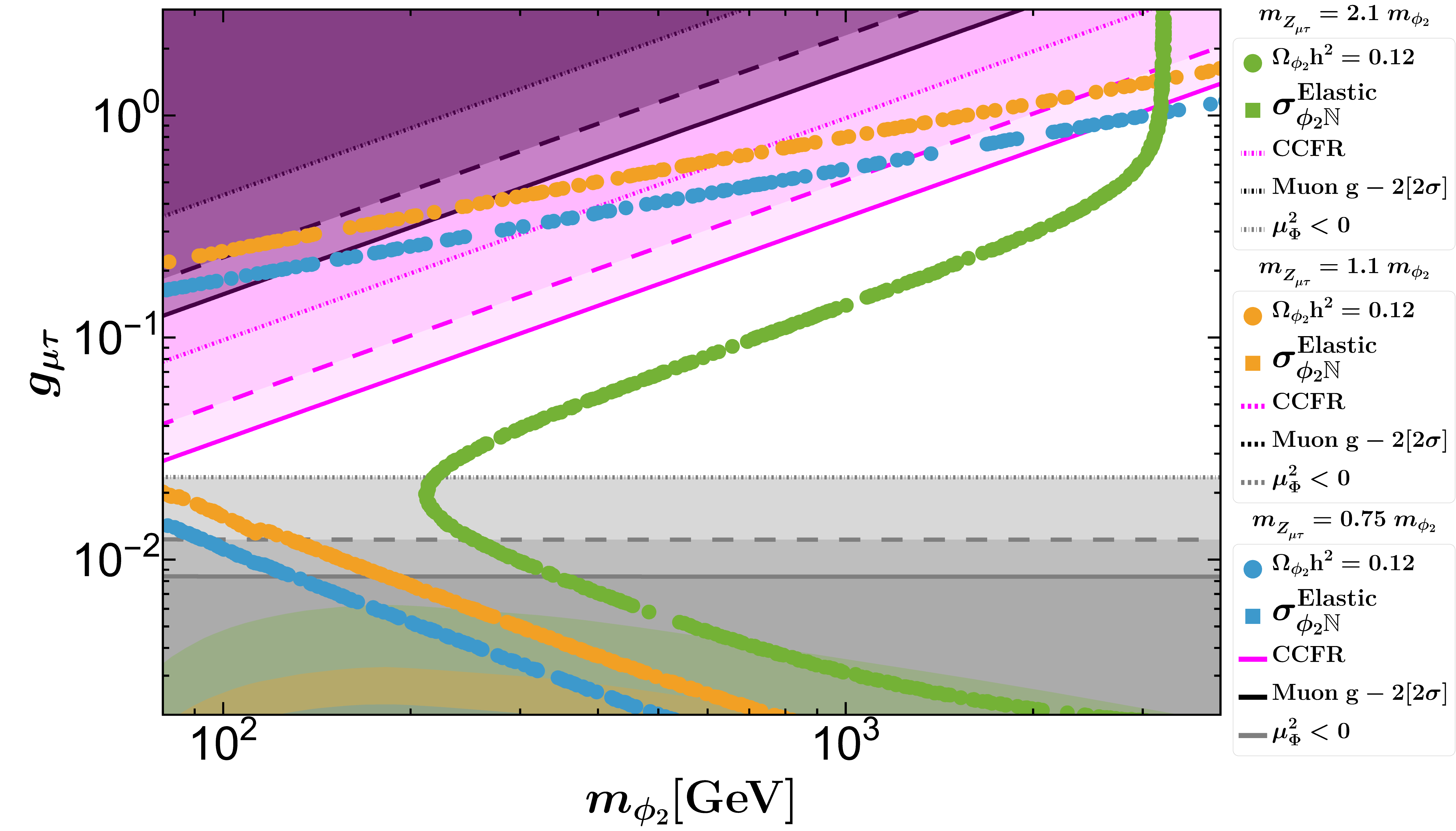}
\caption{The sky-blue, yellow, and green points satisfying the relic density constraint correspond to $\mzmuta/\mtwo =( 0.75$, $1.1$, and $2.1$), respectively. The parameters are fixed to $m_{h_2} = 2.2\,\mtwo$, $\delta m = 150$ keV, $\sin\alpha = 0.05$, and $\lambda_{\Phi S} = \lambda_{\Phi H} = 10^{-3}$. The magenta, grey, and black regions represent the exclusion limits from CCFR neutrino trident production, the Muon $g-2$ measurements, and DM stability criteria, respectively. Additionally, the sky-blue, yellow, and green shaded regions show the spin-independent DD limits for $\mzmuta/\mtwo = (0.75$, $1.1$, and $2.1$).}
\label{fig:scan1}
\end{figure}

\begin{figure}[htb!]
\centering
\includegraphics[width=0.475\linewidth]{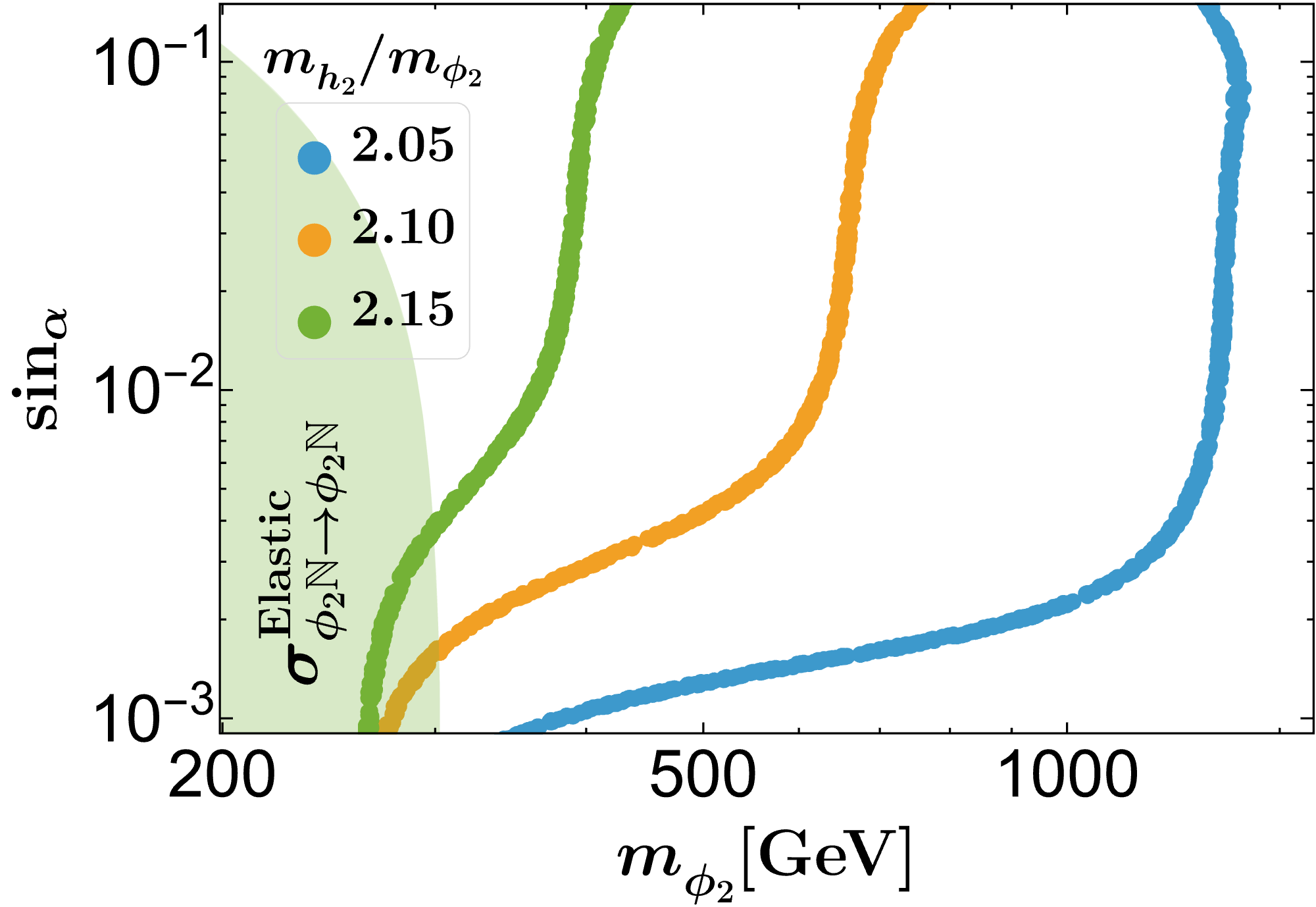}
\includegraphics[width=0.475\linewidth]{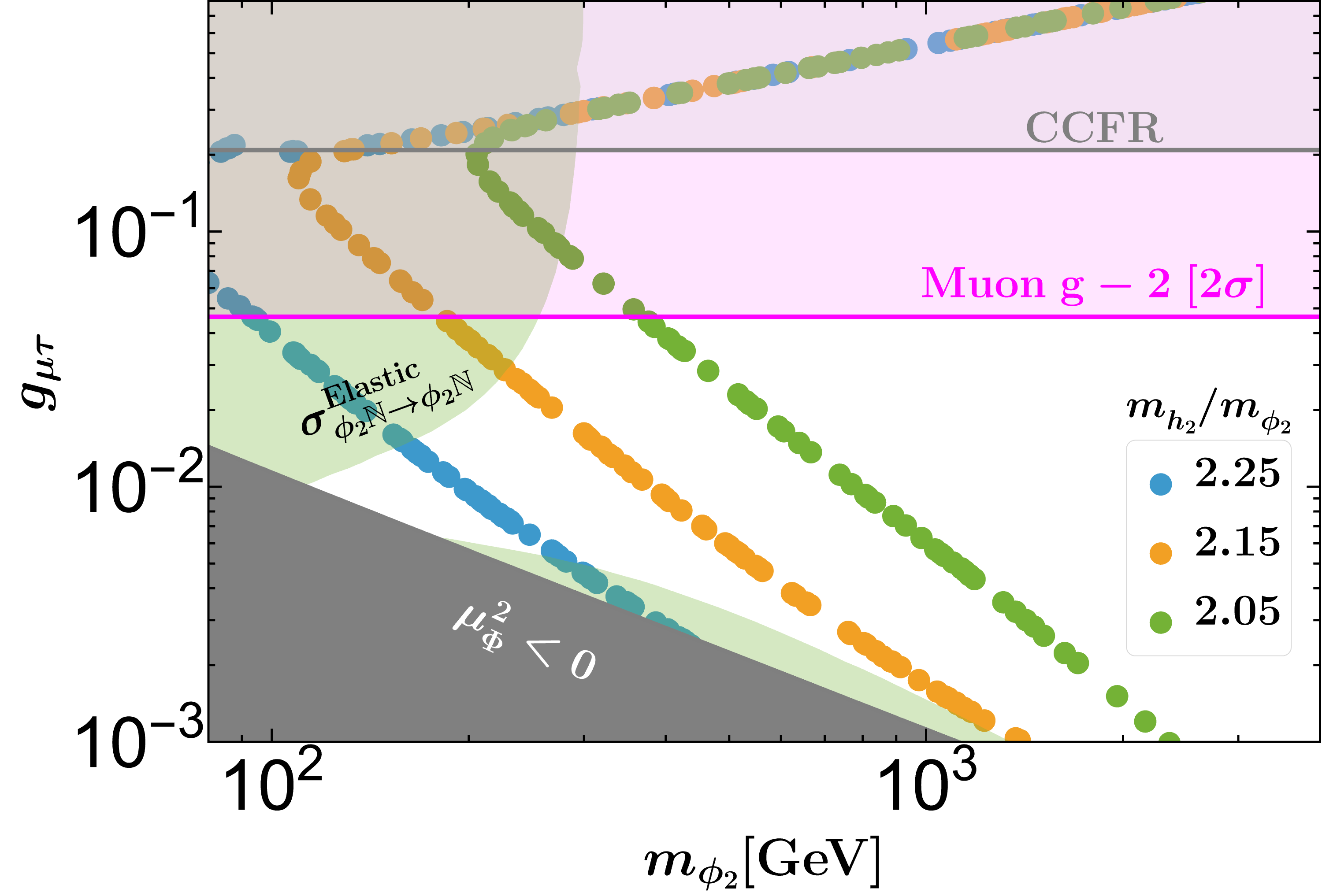}
\caption{The relic abundance-allowed points are shown in planes: $\mtwo-\salpha$ (\textit{left}) and $\mtwo-\gmuta$ (\textit{right}), with the sky-blue, orange, and green points corresponding to different choices of $(\mhtwo/\mtwo)$, as indicated in the legend. We fix $\mzmuta=2.2\,\mtwo$, $\gmuta=0.1$ for the \textit{left} plot, and $\mzmuta=100\,\GeV$, $\salpha=0.1$ for the \textit{right} plot. The remaining parameters are fixed to $\lphiS=5\times10^{-3}$, $\lphiH=10^{-2}$, and $\deltam=150~\keV$. The green-shaded region is excluded by the elastic DD limit from LZ. The magenta, light-grey, and dark-grey shaded regions (in the \textit{right} panel) are excluded by the muon $(g-2)$, CCFR, and DM stability constraints, respectively.}
\label{fig:scan2}
\end{figure}

Finally, the general expression for the differential recoil rate, allowing for different DM couplings to protons and neutrons, can be written as \cite{Lin:2019uvt,Bhattacharya:2025eym},
\begin{align}
\frac{dR^{\rm Elastic}}{dE_R} \simeq N_T\frac{\rho_{\phi_2}}{m_{\phi_2}}\dfrac{\sigma_{\phi_2\rm N}^{\rm Elastic}m_A}{2\mu_{\phi_2\rm N}^2}\left(Z+(A-Z)\dfrac{b_n}{b_p}\right)^2F^2_A(E_R)g(v^{\rm el}_{\rm min})\,,
\end{align}
where $b_p ~(b_n)$ is the corresponding DM coupling strength to the protons (neutrons), and the velocity-dependent integration factors out as
\begin{align}
g(v^{\rm el}_{\rm min})=\int d^3{\bf v} \dfrac{f({\bf v})}{v}\Theta\left(v-v^{\rm el}_{\min}\right)\,,
\end{align}
where $f({\bf v})$ denotes the DM velocity distribution function in the detector frame (properly normalized such that $\int d^3{\bf v} \, f({\bf v}) = 1$), $v = |{\bf v}|$, and $v^{\rm el}_{\min}$ is the minimum incoming speed required for a DM particle of mass $\mtwo$ to impart a nuclear recoil energy $E_R$ in elastic scattering \cite{Lewin:1995rx}. The function $g(v_{\min}^{\text{el}})$ encapsulates the astrophysical velocity distribution of the galactic halo, governing the kinematic threshold of the scattering process. By evaluating this distribution, we translate our predicted event rates into an effective spin-independent cross-section. To test the viability of our framework, we then compare these projections against the latest exclusion limits set by the LZ experiment in the $(\sigma_{\phi_2 \rm N}^{\rm SI}, m_{\text{DM}})$ plane.
In this work, we have calculated the $\phi_2 N \to \phi_2 N$ elastic spin-independent scattering cross sections using \texttt{MicrOmegas6}, and the constrained parameter spaces are shown in~\autoref{fig:scan1} and~\autoref{fig:scan2}.

In \autoref{fig:scan1}, we show the relic abundance satisfied parameter space points in the $\mtwo-\gmuta$ plane for different $\mzmuta/\mtwo$ benchmarks while other parameters, $\mhtwo=2.2\,\mtwo$, $\deltam=150~\keV$, $\sin\alpha=0.05$, $\lphiS=\lphiH=10^{-3}$, are fixed for the scan. For a fixed mass ratio $\mzmuta/\mtwo$ and DM mass $\mtwo$, the relic density constraint yields two degenerate solutions for the gauge coupling $\gmuta$: a lower branch predominantly governed by scalar-mediated annihilation processes, and an upper branch driven by $\zmuta$ exchange. As the mass ratio $\mzmuta/\mtwo$ varies, the enclosed area of this parameter space evolves significantly due to the resonant enhancement of the annihilation cross-section near the $\mzmuta \approx 2\mtwo$ threshold.
The most stringent limits arise from the DM stability criterion ($\mu_\Phi^2 > 0$), which is represented by the grey-shaded region. Following \eq\eqref{eq:muphi}, $\mu_\Phi^2$ can become negative when $\vmuta$ is large, irrespective of the DM mass, scaling as $\mu_\Phi^2 \sim m_{\phi_2}^2 \left(2 -  \lphiS x^2/(4 g^2_{\mu\tau})-\lphiH v^2_{\rm EW}\right)/2$ with $\mzmuta = x\,\mtwo$ for smaller $\lambda_{\phi H}$. Consequently, this bound becomes stronger for smaller $x$, which is also reflected in the plot by the upward shift of the grey-shaded region for larger $x$.
The sky-blue, yellow, and green points are relic density-satisfied points corresponding to $(\mzmuta/\mtwo)= (0.75,~1.1,~2.1)$, respectively. Simultaneously, the same colour shade indicated the elastic DD exclusion region, which already falls into the exclusion region of the DM stability limit. However, in this case, because DM-nucleon scattering is governed by the scalar ($h_{1,2}$) portal interactions with $\lphiS = \lphiH = 10^{-3}$, this process is suppressed for large $\gmuta \gtrsim 10^{-2}$ owing to the relation $\vmuta = \mzmuta/(2\gmuta)$. Consequently, a smaller $\mzmuta$ leads to a smaller $\gmuta$, which is also reflected in the hierarchy of these three colored regions.
On the contrary, CCFR (magenta) and Muon g-2 (black) constraints are excluded at higher $\gmuta$ as the corresponding interactions are proportionally dependent on it and inversely on $\mzmuta$.
Therefore, the entire white region probed by elastic DD lies within the parameter space excluded by the DM stability condition, $\mu^2_\Phi < 0$, rendering the elastic DD limits ineffective.

Similarly, we illustrate two additional plots in \autoref{fig:scan2} in the $\mtwo-\sin\alpha$ (left) and $\mtwo-\gmuta$ (right) planes for three representative choices of the mass ratio ($y \equiv \mhtwo/\mtwo$), for \textit{left} panel $y = (2.05,~2.10,~2.15)$ and for \textit{right} panel $y = (2.05,~2.15,~2.25)$. Additionally, for the \textit{left} panel, we fix $\mzmuta=2.2\,\mtwo$ and $\gmuta=0.1$, while in the \textit{right} panel we take $\mzmuta=100\,\GeV$ and $\salpha=0.1$. For both panels, we set $\lphiS=5\times10^{-3}$, $\lphiH=10^{-2}$, and $\deltam=150~\keV$. In both plots, the relic density-allowed regions are indicated by sky-blue, orange, and green colored points corresponding to the different values of $y$ mentioned in the figure caption.
In the \textit{left} panel, the relic density is predominantly governed by $h_2$ and $\zmuta$-mediated annihilation channels when $\sin\alpha \lesssim 0.01$, whereas larger values of the mixing angle $\sin\alpha$ introduce significant contributions from the $h_1$-mediated process. Notably, an increase in $\sin\alpha$ enhances the DM annihilation cross-section, driving the system toward an underabundant regime that can be compensated by increasing the DM mass $\mtwo$.
Since we have fixed $\gmuta = 0.1$, which consequently fixes the contribution from $\zmuta$-mediated processes, the enhancement of $\mtwo$ (i.e., $\mzmuta$) reduces its contribution to the DM relic density.
In contrast, for the right panel, with fixed values of $\lphiS = 5 \times 10^{-3}$, $\lphiH = 10^{-2}$, $\sin\alpha = 0.1$, and $\mzmuta = 100~\GeV$, the underlying process dependencies are distinctly visible: the upper branch above the inflection point is governed by $\zmuta$-mediated processes, whereas the lower branch is driven by scalar-mediated annihilation. The onset of this bending behaviour is controlled by the mass ratio $\mhtwo/\mtwo$, which dictates the proximity to the $h_2$ resonance region, mirroring the shift away from the $h_2$ resonance observed in the left panel.
As before, some part of the parameter space is disfavored by CCFR (light grey), Muon $g-2$ (magenta), DM stability (dark grey), and elastic DD by LZ (green). Notably, in the \textit{right} plot, we observe a double-bump structure in the DD-excluded green region owing to destructive interference between $h_1$ and $h_2$-mediated elastic DD scattering. The constraints in the \textit{left} plot are relaxed in the \textit{right} plot due to a different choice of coupling parameters.
\subsection{Inelastic Nuclear Recoil}
For the inelastic DM-nucleus scattering process, $\phi_2(p) + \mathbb{N} \to \phi_1(p') + \mathbb{N}$, in the non-relativistic limit,
\begin{gather}\label{eq:non-rel}
p = p'+ q \implies
m_2 + \frac{1}{2} m_2 v^2 = m_1 + \frac{1}{2} m_1 v^{\prime 2} + E_R\,,
\end{gather}
where, $q$ is the transfer momentum. In the non-relativistic approximation, $q^2 \simeq 2 m_A E_R$ and from \eq\eqref{eq:non-rel} we obtains~\cite{Lin:2019uvt},
\begin{align}
v \cos\theta = \frac{\delta m}{q} + \frac{q}{2\mu_{\phi_2 A}}\,.
\label{eq:vcostheta}
\end{align}
Therefore, the minimum possible incoming DM velocity is obtained corresponding to the $\cos\theta = 1$ case, i.e.,
\begin{equation}
v_{\min} = \frac{\delta m}{q} + \frac{q}{2\mu_{\phi_2 A}}\,,
\end{equation}
which in terms of $E_R$ reads
\begin{align}
v_{\min}(E_R) = \frac{1}{\sqrt{2m_A E_R}} \left( \frac{m_A E_R}{\mu_{\phi_2 A}} + \delta m \right)\,.
\label{eq:vmin}
\end{align}
It should be noted that in the $\delta m \to 0$ limit, this reduces to the elastic $v_{min}$ discussed in \eq\eqref{eq:elas}. So the generic $v_{\rm min}$ can be written as
\begin{align}
v_{\min} = v_{\min}^{\text{el}} + \frac{\delta m}{\sqrt{2m_A E_R}}\,,
\label{eq:vmininel}
\end{align}
where $\delta m > 0$ corresponds to endothermic (up-scattering) and $\delta m < 0$ corresponds to exothermic scattering processes. The maximum velocity depends on the maximum splitting $(\delta m)_{\max}$ as
\begin{gather}
(\delta m)_{\max} = \sqrt{2m_A E_R} \left( v_{\max} - \sqrt{\frac{m_A E_R}{2\mu_{\phi_2 A}^2}} \right)\,.
\end{gather}
However, the maximum DM velocity can never exceed $v_{\max} = v_{\text{esc}} + v_{\text{earth}}$, where $v_{\text{esc}} = 544~\mathrm{km/s}$ is the escape velocity and $v_{\text{earth}} = 254~\mathrm{km/s}$ is the velocity of the Earth with respect to the galactic frame. By solving \eq\eqref{eq:vcostheta} for a fixed velocity $v \ge v_{\min}(E_R)$, we get
\begin{align}
E_R^\pm(v) = \frac{\mu_{\phi_2 A}^2}{2m_A} \left( v \pm \sqrt{v^2 - v_\delta^2} \right)^2\,,\quad
v_\delta = \sqrt{\frac{2\,\delta m}{\mu_{\phi_2 A}}}\,,
\end{align}
where the condition $v > v_\delta$ sets the kinematic threshold for endothermic inelastic scattering. Thus, for a fixed DM velocity, the recoil energy lies within a finite window, leading to bounds on $\deltam$ from both sides.

Now, the differential DM-nucleus scattering rate per unit target mass, time, and recoil energy can be expressed as \cite{Gondolo:2002np, Hapitas:2021ilr}:
\begin{align}
\dfrac{dR}{dE_R}=\dfrac{\rho_{0}}{m_{\phi_2}m_A}\int^{v_{\rm max}}_{v_{\rm min}(E_R)}v f_E(\mathbf v)\dfrac{d\sigma^{A}_{\phi_2\mathbb{N}\to\phi_1\mathbb{N}}}{dE_R}d^3v\,,
\end{align}
For DD calculations, we adopt the Standard Halo Model (SHM), with a Maxwell-Boltzmann velocity distribution truncated at the galactic escape velocity~\cite{Lewin:1995rx}, assuming an isotropic velocity distribution in the galactic frame:
\begin{align}
f_{\rm gal}(\mathbf v)
&=\frac{1}{N_{\rm esc}}
\left(\frac{1}{\pi v_0^2}\right)^{3/2}
e^{-\mathbf{v}^2/v_0^2}\,
\Theta(v_{\rm esc}-|\mathbf{v}|),
\end{align}
where $v_0\approx 220~{\rm km/s}$ is the local circular speed, with
\begin{align}
N_{\rm esc}
={\rm erf}(z)-\frac{2z}{\sqrt{\pi}}e^{-z^2},
\qquad
z=\frac{v_{\rm esc}}{v_0}.
\end{align}
In the Earth (lab) frame, the distribution is obtained by boosting
\begin{align}
f_E(\mathbf v)=f_{\rm gal}\left(\mathbf v+\mathbf v_E(t)\right),
\end{align}
where $\mathbf v_E(t)$ denotes the Earth's velocity with respect to the Galactic frame, with $|\mathbf v_E|\sim 240~{\rm km/s}$. The differential DM-nucleus ($\phi_2-\mathbb{N}$) scattering cross section is given by~\cite{Hapitas:2021ilr},
\begin{align}\label{eq:elxs}
\dfrac{d\sigma^A_{\phi_2\mathbb{N}\to\phi_1\mathbb{N}}}{dE_R}=\dfrac{\overline{|\mathcal{M}|^2}_{\phi_2 \mathbb{N}\to \phi_1 \mathbb{N}}}{32\pi m^2_{\phi_2}m_N v^2}\simeq\dfrac{m_A}{2\pi v^2}\left[\dfrac{Ze\,\epsA\gmuta}{m_{\zmuta}^2+2m_AE_R}\right]^2F^2_A(E_R)\,,
\end{align}
where $m_A$ and $Ze$ are the mass and electric charge of the nucleus, and $F_A(E_R)$ is the Helm form factor~\cite{Lewin:1995rx} given by
\begin{align}
F_A(E_R) = \frac{3j_1(qr_n)}{qr_n} \exp\left(-\frac{q^2 s^2}{2}\right)\,,
\end{align}
with $q = \sqrt{2m_AE_R}$, $r_n = 1.14\,A^{1/3}~\mathrm{fm}$, and $s \simeq 0.9~\mathrm{fm}$. The expression \eq\eqref{eq:elxs} assumes that the $\gamma$-mediated scattering dominates over the $Z$-mediated one, and the corresponding expressions of the effective couplings are already discussed in \eq\eqref{eq:loop-mixing}. The expected number of events is obtained by integrating the recoil spectrum over the relevant recoil energy window:
\begin{align}
\mathcal{N} = \varepsilon \int_{E_{R}^{\min}}^{E_{R}^{\max}} dE_R \, \epsilon(E_R)\, \frac{dR}{dE_R},
\end{align}
where $\varepsilon$ denotes the total exposure, corresponding to $2.84~\mathrm{tonne\,-\,yr}$ for the LZ experiment, and $\epsilon(E_R)$ is the detector efficiency. Following~\cite{LZ:2026axp}, the total exposure is $1.037 \times 10^6~\mathrm{kg\,day}$, with the reconstructed energy region of interest (ROI) being $5.4~\mathrm{keV} \leq E_{R} \leq 269.9~\mathrm{keV}$, where $\epsilon(E_R)$ never drops below $50\%$. The observed single event has a reconstructed recoil energy of $E_{R}^{\mathrm{obs}} = 248 \pm 23\,(\mathrm{stat.}) \pm 23\,(\mathrm{syst.})~\mathrm{keV}$. 
\begin{figure}
\centering
\includegraphics[width=0.475\linewidth]{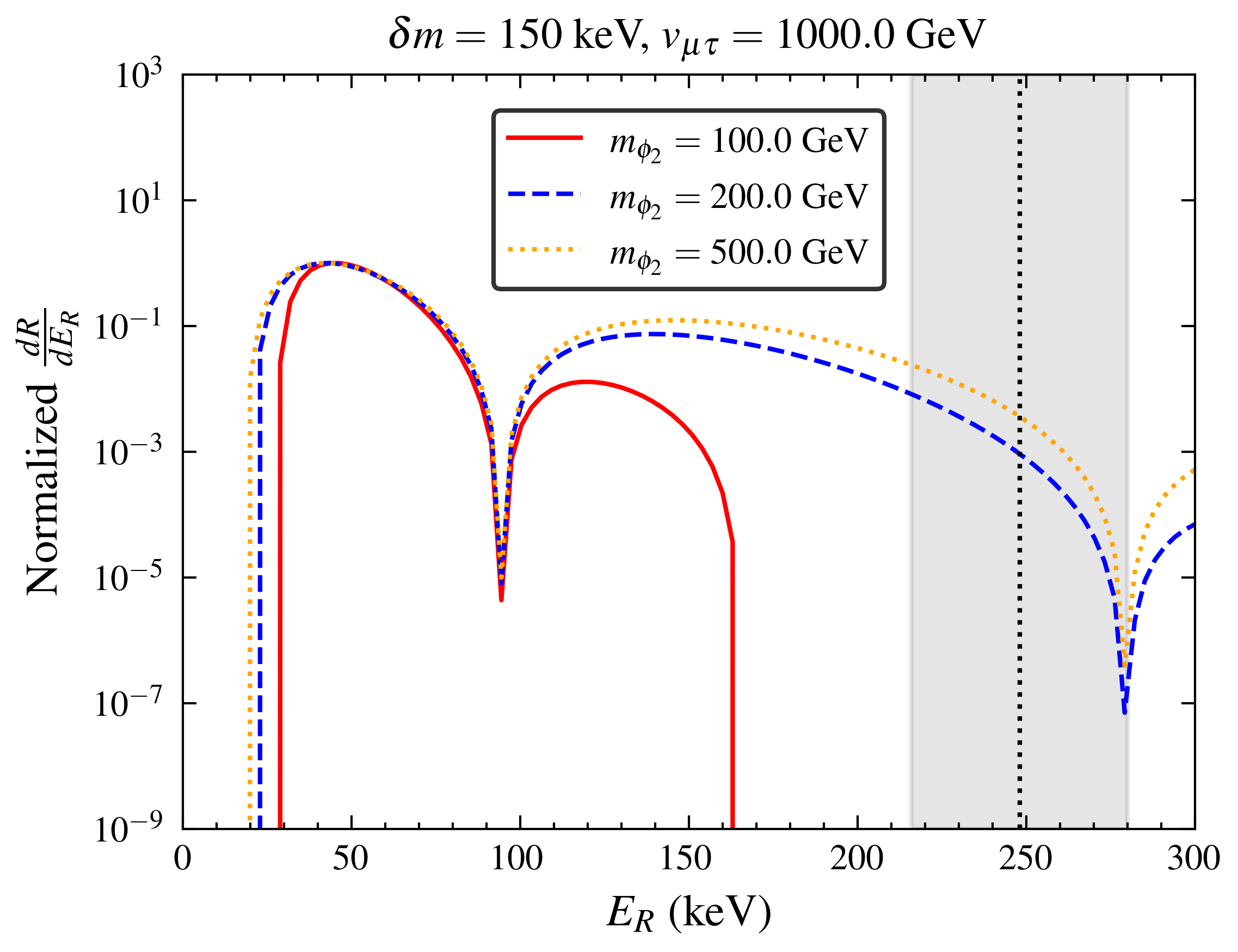}
\includegraphics[width=0.475\linewidth]{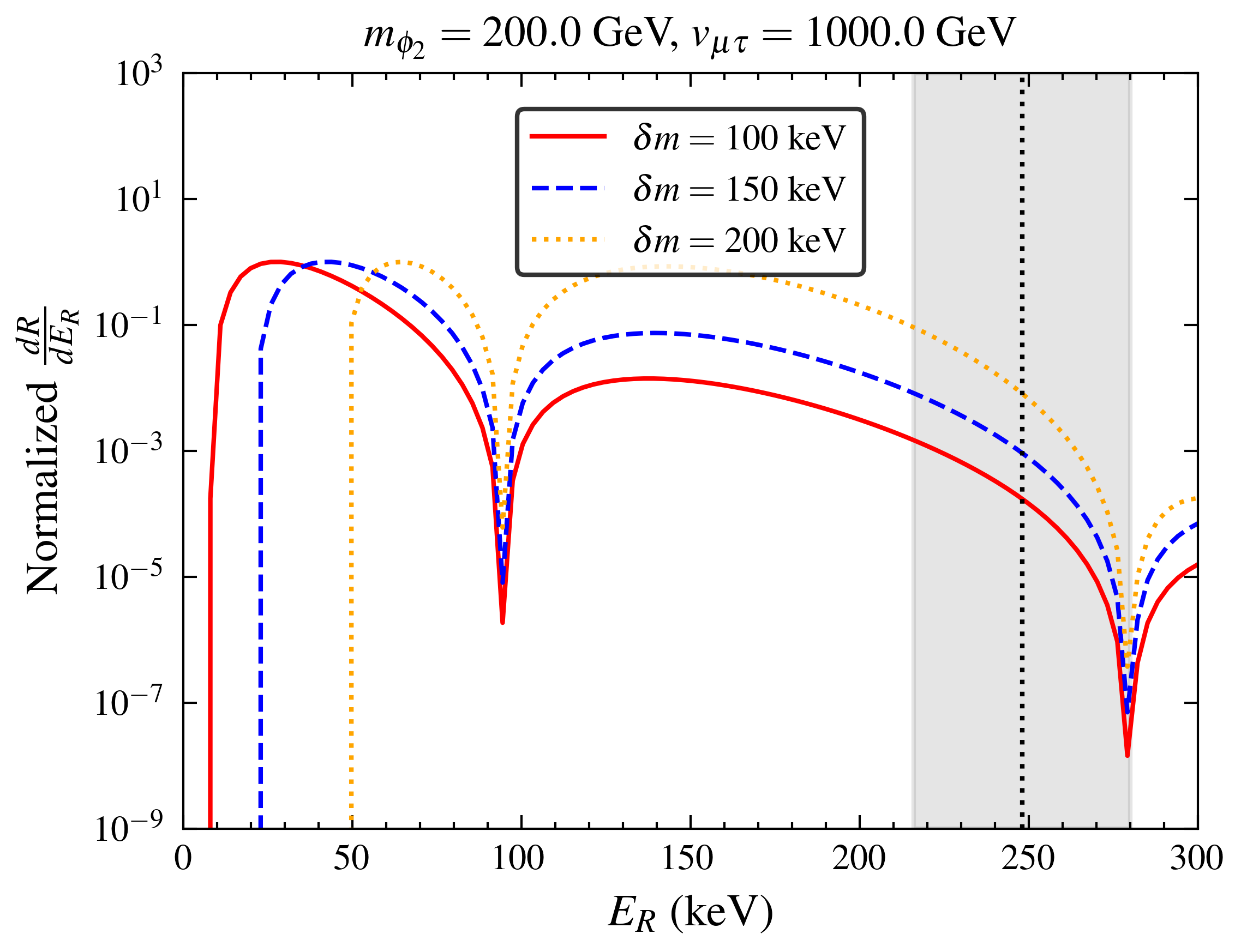}
\caption{Normalized $(dR/dE_R)$ vs. $E_R$ plots varying $\mtwo$ (\textit{left}) and $\delta m$ (\textit{right}). The vertical band corresponds to $E_R = 248 \pm 32\,\keV$, where the uncertainty is obtained by adding the statistical and systematic uncertainties in quadrature.}
\label{fig:recoil}
\end{figure}

\autoref{fig:recoil} shows the normalised recoil rate as a function of recoil energy in our model, highlighting the variations with $\mtwo$ (left panel) and $\deltam$ (right panel). In the left-hand panel, for $E_R > 200\text{ keV}$, the normalised event number increases with $\mtwo$. However, once $\mtwo$ becomes sufficiently large compared to the nuclear mass $m_A$, the reduced mass, and consequently the kinematic reach and event rate, saturates and becomes nearly independent of $\mtwo$. In the right panel, a larger mass splitting ($\deltam > 0$) raises the kinematic threshold for endothermic upscattering, shifting the spectrum and modifying the high-energy event rate, as illustrated by the transition from the thick red to the dashed yellow curves.
\subsection*{LZ Event Likelihood Analysis}
To statistically quantify the observed excess within the parameter space of our model, we perform a likelihood analysis comparing the predicted nuclear recoil spectrum with the observed event rate. This analysis identifies the regions of parameter space compatible with the signal and quantifies the corresponding statistical uncertainties. A similar likelihood treatment of the LZ excess was recently performed in Ref\,.~\cite{Zhu:2026dag}. The extended likelihood function is defined as
\begin{equation}\label{eq:likely}
\mathscr{L}(\vec{\theta}) = \left[\prod_{i}^{n} P(\vec{\theta}) \right] e^{-N(\vec{\theta})}\,,
\end{equation}
where $P(\vec{\theta})$ and $N(\vec{\theta})$ are the probability density and the expected number of events, corresponding to model parameters, $\vec{\theta} = \{m_{\phi_2}, \deltam, \vmuta, \cdots\}$. Since the experiment measures events in the recoil-energy spectrum, \eq\eqref{eq:likely} can be written as
\begin{equation}
    \mathscr{L}(\vec{\theta}) =\left[\prod_{i=1}^{n}\left.\frac{dN(\vec{\theta})}{dE_R}\right|_{E_R=E_i}\right] e^{-N(\vec{\theta})}\,,
\end{equation}
where, $E_i$ denotes the recoil energy of the $i$-th observed event. For single observable event at $E_{R}=248$ keV, the likelihood function reduces to
\begin{equation}
    \mathscr{L}(\vec{\theta}) =\left.\frac{dN(\vec{\theta})}{dE_R}\right|_{E_R=248\text{ keV}} e^{-N(\vec{\theta})}\,,
\end{equation}
To obtain the uncertainty intervals, we employ the maximum-likelihood estimator method. We evaluate $\mathscr{L}(\vec{\theta})$ over a range of parameter  space points and numerically determine its maximum, given by
\begin{equation}\label{eq:4.25}
    \ln{\mathscr{L}(\hat{\theta})} = \max_{\vec{\theta}} \ln{\mathscr{L}(\vec{\theta})} = \max_{\vec{\theta}} \left[-\mathcal{N}(\vec{\theta}) + \ln{\left(\left.\frac{dN(\vec{\theta})}{dE_R}\right|_{E_R=248~\mathrm{keV}}\right)}\right]\,,
\end{equation}
where $\hat{\theta}$ denotes the best-fit values of the model parameters. To simplify the analysis, we construct conditional likelihoods by varying two parameters while fixing all other parameters to arbitrary values. The test statistic is defined in terms of the likelihood ratio as
\begin{equation}\label{eq:4.26}
    \Delta\chi^2 = -2\left[\ln{\mathscr{L}(\vec{\theta})} - \ln{\mathscr{L}(\hat{\theta})}\right]\,.
\end{equation}
In the asymptotic regime, Wilks' theorem~\cite{Wilks:1938dza} would identify the $1\sigma$ and $2\sigma$ regions of a two-parameter analysis with $\Delta\chi^2 < 2.30$ and $6.18$, respectively. With a single observed event, however, the expected number of signal events in the favoured parameter region is of order one or smaller, and the conditions underlying Wilks' theorem are not satisfied (see~\autoref{app:stat}). We therefore determine the confidence regions with the Feldman-Cousins unified approach~\cite{Feldman:1997qc}, a Neyman construction~\cite{Neyman:1937uhy} with likelihood-ratio ordering. For every parameter point $\vec\theta$ we generate pseudo-experiments from the model prediction at $\vec\theta$: the number of events is drawn from a Poisson distribution with mean $N(\vec\theta)$, and their recoil energies from the predicted spectrum. For each pseudo-experiment we compute $\Delta\chi^2$ exactly as for the data, with the maximum in \eq\eqref{eq:4.25} taken over the same parameter range. A point $\vec\theta$ belongs to the $p\%$ confidence region if the $\Delta\chi^2$ of the observed event is smaller than the $p\%$ quantile of this distribution. By construction, the resulting regions have the stated coverage independently of the number of observed events. Throughout, we assume a background-free measurement and neglect the detector energy resolution.

\begin{figure}
\centering
\includegraphics[width=0.475\linewidth]{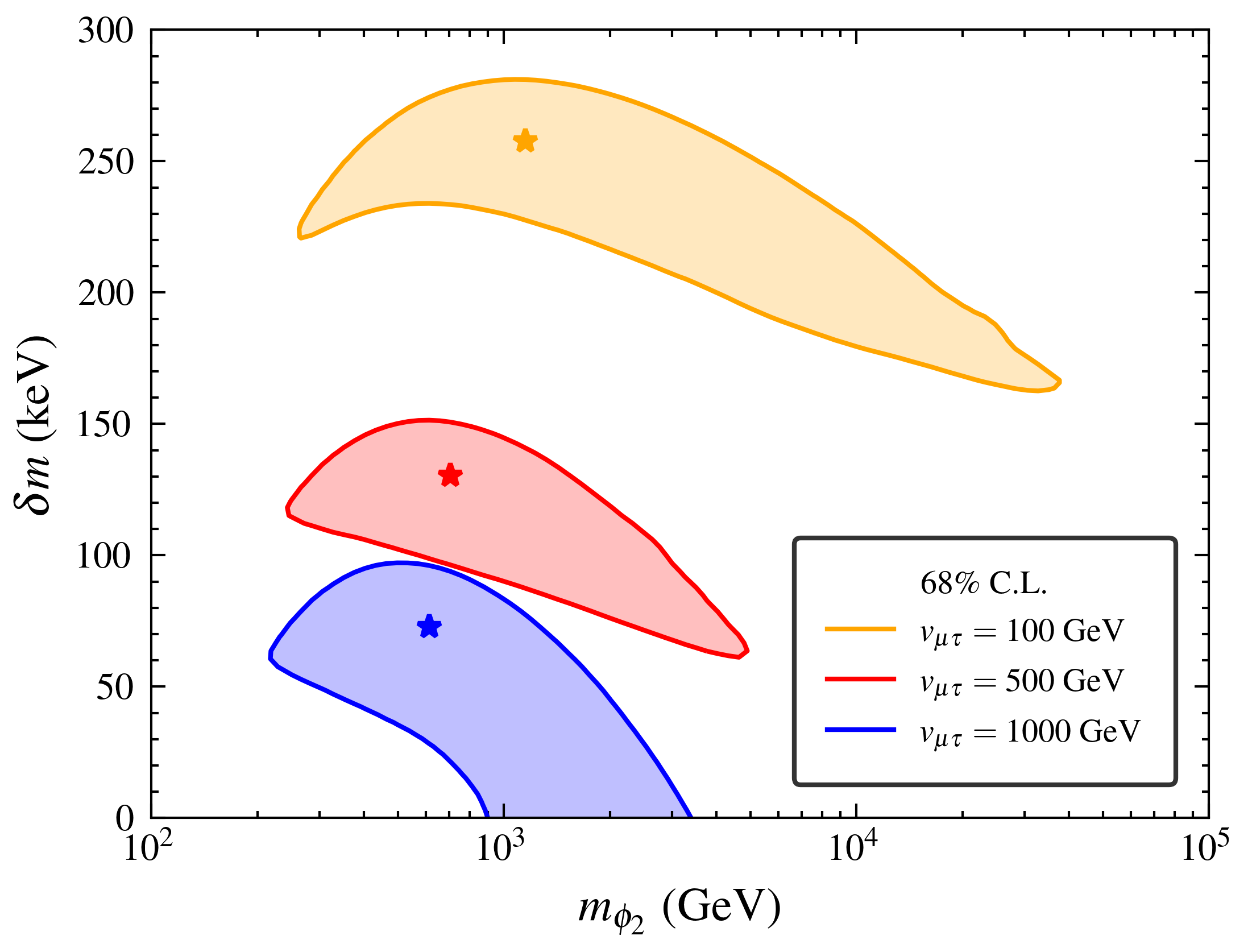}
\includegraphics[width=0.475\linewidth]{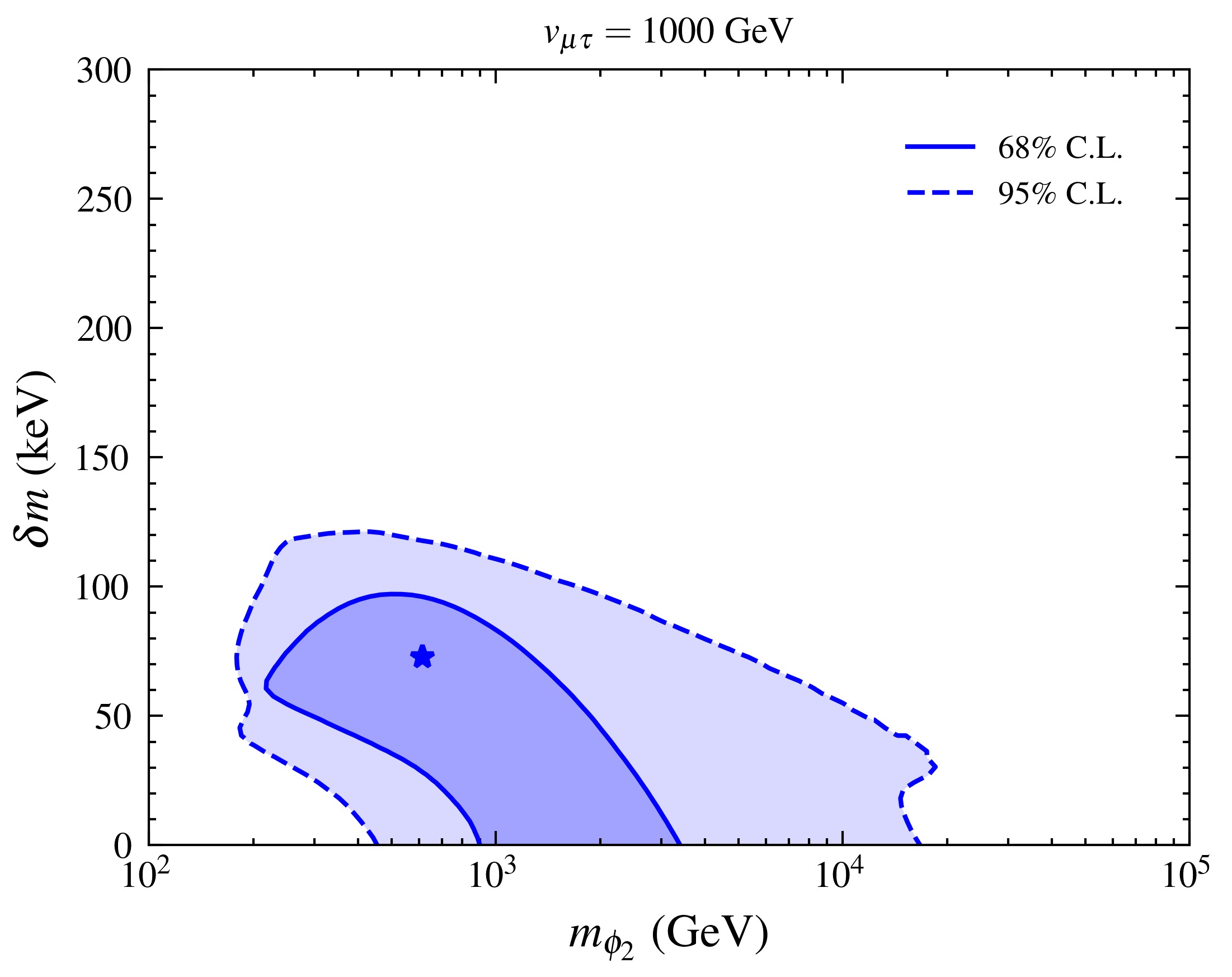} \\
\includegraphics[width=0.475\linewidth]{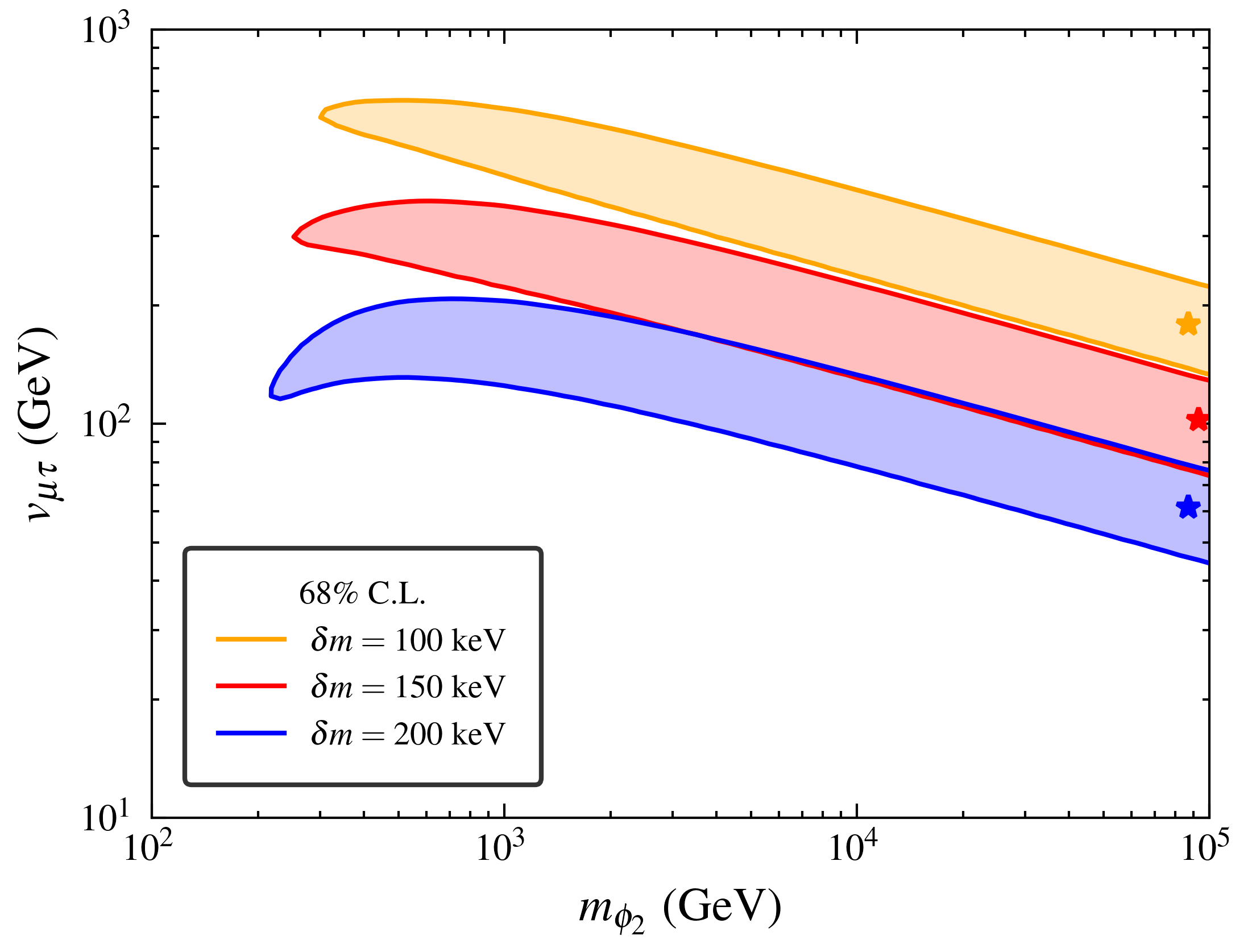}
\includegraphics[width=0.475\linewidth]{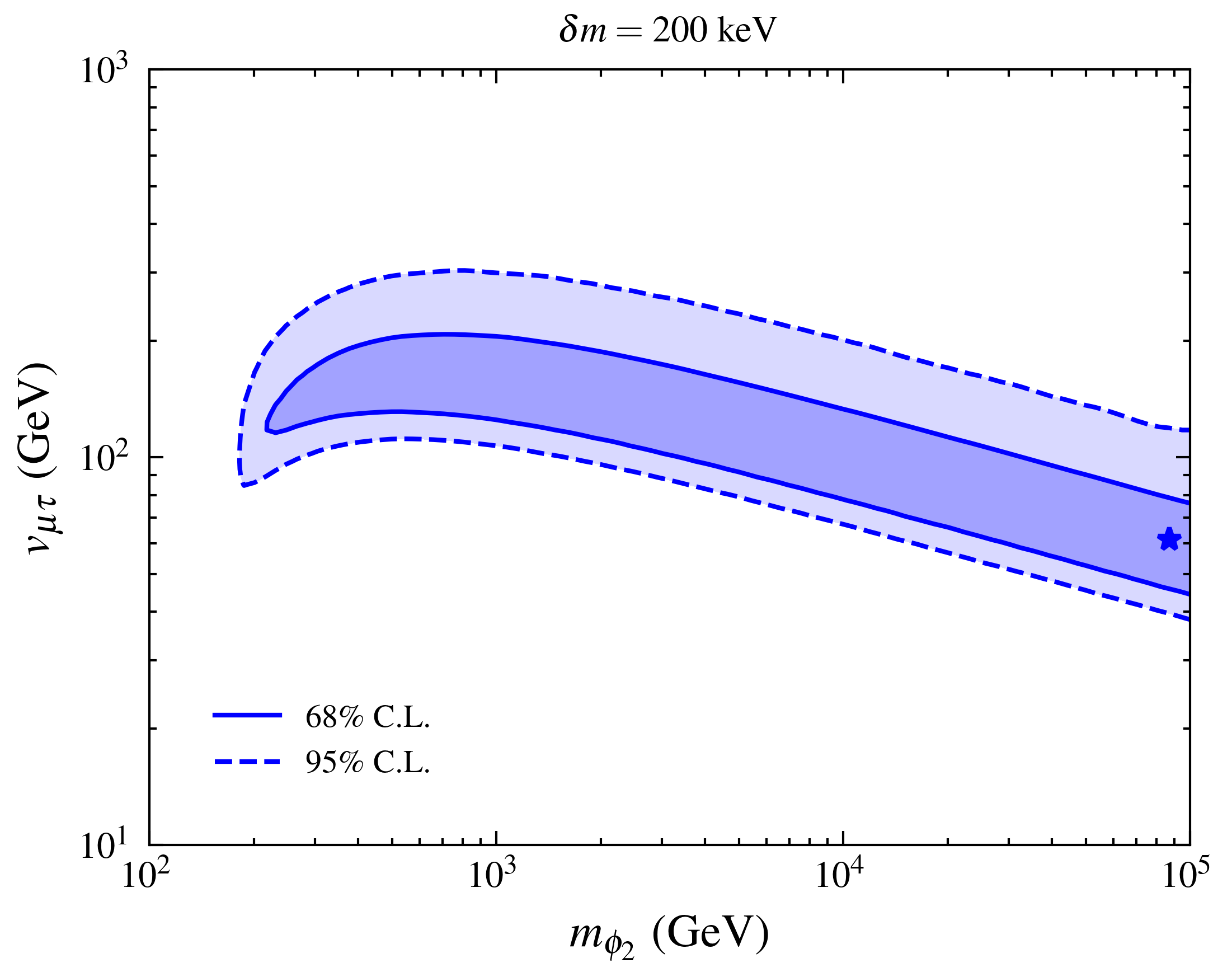}
\caption{$68\%$ C.L. contours that could account for the LZ event in the $m_{\phi_2}-\delta m$ (\textit{top left}) and $m_{\phi_2}-\vmuta$ (\textit{bottom left}) planes. $68\%$ C.L. and $95\%$ C.L. contours for $\vmuta = 1000$ GeV (\textit{top right}) and $\delta m = 200$ keV (\textit{bottom right}) cases. The stars $(\star)$ indicate the best fit values.}
\label{fig:loglike}
\end{figure}

\autoref{fig:loglike} (\textit{left} panels) shows the $68\%$ confidence level (C.L.) contours in the $m_{\phi_2}-\deltam$ (\textit{top}) and $m_{\phi_2}-\vmuta$ (\textit{bottom}) planes, respectively, where the other parameter is varied to different values, detailed in plot legends. The \textit{right} panels shows comparison between the $68\%$ C.L. and $95\%$ C.L. contours in the $m_{\phi_2}-\deltam$ (\textit{top}) and $m_{\phi_2}-\vmuta$ (\textit{bottom}) planes, respectively. The best-fitted values are indicated using stars $(\star)$ in the $m_{\phi_2}-\deltam$ plane, whereas in the case of the $m_{\phi_2}-\vmuta$ plane, the best-fitted values lie beyond the parameter space visible in the plots.
We observe that smaller values of $\vmuta$ allow for relatively larger mass splittings, $\deltam$. As the mass splitting increases, the likelihood contour shrinks along the $\deltam$ direction while extending along the $m_{\phi_2}$ direction, indicating greater freedom in the DM mass but a more restricted range of the mass splitting for smaller values of $\vmuta$. In the $m_{\phi_2}-\vmuta$ plane, the likelihood maximum lies along a negatively sloped linear extension, indicating that larger DM masses favour smaller values of $\vmuta$. Similarly, as the mass splitting increases, smaller values of $\vmuta$ are favoured. However, it should be noted that $\vmuta$ determines the mass of the $\zmuta$ boson. Therefore, lowering $\vmuta$ too much may result in a lighter $\zmuta$ that is excluded by existing experimental constraints.
\section{Other Detection Prospects}
\label{sec:collider}
In this section, we discuss complementary avenues for probing the extended scenario, providing additional constraints and search prospects for the model. We first investigate constraints from ID experiments, where DM annihilation can lead to excesses in the expected SM fluxes. We then consider existing LHC constraints on the $\zmuta$ boson, particularly in the low-mass regime, followed by future search prospects at muon colliders, which can provide a sensitive probe of the model due to its characteristic $\ULmuLtau$ charge assignments.

\subsection{Indirect Detection}
\label{sec:ID}
An important avenue for probing DM indirectly is measuring SM particle fluxes, where any excess over SM predictions could serve as a potential signature of DM~\cite{Gaskins:2016cha, Cirelli:2010xx}. In this section, we focus on indirect searches for thermal DM with masses in the GeV–TeV range. In dense regions of the galactic halo, DM can annihilate or decay into SM particles, such as photons, neutrinos, antiprotons, and positrons, via prompt emission or final-state radiation. Several ground and space-based detectors target these signatures, including Fermi-LAT, HAWC, H.E.S.S., and MAGIC ($\gamma$-rays) \cite{Fermi-LAT:2025gei}, AMS-02 (antiprotons and positrons) \cite{Calore:2022stf}, and CTA (Cherenkov Telescope Array) \cite{CTA:2020qlo}. Furthermore, measurements of Cosmic Microwave Background (CMB) anisotropies provide robust constraints on DM annihilation into SM particles during the recombination epoch, as measured by the \texttt{Planck} collaboration \cite{Planck:2018vyg}.

In this model framework, the ID constraints are primarily driven by $s$-wave annihilation processes. Specifically, the annihilation channel $\one\one \to \zmuta\zmuta$, mediated by the scalar and gauge bosons, followed by subsequent leptonic decays $\zmuta \to \mu^+\mu^-\,,\tau^+\tau^-$, is heavily constrained by Fermi-LAT $\gamma$-ray observations. This same channel is subject to stringent CMB bounds arising from energy injection during recombination, which are conventionally parametrised by: $p_{\text{ann}} = ({f_{\text{eff}}\langle\sigma v\rangle}/{\mtwo })$~\cite{Planck:2018vyg}. In contrast, alternative channels such as $\phi_1\phi_2 \to \zmuta \to \ell^+\ell^-$ are $p$-wave suppressed (velocity-suppressed) and consequently yield negligible contributions to late-time ID constraints. Additionally, Higgs-mediated processes are suppressed by DD bounds and can be safely neglected in this analysis. The exclusion limits from indirect searches are estimated using \texttt{micrOMEGAs6}, based on data from Fermi-LAT and CMB. Finally, the exclusion bounds are shown in~\autoref{fig:scan3}. The cyan shaded region illustrates the CMB exclusion limit for $\mzmuta = 0.75 ~\mtwo$, whereas other mass combinations yield no constraints owing to off-shell leptonic decays. Comparable bounds are also obtained from Fermi-LAT observations, which are omitted from the figure to avoid clutter.
\subsection{Limits from LHC Searches}
\begin{figure}[htb!]
\centering
\includegraphics[width=0.475\linewidth]{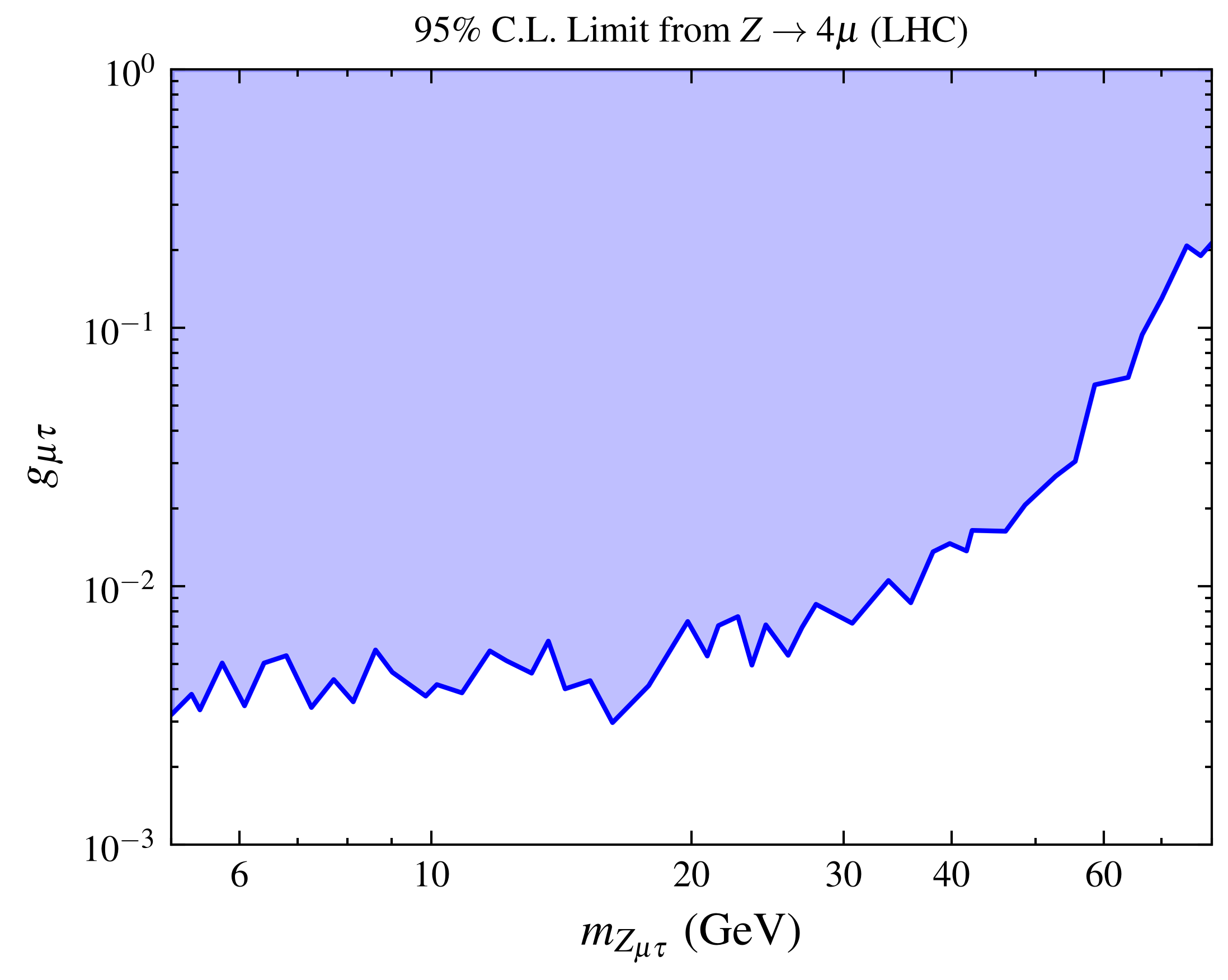}
\includegraphics[width=0.475\linewidth]{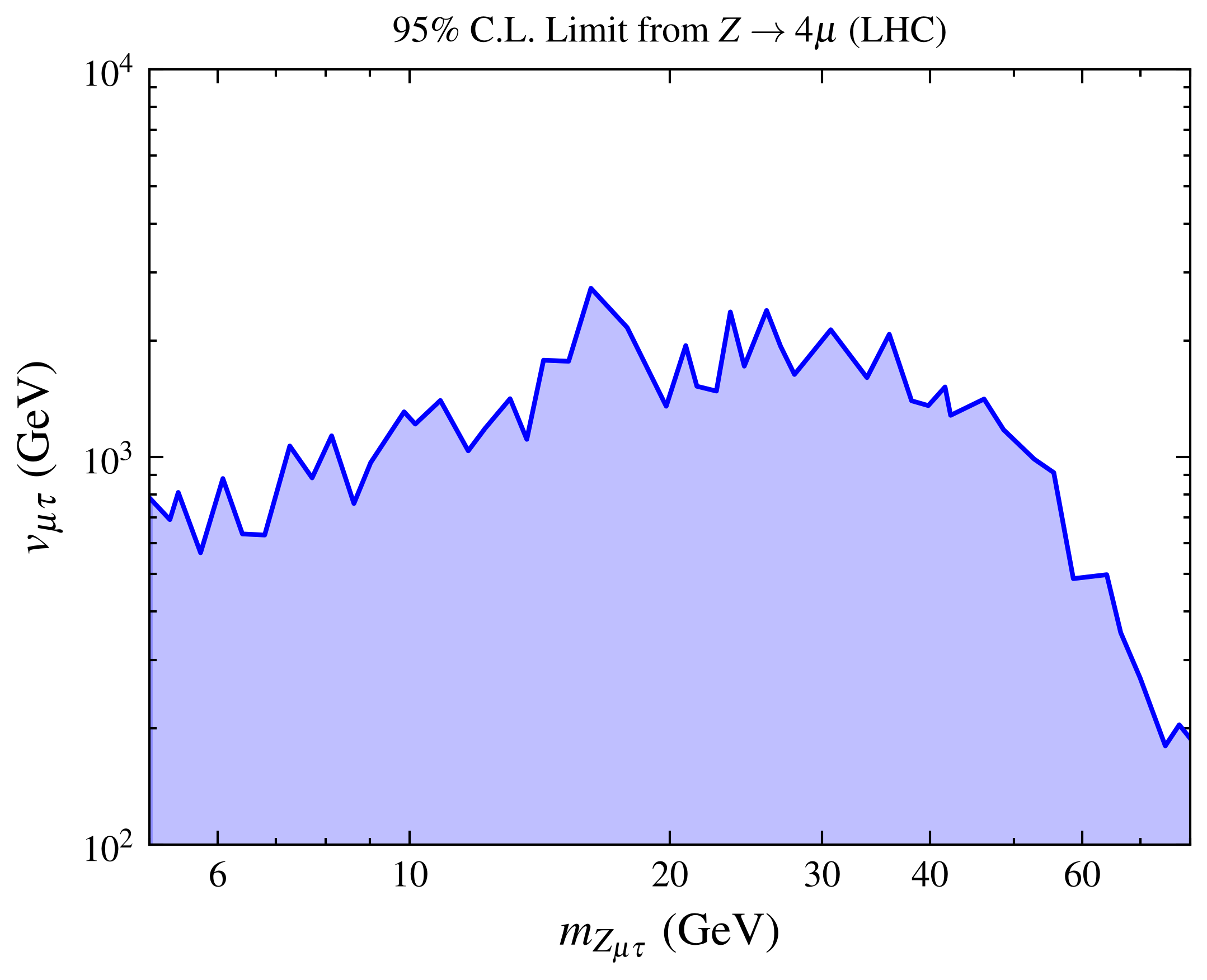}
\caption{LHC exclusion limits from $Z\to 4\mu$ decay in the $\mzmuta-\gmuta$ (\textit{left}) and $\mzmuta-\vmuta$ (\textit{right}) planes, respectively.}
\label{fig:muC1}
\end{figure}

The LHC has extensively studied the $\ULmuLtau$ model, particularly in the low-$\zmuta$ mass regime~\cite{ATLAS:2023vxg, CMS:2018yxg}, owing to its tree-level coupling to muons. In particular, the decay $Z\to4\mu$ provides a clean probe of this scenario. The process proceeds through $Z\to\mu^+\mu^-$, followed by the radiation of a $\zmuta$ boson from one of the muon final-states, which subsequently decays into a $\mu^+\mu^-$ pair, resulting in the characteristic $4\mu$ final state. The invariant mass of the $4\mu$ system peaks at the $Z$ boson mass. The resulting constraints in the $\mzmuta-\gmuta$ plane, extracted from~\cite{ATLAS:2023vxg}, are shown in~\autoref{fig:muC1} (\textit{left} panel), while the corresponding limits in the $\mzmuta-\vmuta$ plane are shown in the \textit{right} panel. The constraints weaken as the $\zmuta$ mass approaches the $Z$ boson mass, where the decay proceeds increasingly off shell and is consequently suppressed by phase-space effects.

Similar effects can also arise in the $Z\to2\mu2\tau$ and $Z\to4\tau$ decay modes. However, the reconstruction and identification of $\tau$ final states are comparatively more challenging than those of muons. Drell-Yan processes, $pp\to\mu^+\mu^-/\tau^+\tau^-$, can also receive contributions from $\zmuta$-mediated processes. However, these effects are subdominant due to the absence of tree-level couplings between $\zmuta$ and quarks, while the relatively large SM cross sections make deviations difficult to resolve within the available statistical precision. In our mass region of interest, $h_2$ is sufficiently heavy to evade current searches, with its interactions resembling those of the SM Higgs boson, $h_1$, but with couplings suppressed by the scalar mixing. DM production can proceed through all three mediators, $h_1$, $h_2$, and $\zmuta$. However, given the relatively heavy mediator masses favoured by other constraints, together with the suppressed couplings of these mediators to initial-state quarks and gluons, the resulting production cross section is too small to yield an observable signal in conventional DM searches like mono-jet plus missing transverse energy.

\subsection{Search at Muon Colliders}
\label{sec:muc}
Future muon colliders~\cite{Accettura:2023ked,MuonCollider:2022xlm} offer a promising opportunity for probing the $U(1)_{L_\mu-L_\tau}$ model~\cite{Huang:2021nkl,Huang:2021biu}, owing to the tree-level coupling of the initial-state muons to the $\zmuta$ boson, which plays a central role in this $U(1)$ extension. The presence of the $\zmuta$ boson leads to deviations in dilepton production processes such as $\mu^+\mu^-\to\mu^+\mu^-$ and $\mu^+\mu^-\to\tau^+\tau^-$, which could be probed with high precision at future muon colliders. Since the $\zmuta$ couples to muons and taus with opposite charges, these deviations carry a characteristic interference pattern that distinguishes the $L_\mu-L_\tau$ structure from other neutral gauge bosons. The sensitivity to these deviations, however, depends both on their interference with the SM contributions and on the relevant SM background processes.

\subsubsection*{Dilepton Production Processes}
We consider the two dilepton channels $\mu^+\mu^-\to\mu^+\mu^-$ and $\mu^+\mu^-\to\tau^+\tau^-$ at centre-of-mass (CM) energies $\sqrt{s}=3$ and $10$~TeV, the benchmark collider runs suggetsed in~\cite{Accettura:2023ked,MuonCollider:2022xlm}. The analysis presented here aims to identify the observables that carry the characteristic $L_\mu-L_\tau$ pattern. These are subject to the experimental viability, which is beyond the scope of this study and is discussed in other works like~\cite{Huang:2021nkl,Huang:2021biu}. We compute all quantities analytically at tree level, considering unpolarised beams and massless leptons ($m_{\mu,\tau}\ll\sqrt{s}$), and we restrict the final-state leptons to the angular acceptance $|\cos\theta|< \cos{\theta_c}$, where $\theta_c=10^\circ$, to avoid divergences from $t$-channel processes. This is also mimics the shielding nozzles required against beam-induced backgrounds in muon-collider detectors~\cite{Bartosik:2020xwr}. The SM contribution includes $\gamma$ and $Z$ exchange. We assume that $\zmuta$ decays only into leptons, at tree-level, as other decay modes are one-loop or phase space suppressed.

Before discussing the prospects, address the limitations of this analysis and these must be addressed in a dedicated analysis. First, we do not estimate backgrounds. For $\mu^+\mu^-$ final states, the dominant contaminations arise from $\mu^+\mu^- Z(\nu\bar\nu)$, $W^+(\mu^+\nu_\mu)W^-(\mu^-\bar\nu_\mu)$ and $\tau^+(\mu^+\nu_\mu\bar\nu_\tau)\tau^-(\mu^-\bar\nu_\mu\nu_\tau)$, which can be suppressed with missing-energy requirements. The $\tau^+\tau^-$ channel is more challenging, since $\tau$ reconstruction relies on leptonic and hadronic decay modes with finite efficiencies, and processes such as $W^+W^-$, $e^+e^-$ and $\mu^+\mu^-\gamma$ can contaminate the signal through misidentification or undetected final-state particles~\cite{Huang:2021nkl}. Second, we neglect initial-state radiation, beam-energy spread and electroweak radiation, which are sizeable at multi-TeV lepton colliders~\cite{Han:2020uid}. They mainly affect the resonance region, while ratios and asymmetries are partly protected against them. Third, the $\tau$ polarisation is obtained from the energy spectra of the $\tau$ decay products, as at LEP~\cite{ALEPH:2001uca}, and we do not simulate this step. Our results should therefore be read as the size and pattern of the expected deviations from the SM, not as sensitivity projections.

\subsubsection*{Amplitudes and Observables}
We write the neutral-current interactions of the charged leptons $\ell=\mu,\tau$ as
\begin{equation}
  \mathcal{L} \supset \sum_{V=\gamma,Z,\zmuta} V_\alpha\, \bar \ell \gamma^\alpha \left[g_L^\ell(V)\,P_L + g_R^\ell(V)\,P_R\right] \ell ,
  \label{eq:muc_L}
\end{equation}
with 
\begin{equation}
    g^\ell_{L,R}(\gamma) = -e\,, \quad g_L^\ell(Z) = \frac{e}{s_W c_W}\left(-\frac12 + s_W^2\right)\,, \quad g_R^\ell(Z) = \frac{e}{s_W c_W}\, s_W^2\,,
\end{equation}
and, from \eq\eqref{eq:kinetic_term},
\begin{equation}
  g^\mu_{L,R}(\zmuta) = -\,g^\tau_{L,R}(\zmuta) \equiv g_\ell = Q^\ell_{\mu\tau}\, g_{\mu\tau} = \frac{Q^\ell_{\mu\tau}}{Q^S_{\mu\tau}}\,\frac{m_{\zmuta}}{v_{\mu\tau}} = \frac{m_{\zmuta}}{v_{\mu\tau}} ,
  \label{eq:muc_gl}
\end{equation}
where we used $m_{\zmuta} = Q^S_{\mu\tau}\, g_{\mu\tau}\, v_{\mu\tau}$ and $Q^\ell_{\mu\tau} = Q^S_{\mu\tau} = 2$. The $\zmuta$ coupling is purely vector in nature, and opposite in sign for $\mu$ and $\tau$. Its leptonic width, from decays into $\mu^+\mu^-$, $\tau^+\tau^-$, $\nu_\mu\bar\nu_\mu$ and $\nu_\tau\bar\nu_\tau$, is given by
\begin{equation}
  \Gamma_{\zmuta} = \frac{g_\ell^2}{4\pi}\, m_{\zmuta} = \frac{m_{\zmuta}^3}{4\pi v_{\mu\tau}^2} .
  \label{eq:muc_width}
\end{equation}
The fixed-width treatment below is reliable only for $\Gamma_{\zmuta}\ll m_{\zmuta}$, i.e.\ $v_{\mu\tau}\gg m_{\zmuta}/\sqrt{4\pi}$. For $\mu^-(p_1)\mu^+(p_2)\to \ell^-(p_3) \ell^+(p_4)$, let $\theta$ be the angle between the incoming $\mu^-$ and the outgoing $f^-$, with Mandelstem variables, $t=-\frac{s}{2}(1-\cos\theta)$ and $u=-\frac{s}{2}(1+\cos\theta)$. For massless fermions only helicity-conserving amplitudes survive. Labelling them by the chiralities $i,j\in\{L,R\}$ of the incoming $\mu^-$ and the outgoing $f^-$, we define the reduced $s$ and $t$-channel amplitudes
\begin{equation}
  G^{(s)}_{ij} = \sum_{V} \frac{g^\mu_i(V)\, g^f_j(V)}{s - m_V^2 + i\, m_V \Gamma_V} , \qquad
  G^{(t)}_{ij} = \sum_{V} \frac{g^\mu_i(V)\, g^\mu_j(V)}{t - m_V^2} .
  \label{eq:muc_G}
\end{equation}
The process $\mu^+\mu^-\to\tau^+\tau^-$ proceeds only through the $s$-channel, and the differential cross section is
\begin{equation}
  \frac{d\sigma_{\tau\tau}}{d\cos\theta} = \frac{s}{128\pi}\Big[(1+\cos\theta)^2\left(|G^{(s)}_{LL}|^2 + |G^{(s)}_{RR}|^2\right) + (1-\cos\theta)^2\left(|G^{(s)}_{LR}|^2 + |G^{(s)}_{RL}|^2\right)\Big] .
  \label{eq:muc_dsig_tau}
\end{equation}
For $\mu^+\mu^-\to\mu^+\mu^-$ the $t$-channel exchange of the same bosons contributes as well, giving the differential cross section,
\begin{equation}
\begin{split}
  \frac{d\sigma_{\mu\mu}}{d\cos\theta} &= \frac{1}{32\pi s}\Big[u^2\left(|G^{(s)}_{LL}+G^{(t)}_{LL}|^2 + |G^{(s)}_{RR}+G^{(t)}_{RR}|^2\right) \\ &+ t^2\left(|G^{(s)}_{LR}|^2 + |G^{(s)}_{RL}|^2\right) + s^2\left(|G^{(t)}_{LR}|^2 + |G^{(t)}_{RL}|^2\right)\Big].    
\end{split}
\label{eq:muc_dsig_mu}
\end{equation}
Within the polar acceptance cut, the angular integrals of \eq\eqref{eq:muc_dsig_tau} are elementary, and the different $\tau\tau$ observables are listed below:
\begin{itemize}
\item Total cross section, $\sigma_{\tau\tau}$, given by
\begin{align}
  \sigma_{\tau\tau} &= \frac{s \cos{\theta_{c}}}{128\pi}\left(2 + \tfrac{2}{3} \cos^2{\theta_{c}}\right) \sum_{i,j}|G^{(s)}_{ij}|^2\,, \label{eq:muc_sig}
\end{align}
\item Forward-Backward asymmetry, $A_{FB}^{\tau\tau}$, given by
\begin{align}
  A_{FB}^{\tau\tau} &= \frac{\sigma(\cos\theta>0)-\sigma(\cos\theta<0)}{\sigma(\cos\theta>0)+\sigma(\cos\theta<0)} \nonumber \\ &= \frac{3 \cos{\theta_{c}}}{3+\cos^2{\theta_{c}}}\;\frac{|G^{(s)}_{LL}|^2 + |G^{(s)}_{RR}|^2 - |G^{(s)}_{LR}|^2 - |G^{(s)}_{RL}|^2}{\Sigma}\,, \label{eq:muc_afb}
\end{align}
\item Longitudinal polarization of $\tau^-$, $P_{\tau}$, given by
\begin{align}
  P_\tau &= \frac{\sigma(\tau^-_R)-\sigma(\tau^-_L)}{\sigma(\tau^-_R)+\sigma(\tau^-_L)} \nonumber \\ &= \frac{|G^{(s)}_{RR}|^2 + |G^{(s)}_{LR}|^2 - |G^{(s)}_{LL}|^2 - |G^{(s)}_{RL}|^2}{\sum_{i,j}|G^{(s)}_{ij}|^2}\,, \label{eq:muc_ptau}
\end{align}
\end{itemize}
For $\mu\mu$, the $t$-channel makes the angular integration non-polynomial, and we integrate \eq\eqref{eq:muc_dsig_mu} numerically. To quantify the effect of the $\zmuta$, we use the following quantities:
\begin{equation}
\begin{split}
\delta_{\ell\ell} &= \frac{\sigma_{\ell\ell}}{\sigma_{\ell\ell}^{\rm SM}} - 1\,, \qquad \Delta A_{FB}^{\ell\ell} = A_{FB}^{\ell\ell} - \left[A_{FB}^{\ell\ell}\right]^{\rm SM}\,,\\
\delta_R &= \frac{R_{\tau\mu}}{R_{\tau\mu}^{\rm SM}} - 1\,, \qquad \Delta P_\tau = P_\tau - P_\tau^{\rm SM}\,, \\ 
\end{split}
\label{eq:muc_obs}
\end{equation}
Here, $R_{\tau\mu} = ({\sigma_{\tau\tau}}/{\sigma_{\mu\mu}})$. It should be noted that, in the ratio $R_{\tau\mu}$ and in the asymmetries, normalisation uncertainties such as the luminosity cancel.

\subsubsection*{Relative Deviations}
We observe that, $\zmuta$ adds the same term to every $s$-channel amplitude, i.e.
\begin{equation}
  \delta G^{(s)}_{ij} = \pm\, \frac{g_\ell^2}{s - m_{\zmuta}^2 + i\, m_{\zmuta}\Gamma_{\zmuta}}\,,
  \label{eq:muc_dG}
\end{equation}
with $(+)$ for $\mu\mu$ and $(-)$ for $\tau\tau$, while the photon contributes $+(e^2/s)$. In the $\tau\tau$ channel, the $\zmuta$ therefore interferes constructively with the photon below the resonance ($s<m_{\zmuta}^2$) and destructively above it. For a universal $Z'$ with equal-sign couplings to $\mu$ and $\tau$, the pattern would be reversed. In the weak-coupling regime, and keeping only the photon, the leading term gives
\begin{equation}
\begin{split}
  \delta_{\tau\tau} \simeq +\frac{2s}{e^2 v_{\mu\tau}^2}\;\left(m_{\zmuta}^2\gg s\right)\,, \quad \text{and,} \quad
  \delta_{\tau\tau} \simeq -\frac{2g_\ell^2}{e^2} = -\frac{2\,m_{\zmuta}^2}{e^2 v_{\mu\tau}^2} \; \left(m_{\zmuta}^2\ll s\right) .    
\end{split}
\label{eq:muc_limits}
\end{equation}
For a heavy $\zmuta$, the deviation thus depends on $v_{\mu\tau}$ alone, the same parameter that controls the LZ rate. When $g_\ell^2\gtrsim e^2$, the squared pure $\zmuta$ term dominates and the deviation becomes positive irrespective of the interference sign. In the $\mu\mu$ channel, the $\zmuta$ also contributes to the $t$-channel, through $G^{(t)}_{ij}\supset g_\ell^2/(t-m_{\zmuta}^2)$. Since $t<0$, this term has the same sign as the $t$-channel photon contribution, $e^2/t$, so the interference is always constructive. As $\sigma_{\mu\mu}$ is dominated by the $t$-channel, we have $\delta_{\mu\mu}>0$ almost everywhere. The two channels therefore play complementary roles: $\mu\mu$ measures the strength of the muon coupling $g_\ell^2$, while $\tau\tau$ measures the relative sign of the $\mu$ and $\tau$ couplings.

\begin{figure}[htb!]
  \centering
  \includegraphics[width=\textwidth]{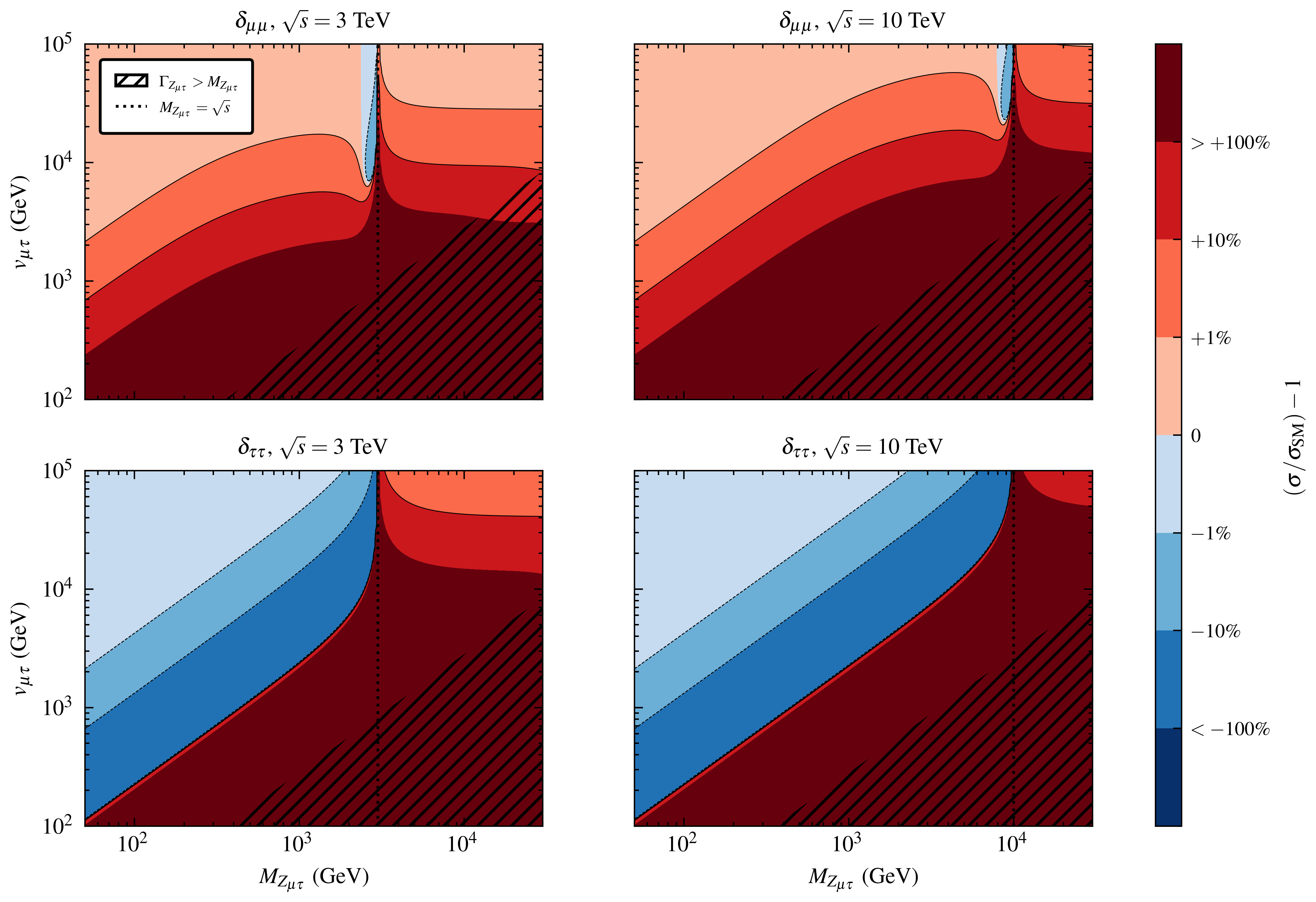}
  \caption{Relative deviations $\delta_{\mu\mu}$ (\textit{top}) and $\delta_{\tau\tau}$ (\textit{bottom}) from the SM cross sections in the $m_{\zmuta}-v_{\mu\tau}$ plane for $\sqrt{s}=3$ TeV (\textit{left}) and $10$ TeV (\textit{right}), with $|\cos\theta|<\cos 10^\circ$. \textit{Red} (\textit{blue}) regions correspond to an excess (deficit). The dotted vertical line indicates $m_{\zmuta} = \sqrt{s}$, and the hatched region corresponds to where the fixed-width treatment is unreliable i.e. $\Gamma_{\zmuta} > m_{\zmuta}$.}
  \label{fig:muc_delta}
\end{figure}

\autoref{fig:muc_delta} shows $\delta_{\mu\mu}$ and $\delta_{\tau\tau}$ in the $m_{\zmuta}-v_{\mu\tau}$ plane. As per expectation, $\delta_{\mu\mu}$ is positive over the whole plane, and, $\delta_{\tau\tau}$ changes sign at $m_{\zmuta}\simeq\sqrt{s}$. It is negative for a light $\zmuta$ and positive for a heavy one, except in the strong-coupling region at small $v_{\mu\tau}$, where the squared pure $\zmuta$ term dominates. For $m_{\zmuta}\gg\sqrt{s}$, the contours become horizontal, in agreement with the contact interaction limit of \eq\eqref{eq:muc_limits}.
\begin{figure}[htb!]
  \centering
  \includegraphics[width=\textwidth]{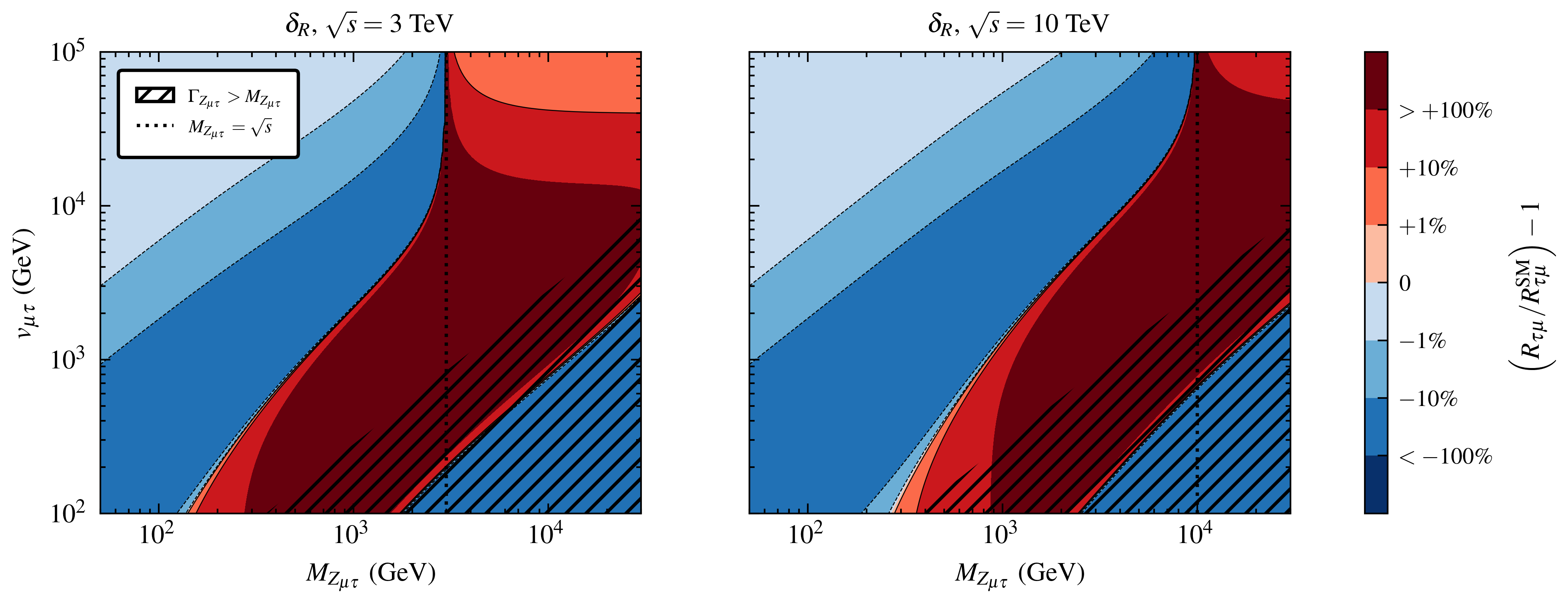}
  \caption{Deviation $\delta_R$ of the flavour ratio $R_{\tau\mu} = (\sigma_{\tau\tau}/\sigma_{\mu\mu})$ from its SM value for $\sqrt{s}=3$ TeV (\textit{left}) and $10$ TeV (\textit{right}). Colour coding is same as in \autoref{fig:muc_delta}.}
  \label{fig:muc_ratio}
\end{figure}
\autoref{fig:muc_ratio} shows the deviation of the flavour ratio $R_{\tau\mu}$. Since $\delta_{\mu\mu}$ and $\delta_{\tau\tau}$ have opposite signs for $m_{\zmuta}<\sqrt{s}$, they add up in the ratio. For $m_{\zmuta}>\sqrt{s}$, $\delta_R$ closely follows $\delta_{\tau\tau}$, since $\delta_{\mu\mu}$ is much smaller there. The ratio thus retains the full sensitivity of the $\tau\tau$ channel, while being independent of the accumulated luminosity.

\begin{figure}[t]
  \centering
  \includegraphics[width=\textwidth]{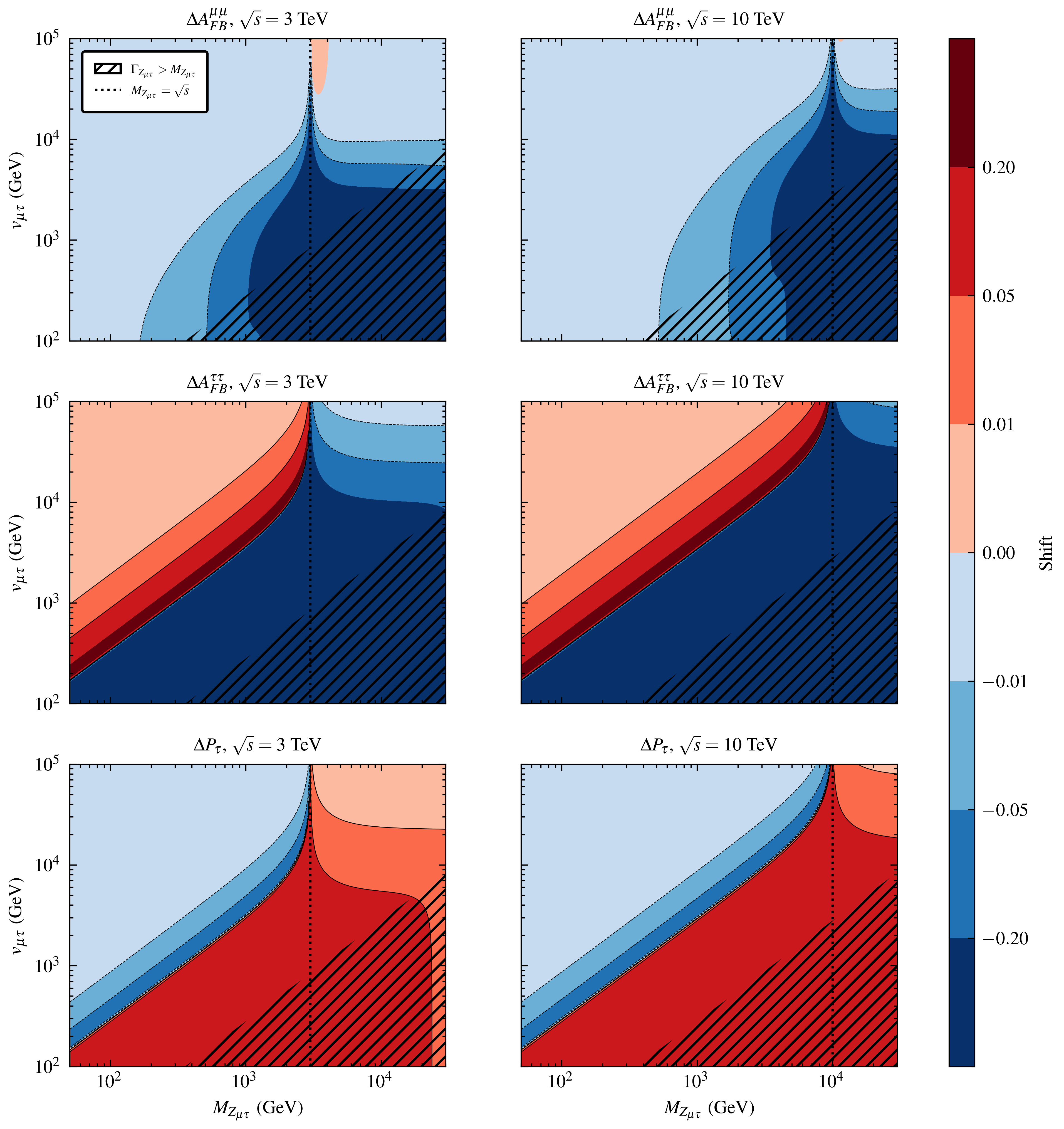}
  \caption{Shifts of the forward-backward asymmetries $\Delta A_{FB}^{\mu\mu}$ (\textit{top}) and $\Delta A_{FB}^{\tau\tau}$ (\textit{middle}) and $\Delta P_\tau$ (\textit{bottom}) with respect to the SM, for $\sqrt{s}=3$ TeV (\textit{left}) and $10$ TeV (\textit{right}). The conventions are same as in \autoref{fig:muc_delta}.}
  \label{fig:muc_asym}
\end{figure}

\subsubsection*{Asymmetries and $\boldsymbol{\tau}$ Polarisation}
In the SM, $A_{FB}^{\tau\tau}$ and $P_\tau$ arise from the chiral structure of the $Z$ couplings, which makes the four amplitudes $G^{(s)}_{ij}$ different. The vector interaction of $\zmuta$ adds the same term to all four, as seen in \eq\eqref{eq:muc_dG}, and therefore shifts these quantities in a characteristic way, shown in~\autoref{fig:muc_asym}. In the weak-coupling regime, $\Delta A_{FB}^{\tau\tau}>0$ and $\Delta P_\tau<0$ for $m_{\zmuta}<\sqrt{s}$, and the signs are reversed for $m_{\zmuta}>\sqrt{s}$, again reflecting the change of sign of the interference. The shift of $A_{FB}^{\mu\mu}$ is much smaller, since the asymmetry of the $\mu\mu$ channel is dominated by the forward-peaked $t$-channel photon exchange i.e. $\left[A_{FB}^{\mu\mu}\right]^{\rm SM}\simeq 1$ within the angular acceptance. These asymmetries complement the total rate-based measurements discussed before and they are insensitive to the overall normalisation, and their sign pattern provides an independent test of the vector structure and the flavour-opposite couplings.

\begin{figure}[htb!]
  \centering
  \includegraphics[width=\textwidth]{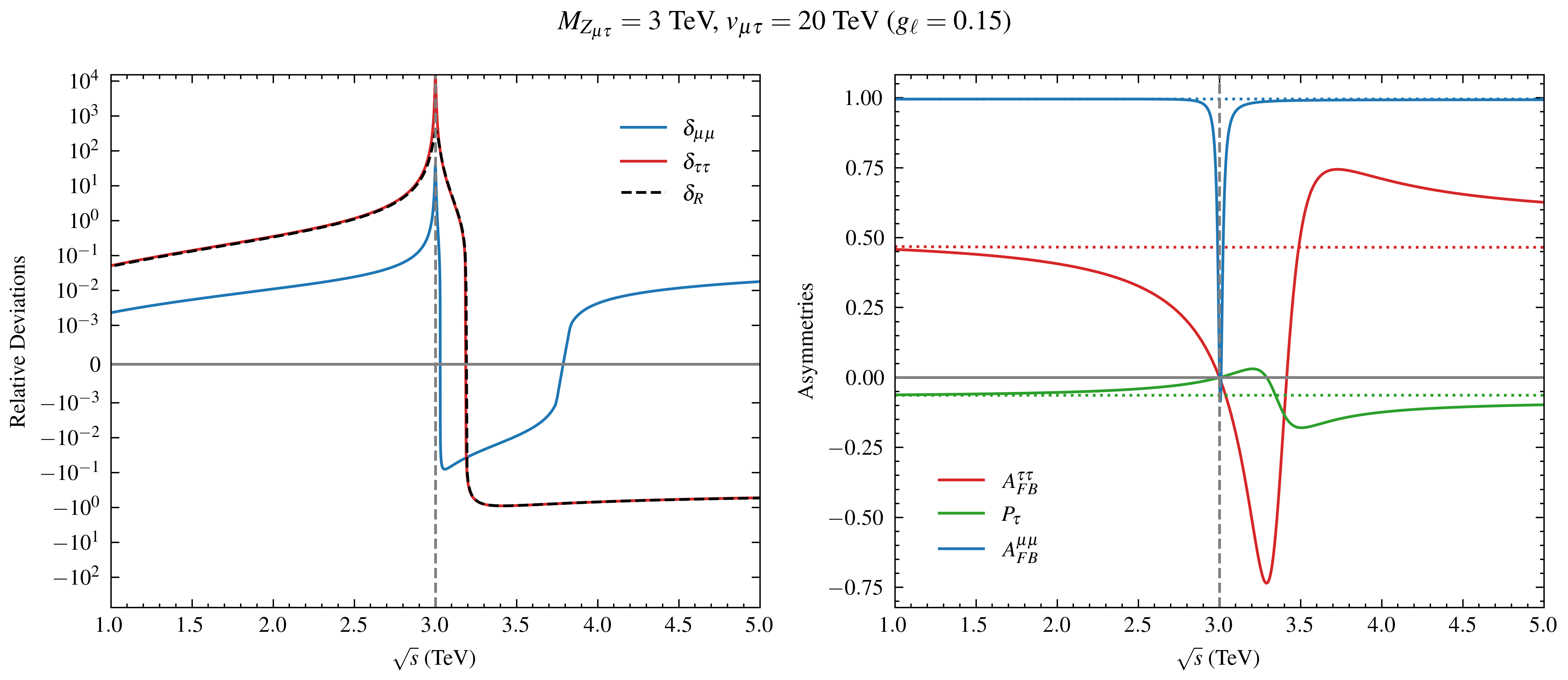}
  \caption{Energy dependence of the observables for a benchmark with $m_{\zmuta} = 3$ TeV and $v_{\mu\tau} = 20$ TeV ($g_\ell = 0.15$, $\Gamma_{\zmuta}\simeq 5.4$~GeV). \textit{Left}: $\delta_{\mu\mu}$, $\delta_{\tau\tau}$ and $\delta_R$. \textit{Right}: $A_{FB}^{\tau\tau}$, $P_\tau$ and $A_{FB}^{\mu\mu}$ (\textit{solid}), with the SM values (\textit{dotted}).}
  \label{fig:muc_escan}
\end{figure}

\autoref{fig:muc_escan} shows the observables as a function of $\sqrt{s}$ for a weakly coupled benchmark, $m_{\zmuta}=3$ TeV and $v_{\mu\tau}=20$ TeV. Below the resonance, $\delta_{\tau\tau}$ is positive and grows from about $5\%$ at $\sqrt{s}=1$ TeV to $35\%$ at $2$ TeV. Above the resonance it turns into a deficit of $\mathcal{O}(50-90\%)$, because the $\zmuta$ and photon amplitudes nearly cancel. $\delta_{\mu\mu}$ stays at the per-cent level and positive, except in a narrow window just above the peak, where the $s$-channel interference briefly makes it negative. On the peak, the $\zmuta$ dominates all four $s$-channel amplitudes equally, so $A_{FB}^{\tau\tau}\to 0$ and $P_\tau\to 0$. These are direct consequences of its purely vector coupling structure. Moreover, $A_{FB}^{\mu\mu}$ drops from its $t$-channel-dominated value from $\sim 1$ to $\sim 0$. Just above the peak, $A_{FB}^{\tau\tau}$ swings from about $-0.5$ to $+0.7$ before slowly returning toward the SM value, driven by the interference of the vector $\zmuta$ with the axial $Z$ couplings. At a muon collider operating at a fixed energy, the resonance region can be accessed through radiative return processes like $\mu^+\mu^-\to\zmuta\gamma$ with $\zmuta\to\ell^+\ell^-$, where the peak in the dilepton invariant mass traces the pole mass of the $\zmuta$. The vanishing of $A_{FB}^{\tau\tau}$ and $P_\tau$ and the equal rates into $\mu^+\mu^-$ and $\tau^+\tau^-$ on the peak would then identify the coupling structure.

\subsubsection*{Model Discrimination}
\begin{figure}[htb!]
  \centering
  \includegraphics[width=\textwidth]{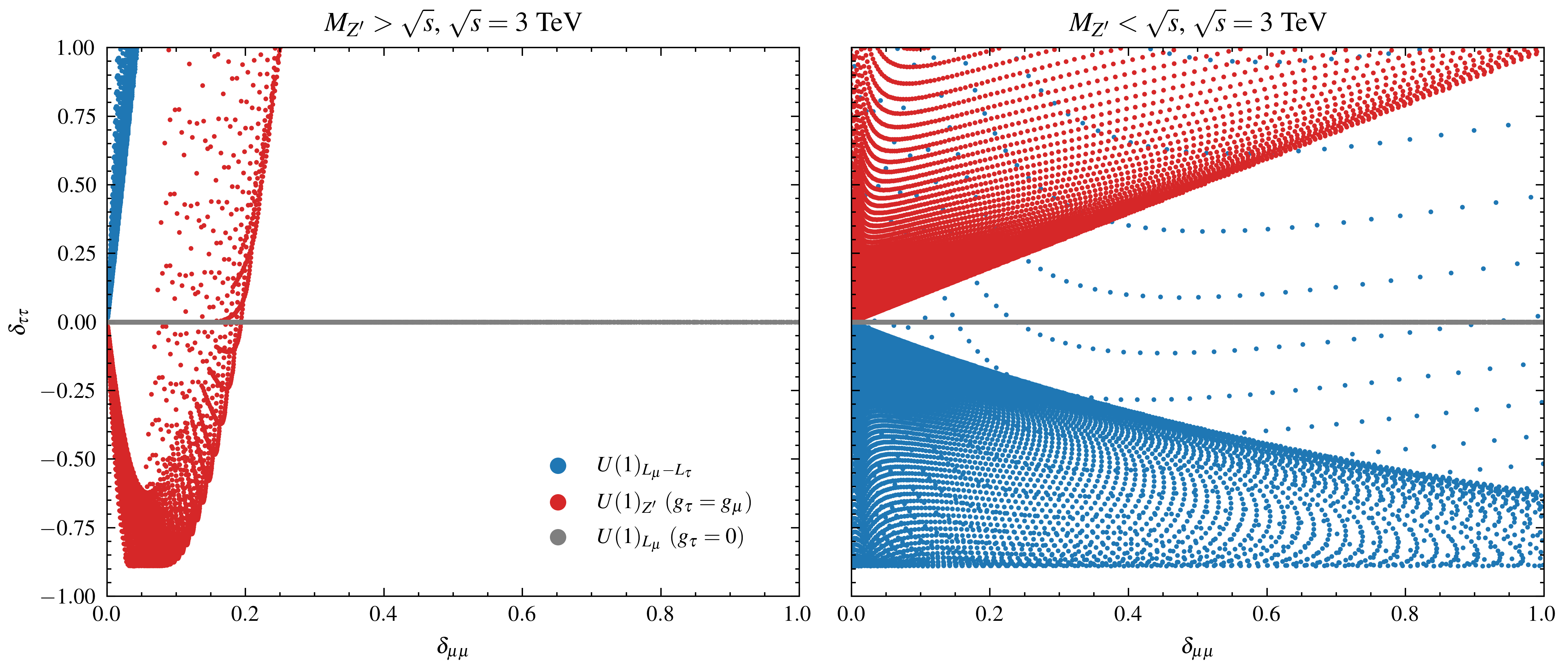}
  \caption{Correlation between $\delta_{\mu\mu}$ and $\delta_{\tau\tau}$ at $\sqrt{s}=3$ TeV for $m_{\zmuta}>\sqrt{s}$ (\textit{left}) and $m_{\zmuta}<\sqrt{s}$ (\textit{right}), obtained by scanning the $m_{\zmuta}-v_{\mu\tau}$ plane with $\Gamma_{Z'}<m_{Z'}$, for different models indicated in the legend. All three models have the same coupling to muons.}
  \label{fig:muc_disc}
\end{figure}

To illustrate how the $L_\mu-L_\tau$ structure can be identified,~\autoref{fig:muc_disc} compares the correlation between $\delta_{\mu\mu}$ and $\delta_{\tau\tau}$ for three $Z'$ model variations with the same coupling to muons ($g_\mu$): (\textbf{A}) the $L_\mu-L_\tau$ scenario, with $g_\tau = -g_\mu$, (\textbf{B}) a universal $Z'$ with $g_\tau = g_\mu$, and (\textbf{C}) a pure $L_\mu$ muonic $Z'$ with $g_\tau = 0$. Since the $t$-channel interference is constructive in all three cases, $\delta_{\mu\mu}>0$ throughout, and $\mu\mu$ alone cannot distinguish them, however, the $\tau\tau$ channel can. For the pure muonic $Z'$ case, $\delta_{\tau\tau}=0$. For a heavy $Z'$ ($m_{Z'}>\sqrt{s}$), the $L_\mu-L_\tau$ neutral boson interferes constructively in the $\tau\tau$ channel and always gives $\delta_{\tau\tau}>0$. The universal $Z'$ interferes destructively and gives $\delta_{\tau\tau}<0$, except at large couplings, where the squared $Z'$ term dominates, and the deviation turns positive, accompanied by a larger $\delta_{\mu\mu}$. For a light $Z'$ ($m_{Z'}<\sqrt{s}$), the roles are swapped. The universal $Z'$ always gives $\delta_{\tau\tau}>0$, while $L_\mu-L_\tau$ gives $\delta_{\tau\tau}<0$ except at large couplings. A deviation observed in both channels thus determines the relative sign of the $\mu$ and $\tau$ couplings, once it is known whether the $Z'$ lies above or below $\sqrt{s}$. This can be established either by comparing the two collider stages, or from the sign pattern of $\Delta A_{FB}^{\tau\tau}$ and $\Delta P_\tau$ in~\autoref{fig:muc_asym}. Together, these observables provide a handle on the flavor structure of the extended gauge sector.

DM searches can proceed along similar lines, with DM produced through processes such as $\mu^+\mu^-\to\phi_1\phi_1/\phi_1\phi_2/\phi_2\phi_2$ via $s$-channel $\zmuta$ mediation. Since these processes result in invisible final states, an additional $\gamma$ or $Z$ radiated from the initial-state muons provides a visible recoil signature. However, the dominant SM backgrounds, $\nu\bar\nu\gamma$ and $\nu\bar\nu Z$, can themselves receive contributions from $\zmuta$-mediated processes, making a straightforward sensitivity estimate challenging. A dedicated analysis including backgrounds and detector effects will therefore be necessary to assess the discovery and exclusion prospects of both the dilepton and the mono-$\gamma/Z$ channels.
\section{Summary and Conclusion}
\label{sec:summary}
The high-energy recoil event observed at the LZ experiment has opened up a new avenue for BSM scenarios involving inelastic DM nuclear scattering. In this work, we explore one such possibility by extending the $\ULmuLtau$ model with a pair of complex scalars carrying different charges under the $\ULmuLtau$ gauge group. This charge assignment allows for a trilinear interaction involving the two scalar fields, which, following spontaneous $\ULmuLtau$ symmetry breaking, generates a mass splitting between the components of the scalar field that do not acquire a VEV. This small mass splitting renders one component slightly heavier than the other, while the lighter component remains stable due to the remnant $\Ztwo$ symmetry and serves as the stable DM candidate, thereby providing a possible explanation of the inelastic recoil excess observed at LZ.

We first discuss the relic density-compatible parameter space for different combinations of model parameters, subject to constraints from DM stability and low-energy observables, including the muon $(g-2)$ and neutrino trident production limits from CCFR. We then consider the DD constraints from elastic scattering and find that the existing limits do not significantly restrict the viable parameter space. We subsequently turn to the central aspect of this work, namely, the possibility of explaining the single-event excess observed in the LZ experiment. We demonstrate the model's viability in the inelastic scattering regime and perform a two-parameter likelihood analysis to identify regions of parameter space consistent with the observed event. We further discuss constraints from ID and existing collider searches, and outline the prospects for probing the model at future muon colliders. Taken together, our analysis provides a comprehensive picture of the parameter space that can accommodate the LZ excess while remaining consistent with existing experimental constraints and observations.

\begin{figure}[htb!]
\centering
\includegraphics[width=0.75\linewidth]{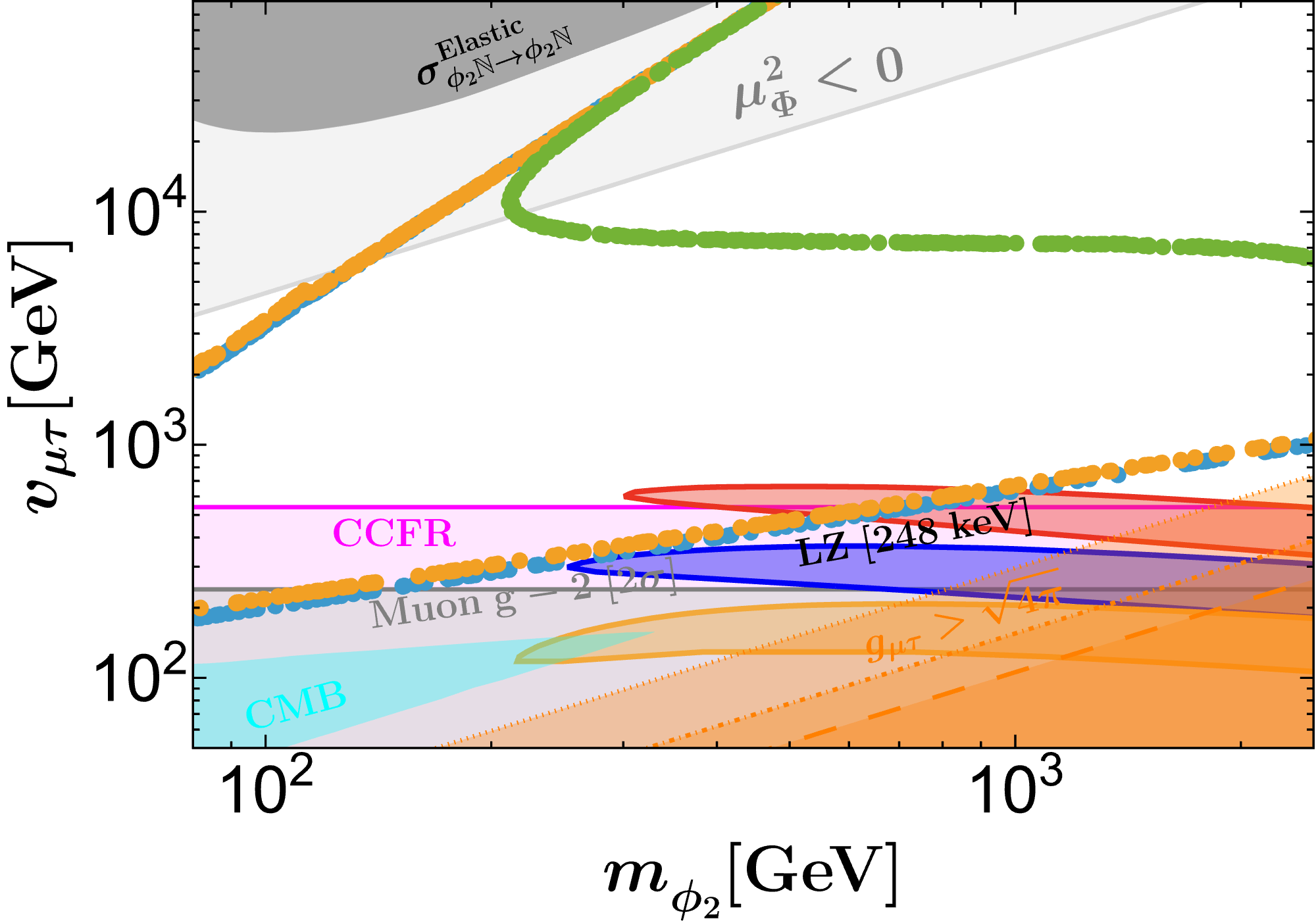}
\caption{The sky-blue, yellow, and green points satisfying the observed DM relic density correspond to $\mzmuta = (0.75, 1.1, 2.1)\,\mtwo$, respectively, for fixed benchmark parameters $\mhtwo = 2.2\mtwo$, $\deltam = 150~\text{keV}$, $\sin\alpha = 0.05$, and $\lphiS = \lphiH = 10^{-3}$. The red, blue, and orange shaded regions denote the favored $68\%$ C.L. contours corresponding to the LZ single-event observation for $\deltam = (100,~150,~200)~\text{keV}$, respectively. The CMB exclusion limit, indicated by the cyan shading, applies exclusively to $\mzmuta = 0.75\mtwo$ and is relaxed for the other choices. The light-grey, medium-grey, deep-grey, and magenta shaded regions are disfavored by DM stability, muon $g-2$ measurements, LZ elastic scattering, and neutrino trident production constraints from CCFR. For each observable, only the most conservative exclusion limits in the $\mtwo\text{--}\vmuta$ plane are displayed. The orange-shaded regions bounded by the dashed, dot-dashed, and dotted lines, corresponding to $\mzmuta=(0.75,\,1.1,\,2.1)\,\mtwo$, respectively, are excluded by the perturbativity constraint $\gmuta$.}
\label{fig:scan3}
\end{figure}
Finally, we summarise our results in \autoref{fig:scan3}, which combines all relevant bounds applicable to the $\mtwo-\vmuta$ plane. The other parameters are fixed as $\mzmuta = x~\mtwo$, $\mhtwo = 2.2\mtwo$, $\deltam = 150~\text{keV}$, $\sin\alpha = 0.05$, and $\lphiS = \lphiH = 10^{-3}$. The relic density satisfied regions are shown by sky blue, yellow, and green points for $x = (0.75, 1.1, 2.1)$, respectively. A similar explanation applies to the relic density, which is divided into two regimes: the upper half is driven by scalar domination, and the lower half corresponds to $\zmuta$ dominance. With an increase in $x$ (i.e., an enhancement of $\mzmuta$), the $\zmuta$-mediated processes become suppressed, requiring a larger value of $\vmuta$ to achieve the correct relic abundance via scalar-mediated processes. Regarding the LZ bounds, higher values of $\vmuta$ ($\gtrsim 20~\TeV$) face the LZ exclusion limit (dark grey shade) following \eq\eqref{eq:phi2N}, considering $\mhtwo = 2.2\mtwo$ and $\sin\alpha = 0.05$. However, this region is already disallowed by DM stability criteria. Beyond these, several other limits apply to this plane, including ID limits using CMB data (indicated by the cyan shade), Muon $g-2$ (medium grey), and most importantly, neutrino trident production by CCFR (magenta shade). Notably, the LZ excess can only be explained by the region defined by $0.3 \lesssim \mtwo/\TeV \lesssim 2.0$ and $0.55 \lesssim \vmuta/\TeV \lesssim 0.70$ for $\deltam = 100\,\text{keV}$ (represented by the red colour shade), whereas mass splittings of $\deltam = 150$ and $200\,\text{keV}$ are incompatible with the CCFR exclusion. Furthermore, IceCube constraints on dark matter annihilation do not affect the results presented above.

The framework explored in this work provides a well-motivated realisation of inelastic DM, in which the charge assignments of the extended scalar sector naturally allow the small mass splitting required to account for the observed recoil. At the same time, the $\ULmuLtau$ gauge portal connects the dark sector to different SM states with characteristic interaction strengths, leading to complementary probes across low-energy, DD, ID, and collider experiments. We find that the viable parameter space accommodating the LZ excess remains largely unconstrained by existing experimental searches, leaving significant room for further exploration. Importantly, the parameter space identified in this work can be tested with upcoming LZ data, which can provide crucial evidence for or against our inelastic nuclear recoil interpretation, while future muon colliders offer a complementary and powerful probe of the $\ULmuLtau$ gauge sector through its tree-level coupling to muons and taus. Together, these complementary probes make the proposed framework extremely promising for future validations of the LZ excess and the underlying nature of the dark sector.


\appendix
\section{Metastability of Dark Partner}
\label{app:stability}
In this section, we investigate the long-lived nature of the dark partner, $\phi_1$, whose decay is suppressed by the small mass splitting and proceeds through a three-body decay into the DM candidate and neutrinos.
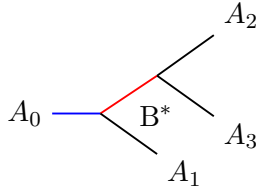
\begin{figure}[htb!]\centering
\begin{tikzpicture}
\begin{feynman}
\vertex (a1){\(A_0\)};
\vertex [right=1.0 cm of a1] (a2);
\vertex [above right=0.5 cm and 0.75 cm of a2] (a3);
\vertex [below right=0.5 cm and 0.75 cm of a2] (a4){$A_1$};
\vertex [above right=0.5 cm and 0.75 cm of a3] (a5){$A_2$};
\vertex [below right=0.5 cm and 0.75 cm of a3] (a6){$A_3$};
\diagram*{
(a1) -- [ line width=0.25mm,plain, arrow size=0.7pt, style=blue, edge label={\(\rm\color{black}{}\)}] (a2), 
(a2)-- [ line width=0.25mm, plain, style=red, arrow size=0.7pt , edge label'={\(\color{black}{\rm B^{*}}\)}] (a3) ,  
(a4)-- [ line width=0.25mm,plain, style=black, arrow size=0.7pt] (a2) ,  
(a5)-- [ line width=0.25mm,plain, arrow size=0.7pt ] (a3)-- [ line width=0.25mm,plain, arrow size=0.7pt] (a6)};
\end{feynman}
\end{tikzpicture}
\caption{Figure show the three-body decay, $A_0\to A_1~A_2~A_3$, via off-shell $\rm B$.}
\label{fig:phi1phinunu}
\end{figure}
We start by discussing a general three-body decay scenario: $A_0(p_0)\to A_1(p_1)~A_2(p_2)~A_3(p_3)$ (shown in~\autoref{fig:phi1phinunu}), in the rest frame of the decaying particle $A_0$ is given in terms of dalitz variable as,
\begin{gather}
\Gamma_{A_0\to A_1~A_2~A_3}=S_k\dfrac{1}{(2\pi)^3}\dfrac{1}{32m_0^3}\int dm_{12}^2\int dm_{23}^2~\overline{|\mathcal{M}|^2}_{A_0\to A_1~A_2~A_3}\,,
\end{gather}
where $m_{12}=(p_1+p_2)^2,~m_{23}=(p_2+p_3)^2,~m_{13}=(p_1+p_3)^2$ and
\begin{subequations}\begin{gather}
m_{23}^2|_{\rm max}=(E_2^*+E_3^*)^2-\left(\sqrt{{E_2^*}^2-m_2^2}-\sqrt{{E_3^*}^2-m_3^2}\right)^2\,,\\
m_{23}^2|_{\rm min}=(E_2^*+E_3^*)^2-\left(\sqrt{{E_2^*}^2-m_2^2}+\sqrt{{E_3^*}^2-m_3^2}\right)^2\,,\\
m_{12}^2|_{\rm max}=(m_0-m_3)^2 {\rm~and ~}  m_{12}^2|_{\rm min}=(m_1+m_2)^2\,,\\
E_2^*=(m_{12}^2-m_1^2+m_2^2)/2m_{12}{\rm~and ~} E_3^*=(m^2_0-m_{12}^2-m_3^2)/2m_{12}\,,
\end{gather}\end{subequations}
and $S_k=1/k!$ is the corresponding symmetry factor, with $k$ the number of identical final state particles. In our work, the heavier real scalar $\phi_1$ can decay into the leptons and the lightest partner $\phi_2$, depending on kinetic accessibility. If the splitting is smaller than twice the electron mass $(\deltam<2m_e)$, then possible decays only into neutrinos ($\nu_\mu$ and $\nu_\tau$). However, the corresponding decay width of $\phi_1$ into leptons ($\ell$ and $\nu_\ell$) can be expressed as follows, 
\begin{gather}
\Gamma(\phi_1\to\phi_2\nu_\ell\bar{\nu}_\ell)=\dfrac{g^4_{\mu\tau}(Q_{\mu\tau}^\ell Q^\Phi_{\mu\tau})^2}{192\pi^3m^3_{\phi_1}}\int^{\deltam^2}_{0}dq^2\dfrac{\lambda^{3/2}(m^2_{\phi_1},m^2_{\phi_2},q^2)}{(q^2-m^2_{\zmuta})^2+m^2_{\zmuta}\Gamma^2_{\zmuta}}\,,\\
\Gamma(\phi_1\to\phi_2\ell\bar{\ell})=\dfrac{g^4_{\mu\tau}(Q_{\mu\tau}^\ell Q^\Phi_{\mu\tau})^2}{96\pi^3m^3_{\phi_1}}\int^{\deltam^2}_{4m^2_\ell}dq^2\dfrac{\lambda^{3/2}(m^2_{\phi_1},m^2_{\phi_2},q^2)}{(q^2-m^2_{\zmuta})^2+m^2_{\zmuta}\Gamma^2_{\zmuta}}\left(1+2\dfrac{m^2_\ell}{q^2}\right)\sqrt{1-4\dfrac{m^2_\ell}{q^2}}\,,
\label{eq:phi1topll}
\end{gather}
where, $\lambda(a,b,c)=a^2+b^2+c^2-2ab-2bc-2ca$ and $\ell$ denotes the SM muon and tau leptons, and 
\begin{gather}
\Gamma_{\zmuta}=\sum_{\ell=\mu,\tau}\Gamma(\zmuta\to\ell\bar\ell)+\Gamma(\zmuta\to\nu_\ell{\bar\nu}_\ell)\,,
\end{gather}
with
\begin{gather}
\Gamma(\zmuta\to\nu_\ell\bar{\nu}_\ell)=\dfrac{Q_{\mu\tau}^2g^2_{\mu\tau}\mzmuta}{24\pi}\,,\\
\Gamma(\zmuta\to\ell\bar\ell)=\dfrac{Q_{\mu\tau}^2g^2_{\mu\tau}\mzmuta}{12\pi}\left(1+2\dfrac{m_\ell^2}{m^2_{\zmuta}}\right)\sqrt{1-4\dfrac{m_\ell^2}{m^2_{\zmuta}}}\,.
\end{gather}
Under the assumptions that ${m_1, m_2, \mzmuta}\gg\deltam$ and $\Gamma_{\zmuta}\ll \mzmuta$, the three-body decay width becomes,
\begin{gather}
\Gamma(\phi_1\to\phi_2\nu_\ell\bar{\nu}_\ell)\simeq\dfrac{(Q_{\mu\tau}^\ell Q^\Phi_{\mu\tau})^2\deltam^5}{60\pi^3(v_{\mu\tau}^2 Q_{\mu\tau}^S)^2}\,,\\
\Gamma(\phi_1\to\phi_2\ell\bar{\ell})\simeq\dfrac{(Q_{\mu\tau}^\ell Q^\Phi_{\mu\tau})^2\deltam^5}{30\pi^3(v_{\mu\tau}^2 Q_{\mu\tau}^S)^2}\,.
\end{gather}
\subsection*{BBN and CMB Consistency from Heavier Dark Scalar Decay:}
For the heavier thermal DM candidate $\phi_1$ with mass $\mone = \mtwo + \deltam$, freeze-out occurs at $T_{\text{FO}} \sim \mone/20$, well before the Big Bang Nucleosynthesis (BBN) epoch for $\mone \sim \mathcal{O}(\GeV)$. Nevertheless, this heavier state can survive post-freeze-out if its decay is sufficiently suppressed. Specifically, for 
$\mone < \mtwo + 2m_e$, $\phi_1$ decays via
\begin{align}
\phi_1 \rightarrow \phi_2 + \nu_\ell\, \bar{\nu}_\ell\,,
\end{align}
For small mass splittings relevant to the DM inelastic DD, the three-body decay width is strongly suppressed i.e. $\Gamma_{\phi_1} \propto (\deltam)^5$, potentially resulting in a lifetime comparable to or longer than the BBN timescale. A surviving $\phi_1$ population can inject non-thermal neutrinos during BBN, modifying primordial light-element abundances \cite{Chao:2022mcu, Fradette:2017sdd, Yeh:2024ors}. Rather than imposing a universal lifetime bound, we assess BBN consistency using the post-freeze-out abundance $Y_{\phi_1}$, the lifetime $\tau_{\phi_1}$, and the mass splitting $\deltam$. If $\phi_1$ decays sufficiently early or its abundance is negligible by the onset of BBN, its impact on primordial predictions remains minimal \cite{Kawasaki:2017bqm, Menestrina:2011mz, Kanzaki:2007pd, Cumberbatch:2007me, Poulin:2015opa, Poulin:2016nat, Hooper:2025fda, Slatyer:2012yq, Alves:2023jlo, Kawasaki:2020qxm, Hambye:2021moy}. Otherwise, the produced neutrinos from the scalar decay can alter primordial abundances by modifying the Hubble expansion rate ($\Delta N_{\text{eff}}$) and, if the neutrino energy exceeds the $782\text{ keV}$ threshold, directly driving charged-current reactions that alter the neutron-to-proton ratio \cite{Escudero:2019gfk, Hambye:2021moy}.
\begin{figure}[htb!]
\centering
\includegraphics[width=0.75\linewidth]{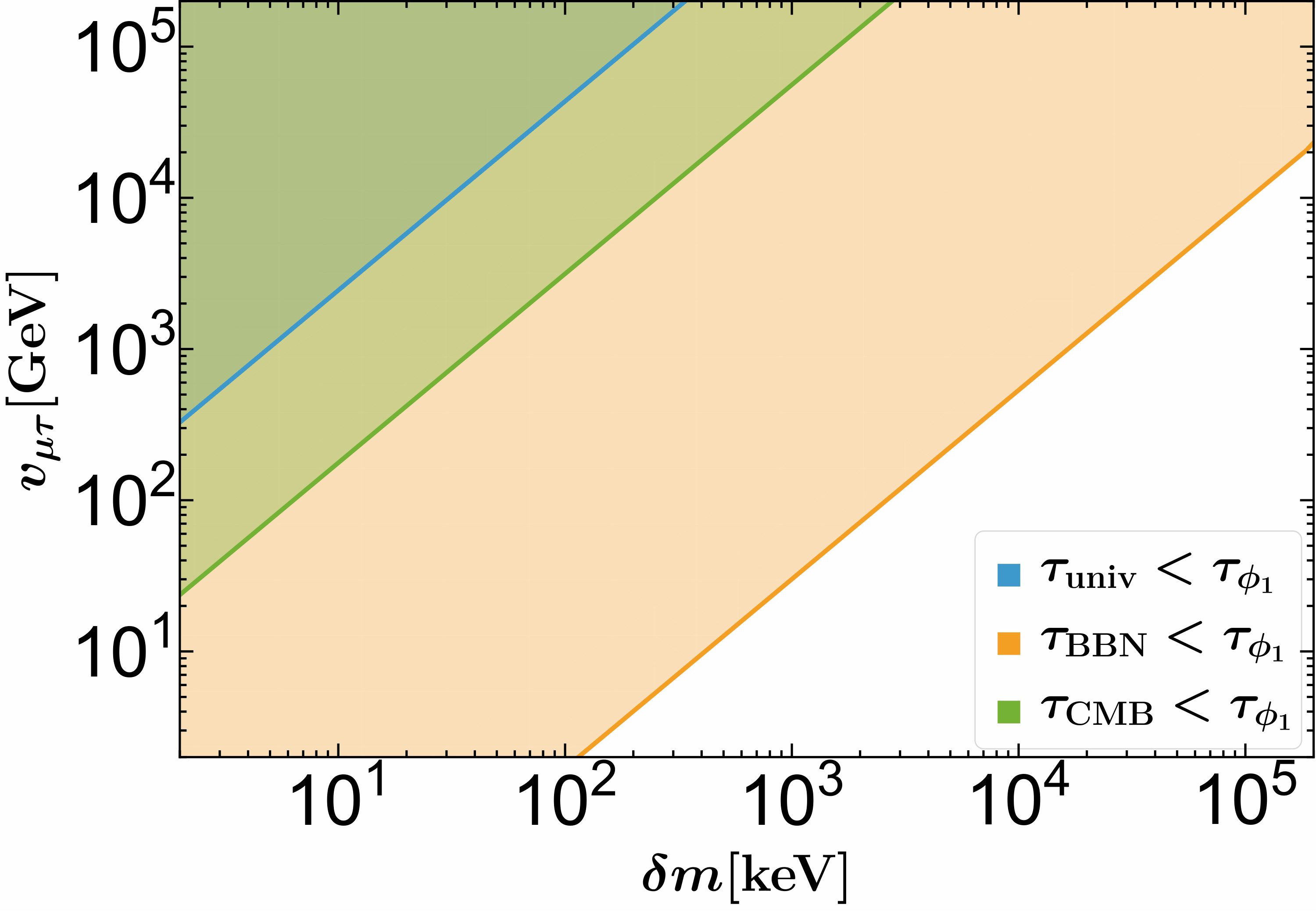}
\caption{The shaded region corresponds to the parameter space in which the $\phi_1$ decay corresponds to a lifetime longer than the corresponding epoch indicated in the legend, in the $\deltam-\vmuta$ plane. We use $\tau_{\rm BBN}\sim 1\text{ sec}\,=\,1.52\times10^{24}~{\rm GeV}^{-1}$ for the BBN epoch, $\tau_{\rm CMB}\simeq 1.2\times10^{13}\text{ sec}$ for the CMB recombination epoch, and $\tau_{\rm univ}\simeq 4.35\times10^{17}\text{ sec}$ for the age of the Universe.}
\label{fig:phi1decay}
\end{figure}
The shaded regions in \autoref{fig:phi1decay} represent the parameter space (in the $\deltam - \vmuta$ plane) beyond the epochs of BBN, CMB, and the present age of the Universe, where $\phi_1$ remains present alongside $\phi_2$ during these cosmological epochs. The dependence on other parameters is very mild and is thus omitted. In our working regime, BBN bounds are weaker for $\deltam \lesssim 782\text{ keV}$, whereas the CMB results remain approximately unaltered due to the choice of $\tau_{\phi_2}$. Nonetheless, precise quantitative statements require a dedicated analysis, which we leave for future work.
\section{IceCube Limit on Solar Capture}
\label{sec:IceCube}
IceCube observations of solar and atmospheric neutrinos provide stringent limits on DM decay and annihilation cross-sections, thereby constraining the production of high-energy secondary neutrinos. For Galactic Center (GC) signals, the diffuse neutrino flux scales directly with $\langle \sigma v \rangle$ and the line-of-sight integrated DM core density profile ($J$-factor); propagating through vacuum, these fluxes span $E_\nu \sim \mathcal{O}(\GeV-\rm PeV)$ depending on the DM mass. In contrast, neutrinos from solar core DM annihilation reflect the scattering, driven capture rate $(C_\odot)$ rather than annihilation dynamics, assuming capture-annihilation equilibrium. Given the abundance of solar Hydrogen, capture is dominated by spin-dependent (SD) DM-proton interactions, as spin-independent (SI) channels are severely constrained by terrestrial direct-detection experiments (such as LZ, XENONnT, etc.). While co-produced hadronic and vector-boson states undergo rapid energy loss and absorption in the dense solar plasma ($\rho_{\text{core}} \sim 150~ {\rm\,gm/ cm^{3}}$), sub-TeV neutrinos escape unattenuated; above $E_\nu \sim 1 \text{ TeV}$, high-energy neutrinos are heavily attenuated by inelastic scattering off solar matter. Consequently, solar DM searches yield a directional, point-like signal moving along the ecliptic plane, accessible via the IceCube DeepCore sub-array in the $10-100 \text{ GeV}$ regime \cite{IceCube:2016dgk}, distinguishable from the stationary, diffuse background of the Galactic halo. The details of solar-captured DM, the consequent neutrino production, and the resulting limits from IceCube observations are discussed below.

Assuming the Sun resides within the galactic DM halo, DM particles approach the solar rest frame with an asymptotic local velocity $u$. As they fall into the Sun's gravitational well, they are accelerated to a local velocity $w = \sqrt{u^2 + v_{esc}^2(r)}$, where $v_{esc}(r)$ is the local escape velocity at a given radius $r$ inside the Sun. During their transit, these DM particles can undergo elastic or inelastic scattering with solar nuclei (such as H, He, O, and Fe). Most particles simply pass through and escape because their kinetic energy remains too high. However, a particle is gravitationally captured if the recoil energy $E_R$ transferred to the target nucleus is sufficient to drop the DM particle's final velocity below the local escape velocity. Kinematically, this capture condition requires the energy loss to satisfy $E_R > \frac{1}{2} m_{\phi_2} u^2$ for elastic scattering \cite{Press:1985ug,Faulkner:1985rm,Kamionkowski:1991nj,Bottino:1994xp}, or $E_R > \frac{1}{2} m_{\phi_2} u^2 + \delta m$ for endothermic inelastic scattering \cite{Nussinov:2009ft,Menon:2009qj}.

Once gravitationally bound, the trapped DM particles follow closed elliptical orbits that repeatedly intersect the Sun. With each subsequent transit, they undergo further scattering events, losing more kinetic energy and causing their orbits to gradually shrink. Through these repeated scatterings (and subsequent decays back to the ground state in the inelastic case), the DM particles continuously lose energy until they sink to the centre and accumulate. Because the timescale for this orbital degradation and thermalisation is typically much shorter than the age of the Sun ($t_\odot \approx 1.5 \times 10^{17}$ s), we can safely assume that the trapped DM population reaches thermal equilibrium, i.e., DM capture rates and annihilation rates are equal, within the solar core.

Eventually, the captured DM particles annihilate or decay into SM particles. However, most of the SM particles produced in these processes cannot escape from the solar core and are instead absorbed by the solar material before reaching the solar surface. In contrast, high-energy neutrinos can escape from the Sun and travel to Earth. These neutrinos can be detected by neutrino telescopes in two ways: (i) by observing events produced through neutrino interactions within the detector, or (ii) by detecting the upward-going muons produced by neutrino interactions with the surrounding material, which subsequently traverse the detector. IceCube is a neutrino observatory capable of probing high-energy neutrinos over a broad energy range, from the GeV scale to the PeV scale. In this work, we use the published IceCube constraints to derive limits on the DM annihilation \cite{IceCube:2021xzo, IceCube:2025fcu, IceCube:2016dgk, IceCube:2012ugg} and decay rates, assuming no statistically significant excess of neutrinos above the expected background.

Notably, the IceCube DeepCore subarray extends sensitivity to neutrinos with energies of approximately $10\,\mathrm{GeV}$ and above. Consequently, neutrinos produced in the decay $\two\to\one\nu_\ell\bar{\nu}_\ell$ are generally too soft to be constrained by standard high-energy IceCube searches when the mass splitting is small, $\delta m \sim \mathcal{O}(\mathrm{keV})$. In this scenario, the relevant IceCube constraints primarily arise from $\two\two$ annihilations into $W^+W^-$, $\tau^+\tau^-$, $\mu^+\mu^-$, $b\bar{b}$, or $\nu_\ell\bar{\nu}_\ell$ final states. These processes can yield neutrinos with energies comparable to the dark matter mass, $\mtwo\sim\mathcal{O}(100)\,\mathrm{GeV}$, provided the respective channels are kinematically accessible.

The total capture rate, \(C_\odot\), of DM particles by the Sun can be evaluated following \cite{Gould:1987ww, Menon:2009qj, Nussinov:2009ft, Baum:2016oow, Bose:2021cou} as
\begin{align}
C_\odot
&=\sum_i\int_0^{R_\odot} d^3r\, n_i(r)
\int_{u_{\min}}^{u_{\max}} du\,\dfrac{\rho_{\phi_2}}{\mtwo}\ \frac{f(u)}{u}\,w^2
\int_{E_{\min}}^{E_{\max}} dE_R\,
\frac{d\sigma_i}{dE_R},
\label{eq:capture0}
\end{align}
where the local DM density is $\rho_{\phi_2}\simeq\rm 0.4\ GeV/cm^3$, \(i\) labels the elements present in the Sun, \(R_\odot\) is the solar radius, \(n_i(r)\) is the number density of the \(i\)-th target element, \(d\sigma_i/dE_R\) is the differential DM–nucleus scattering cross section with respect to the nuclear recoil energy \(E_R\), and $f_\odot(u)$ is the normalized DM speed distribution in the Solar frame. Here, \(u\) is the DM velocity far from the Sun, \(w=\sqrt{u^2+v_{\rm esc}^2(r)}\) is its local velocity at a radial distance \(r\), and \(v_{\rm esc}(r)\) is the local solar escape velocity. The velocity integration limits are determined by the incident DM velocity distribution and the kinematic requirements for capture. In particular, \(u_{\min}=0\), while \(u_{\max}\) is determined by the maximum velocity for which the DM particle can lose sufficient energy to become gravitationally bound to the Sun. For the standard Maxwellian model for the galactic velocity distribution \cite{Blennow:2018xwu},
\begin{align}
f(u)=\dfrac{1}{N_{\rm esc}}\dfrac{u}{\sqrt{\pi}v_0v_\odot}
\begin{cases}
e^{-(u-v_\odot)^2/v^2_0}-e^{-(u+v_\odot)^2/v^2_0}\,,\quad& u< v_{\rm esc}^{gal}-v_\odot\,,\\
e^{-(u-v_\odot)^2/v^2_0}-e^{-(v^{gal}_{\rm esc}/v_0)^2}\,,\quad& v_{\rm esc}^{gal}-v_\odot\leq u\leq  v_{\rm esc}^{gal}+v_\odot\,,\\
0\,,\quad& u>v_{\rm esc}^{gal}+v_\odot\,,\\
\end{cases}
\end{align}
where $v_0\simeq 220~\rm km/s$ is the galactic velocity dispersion parameter, $v_\odot$ is the Sun's speed relative to the galactic rest frame, $v_{\rm esc}^{\rm gal}\simeq 544~\rm km/s$ is the galactic escape velocity, and 
\begin{align}
N_{\rm esc}={\rm erf}(z)-\dfrac{2z}{\sqrt{\pi}}e^{-z^2}, \text{ with }z=\dfrac{v_{\rm esc}^{gal}}{v_0}\,.
\end{align}
For nonrelativistic inelastic scattering, from energy-momentum conservation, we get the minimum local speed of DM,
\begin{align}
w_{\rm min}=\sqrt{\dfrac{2\deltam}{\mu_{i\phi_2}}}\,.
\label{eq:wmin}
\end{align}
By solving the quadratic equation for the recoil energy derived from energy-momentum conservation, we obtain:
\begin{align}
E_R^{\pm}=\dfrac{\mu_{i\phi_2}^2w^2}{2m_i}\left(1\pm\sqrt{1-\dfrac{2\deltam}{\mu_{i\phi_2} w^2}}\right)^2\,,
\label{eq:ERpm}
\end{align}
Additionally, the condition $\mu_{i\phi_2}w^2>2\deltam$ is required to obtain a physical recoil. Furthermore, after scattering, the DM must lose enough energy to become gravitationally bound; thus, the required capture condition is:
\begin{align}
E_R+\delta m\geq \dfrac{1}{2}m_{\phi_2}u^2\,.
\end{align}
Therefore, the minimum recoil energy required for capture is $E_{\rm cap}=\dfrac{1}{2}m_{\phi_2}u^2-\deltam$. Consequently, the recoil energy integration limits are written as \cite{Gould:1987ir, Catena:2018vzc, Pospelov:2026ewn},
\begin{align}
E_{\rm max}=E_R^+\,,\quad\text{and}\quad E_{\rm min}=\text{max}[E_R^-,\dfrac{1}{2}m_{\phi_2}u^2-\deltam,0]\,.
\label{eq:sunrecoillimit}
\end{align}
The recoil-energy integration limits, \(E_{\min}\) and \(E_{\max}\), are determined by the scattering kinematics and the capture condition. For elastic scattering, the minimum energy transfer required for capture is
\begin{align}
E_{\min}^{\rm el}=\frac{1}{2}m_{\phi_2}u^2,
\end{align}
whereas for endothermic inelastic scattering, \(\phi_2 N\to\phi_1 N\), with mass splitting \(\delta m=m_{\phi_1}-m_{\phi_2}>0\), it becomes
\begin{align}
E_{\min}^{\rm inel}
=\frac{1}{2}m_{\phi_2}u^2-\delta m.
\end{align}
The physical recoil-energy integration range must also satisfy \(E_R\geq 0\) and the kinematic limits of the scattering process. Following the above quadratic equation for $u$ (\Eq\eqref{eq:wmin}), we obtain
\begin{align}
u^2\geq \dfrac{2\deltam}{\mu_{i\phi_2}}-v_{\rm esc}^2(r)\,.
\end{align}
Therefore, the minimum DM speed is given by,
\begin{align}
u_{\rm min}^{\rm kin}(r,i)=\sqrt{\text{max}[0, \dfrac{2\deltam}{\mu_{i\phi_2}}-v_{\rm esc}^2(r)]}\,.
\end{align}
The maximum DM speed in the Sun's rest frame is given by,
\begin{align}
u_{\rm max}^{\rm halo}=v_{\rm esc}^{gal}+v_\odot\,.
\end{align}
For $v_{\rm esc}^{gal}=\rm 544\ km/s$ and $v_\odot=240\rm~km/s$, we obtain $u_{\rm max}^{\rm halo}=784~\rm km/s$. However, this does not guarantee capture. For a guaranteed DM capture, the required condition is given by,
\begin{align}
E_R^+(u) +\delta m\geq \dfrac{1}{2}\mtwo u^2\,,
\end{align}
and by solving this, we obtain the maximum possible capture speed $u_{\rm max}^{\rm cap}(r,i)$. Finally, the effective velocity regime for capture is defined as,
\begin{align}
u_{\rm min}^{\rm kin}(r,i)\leq u\leq \text{min}[u_{\rm max}^{\rm halo},u_{\rm max}^{\rm cap}(r,i)]\,.
\end{align}
Applying these limits, we can rewrite \Eq\eqref{eq:capture0} as,
\begin{align}
C_\odot
=4\pi R_\odot^3\dfrac{\rho_{\phi_2}}{\mtwo}\sum_i\int_0^{1} dx\,x^2\ n_i(x)
\int_{0}^{u_{\max}^{\rm halo}} du\,\ \frac{f(u)}{u}\,w^2 \Theta(\mu_{i\phi_2}w^2-2\deltam)
\int_{E_{\min}}^{E_{\max}} dE_R\,
\frac{d\sigma_{i\phi_2}^{\rm inelastic}}{dE_R},
\label{eq:capture00}
\end{align}
where we have defined $r=x R_\odot$, and $E_R^{\pm}$, $E_{\rm min}$, and $E_{\rm max}$ follow the expressions given previously in \Eqs \eqref{eq:ERpm} and \eqref{eq:sunrecoillimit}, respectively. The differential cross-section with respect to $E_R$ is 
\begin{align}
\frac{d\sigma_{i\phi_2}^{\rm inelastic}}{dE_R}\simeq\dfrac{m_i}{2\pi w^2}\left[\dfrac{Ze\,\epsA\gmuta}{m_{\zmuta}^2+2m_iE_R}\right]^2F^2_i(E_R)\,.
\end{align}
To obtain the present-day captured DM annihilation rate, we need to solve the time evolution of the total number of captured DM particles (\(N\)) inside the Sun, assuming that the depletion by evaporation is negligible, which is applicable for heavy DM \cite{Gould:1987ju}.
The evolution equation can be written as
\begin{align}
\frac{dN}{dt}
=C_\odot-C_A N^2\,,
\label{eq:capture1}
\end{align}
where \(C_A\) is related to the DM annihilation coefficient, and the annihilation rate is given by \(\Gamma_A=\frac{1}{2}C_A N^2\). In \Eq\eqref{eq:capture1}, we have neglected the evaporation term $(C_E N)$ for WIMP masses above a few GeV \cite{Gould:1987ju}. In the presence of a heavier DM state, the decay term should also be included when considering the evolution of the individual DM species. However, when considering the total captured DM population, the decay term cancels because the decay converts the heavier state into the lighter state without changing the total number of captured DM particles, provided that the daughter particle remains gravitationally bound to the Sun. However, the decay of the heavier DM state may contribute to the neutrino signal observed by Earth-based detectors, depending on the decay products, their energy spectrum, and the detector sensitivity. For a spatially thermalised DM population, the annihilation coefficient is generally written as,
\begin{align}
C_A=\langle \sigma v\rangle \dfrac{\int d^3r\ n_{\phi_2}^2(r)}{[\int d^3r\ n_{\phi_2}(r)]^2}\equiv \dfrac{\langle \sigma v\rangle}{V_{\rm eff}}\,.
\end{align}
In the above, the effective volume of the Sun, $V_{\rm eff}$, can be formulated as follows. Starting from the basic definition, the annihilation rate per unit volume is $\dfrac{d\Gamma_A}{dV}=\dfrac{1}{2}\langle \sigma v\rangle n_{\phi_2}^2(r)$, and the total annihilation rate inside the Sun's core is:
\begin{align}
\Gamma_A=\dfrac{1}{2}\langle \sigma v\rangle\int_{V_\odot}n_{\phi_2}^2(r)d^3r\equiv -\dfrac{1}{2}\dfrac{dN}{dt}\bigg|_{\rm ann}\,.
\end{align}
The total number of captured DM particles is, 
\begin{align}
N(t)=\int_{V_\odot}n_{\phi_2}(r,t)d^3r\,.
\end{align}
If we assume that the DM particles are thermalised with the Sun's core temperature through subsequent scattering with core elements, we can write $n_{\phi_2}(r,t)\equiv N(t)f(r)$, where $f(r)$ is the normalised spatial distribution of DM particles such that $\int_{V_\odot} f(r)\ d^3r=1$. Therefore,
\begin{align}
\int_{V_\odot}n^2_{\phi_2}(r,t)d^3r=N^2(t)\int_{V_\odot}f^2(r)d^3r\,,
\end{align}
and consequently, the total annihilation rate becomes:
\begin{align}
\Gamma_A=\dfrac{1}{2}\langle \sigma v\rangle N^2(t)\int_{V_\odot}f^2(r)d^3r\equiv \dfrac{1}{2}C_A N^2(t)\,,
\end{align}
Therefore, the particle depletion term can be written as:
\begin{align}
\dfrac{dN(t)}{dt}\bigg|_{\rm ann}=-C_AN^2(t)\,.
\end{align}
Notably, the effective volume is defined as:
\begin{align}
V_{\rm eff}^{-1}=\int_{V_\odot}f^2(r)d^3r\,.
\end{align}
Therefore, $C_A$ and $\Gamma_A$ can be written in terms of $V_{\rm eff}$ as follows:
\begin{gather}
C_A=\dfrac{\langle \sigma v\rangle}{V_{\rm eff}}\,,\\
\Gamma_A=\dfrac{1}{2}\dfrac{\langle \sigma v\rangle}{V_{\rm eff}} N^2(t)\,.
\end{gather}
Assuming an isothermal DM distribution—i.e., it reaches thermal equilibrium with the solar medium—and a centrally concentrated, nearly constant solar-core density, the gravitational potential near the centre is approximately harmonic:
\begin{align}
n_{\phi_2}(r)=n_{\phi_2}(0)e^{-r^2/r_{\phi_2}^2}\,,
\end{align}
where 
\begin{align}
r_{\phi_2}=\sqrt{\dfrac{3k_BT_c}{2\pi G\rho_c m_{\phi_2}}}\,,
\end{align}
where $T_c$ is the core temperature, $k_B$ is the Boltzmann constant, and $G$ is the gravitational constant. In this scenario,
\begin{align}
V_{\rm eff}=(2\pi)^{3/2}r_{\phi_2}^3=\left(\dfrac{3m_{\rm pl}^2T_c}{\rho_c m_{\phi_2}}\right)^{3/2}\,.
\end{align}
For $T_c=1.39\times 10^7\text{ K}\equiv 1.2\ \keV$ and $\rho_c=150\text{ g/cm}^3$, the effective volume simplifies to \cite{Bottino:1994xp,Jungman:1995df,Profumo:2013yn},
\begin{align}
V_{\rm eff}\simeq 1.09\times 10^{28}\left(\dfrac{100\ \GeV}{\mtwo}\right)^{3/2}\rm cm^3\,.
\end{align}
The time evolution of the WIMP population inside the Sun is given by \Eq\eqref{eq:capture1}:
\begin{align}
\frac{dN}{dt}=C_\odot-C_A N^2\,,
\label{eq:capture1}
\end{align}
Assuming $N(0)=0$, the solution is given by:
\begin{align}
N(t)=\sqrt{\dfrac{C_\odot}{C_A}}\tanh(t/\tau_{\rm eq})\,,
\end{align}
where $\tau_{\rm eq}=1/\sqrt{C_\odot C_A}$ is the timescale required to reach equilibrium between capture and annihilation.
The annihilation rate then becomes:
\begin{align}
\Gamma_A(t)=\dfrac{C_\odot}{2}\tanh^2(t/\tau_{\rm eq})\,,
\end{align}
If $t\ll\tau_{\rm eq}$, the annihilation rate is strongly suppressed, leading to:
\begin{align}
\Gamma_A\approx \dfrac{C_\odot}{2}(t/\tau_{\rm eq})^2\,.
\end{align}
On the other hand, at late times when equilibrium is reached ($t\gg \tau_{\rm eq}$):
\begin{align}
\Gamma_A\simeq\dfrac{C_\odot}{2}\,.
\label{eq:capture3}
\end{align}
Finally, the annihilation rate at the current solar age $t_{\odot}$ is \cite{Bottino:1994xp},
\begin{align}
\Gamma_A(t_\odot)=\dfrac{C_\odot}{2}\tanh^2(t_\odot/\tau_{\rm eq})\,.
\label{eq:solar-capture}
\end{align}
with $\tau_{\rm eq}=1/\sqrt{C_\odot C_A}$ where $C_A=\dfrac{\langle\sigma v\rangle}{V_{\rm eff}}$ and $V_{\rm eff}\simeq 1.09\times 10^{28}\left(\dfrac{\mtwo}{100\ \GeV}\right)^{-3/2}\rm cm^3\,.$

\begin{figure}[htb!]
\centering
\includegraphics[width=1.0\linewidth]{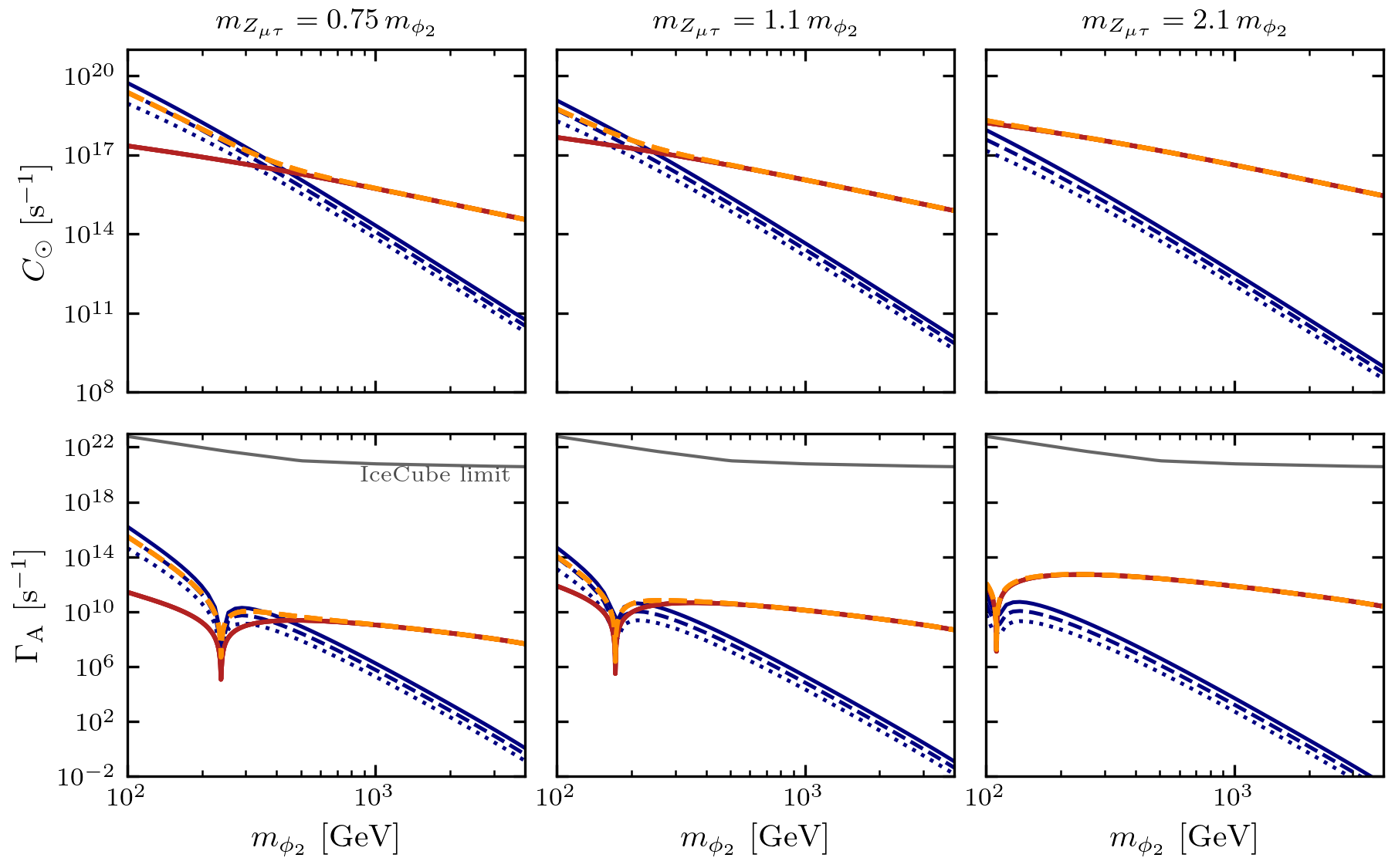}
\caption{The top and bottom panels show the DM capture rate ($C_{\mathrm{cap}}$) and annihilation rate ($\Gamma_{\mathrm{ann}}$) as a function of the DM mass $m_{\phi_2}$, respectively. The solid, dashed, and dotted blue lines correspond to inelastic scattering for mass splittings of $\delta m = [100, 150,\text{ and }200]$ keV, respectively, while the dark-red line denotes the elastic scattering of $\phi_2$. The combined total rates are indicated by the dashed yellow line. The grey line represents the IceCube upper limit for $\Gamma_A(\phi_2\phi_2 \to W^+W^-)$. We have fixed other parameters as $\mhtwo=2.2\ \mtwo$ $\lphiS=\lphiH=10^{-3}$, $\gmuta=0.1$ and $\sin\alpha=0.05$.}
\label{fig:IceCube}
\end{figure}

In this model, Higgs-portal annihilation provides the most effective mechanism for neutrino production, with the \(W^+W^-\) final state dominating over the other possible channels for $\mtwo\gtrsim m_W$. The corresponding cross section for DM annihilation into $W^+W^-$ is written as:
{\small\begin{align}
\begin{split}
\sigma(\two\two\to W^+W^-) =\frac{\left(12m_W^4-4m_W^2s+s^2\right)}{16\pi v^2_{\rm EW}(m_{h_1}^2-s)^2(s-m_{h_2}^2)^2(s-4m_{\phi_2}^2)}\sqrt{\dfrac{(s-4m_{\phi_2}^2)(s-4m_W^2)}{s^2}}
\\\times\bigg[\lambda_{\Phi H}v_{\rm EW}^{}s-m_{h_1}^2\sin\theta\left(\lambda_{\Phi H}\sin\theta\,v_{\rm EW}^{}+\sqrt{1-\sin^2\theta}\left(\lambda_{\Phi S}v_s-\sqrt{2}\,\mu_{\Phi S}\right)\right)
\\+m_{h_2}^2\left(-\lambda_{\Phi H}v_{\rm EW}^{}+\sin\theta\left(\lambda_{\Phi H}\sin\theta\,v_{\rm EW}^{}+\sqrt{1-\sin^2\theta}\left(\lambda_{\Phi S}v_s-\sqrt{2}\,\mu_{\Phi S}\right)\right)\right)\bigg]^2\,.
\end{split}
\end{align}}
Moreover, within this model framework, $\phi_2$ can undergo elastic scattering with solar nuclei via Higgs-portal interactions, providing an additional mechanism for dark matter to be captured and trapped within the solar core. The subsequent annihilation of these states into $W^+W^-$ pairs, mediated by the same portal interactions, generates high-energy neutrinos that can be probed by the IceCube Neutrino Observatory. To comprehensively account for these dynamics, we evaluated the elastic and inelastic scattering components independently. Finally, to determine the inclusive neutrino signal, we incorporate the combined solar capture rate ($C_\odot^{\phi_1} + C_\odot^{\phi_2}$) and the total annihilation rate ($\Gamma_A^{\phi_1} + \Gamma_A^{\phi_2}$), which are represented in \autoref{fig:IceCube}.

In \autoref{fig:IceCube}, we compare the solar capture of dark matter via elastic and inelastic scattering and evaluate the total annihilation rate against the IceCube exclusion limit. The top and bottom panels display the capture and annihilation rates, respectively, across three mass ratios: $\mzmuta/\mtwo = 0.75, 1.1,$ and $2.1$. We find that the IceCube constraints on the annihilation of solar-captured dark matter do not exclude any additional parameter space within our analysis regime. Notably, a kink appears in the annihilation curves for both the elastic and inelastic channels, arising from the interference between $h_1$- and $h_2$-mediated dark matter annihilation into $W^+W^-$ pairs. For a fixed gauge coupling $\gmuta=0.1$, the position of this kink shifts with $\mzmuta$ due to the corresponding variation in the vacuum expectation value $\vmuta$.
\section{Statistics of Single Event Excess}
\label{app:stat}
In this appendix we explain why Wilks' theorem (used in a lot of analyses concerning the LZ excess) cannot be used to derive confidence regions from the single LZ event, describe the Feldman-Cousins (FC) construction used in~\autoref{sec:DD}, and compare the two. Both methods use the same test statistic, the likelihood ratio $\Delta\chi^2(\vec\theta)$ of \eq\eqref{eq:4.26}, and both accept a parameter point if $\Delta\chi^2(\vec\theta)$ lies below a critical value. They differ only in how this critical value is obtained. Wilks' theorem states that, if $\vec\theta$ is the true parameter point, $\Delta\chi^2$ follows a $\chi^2$ distribution with as many degrees of freedom as fitted parameters. For two parameters this gives the familiar thresholds $2.30$ ($1\sigma\sim68\%$ C.L.) and $6.18$ ($2\sigma\sim95\%$ C.L.), applied uniformly over the whole parameter plane. The theorem is, however, an asymptotic result. It requires enough events for the likelihood to be approximately Gaussian in the parameters, and it requires every fitted parameter to be constrained by the data. Neither condition holds for the single event excess observed at LZ. A single event is far from the asymptotic regime, and over most of the favored region the model predicts $N(\vec\theta) \lesssim 1$ events. In addition, the $m_{\phi_2}-v_{\mu\tau}$ plane contains an almost flat direction, i.e. for $m_{\phi_2} \gg m_N$ the spectral shape becomes independent of the DM mass, and the event rate depends only on the joint combination of $m_{\phi_2}$ and $v_{\mu\tau}$. The data thus effectively constrain one parameter, not two. The consequence is easiest to see for a parameter point that predicts few events, say $N(\vec\theta) = 0.1$. If this point were true, about $90\%$ of repeated experiments would observe no event at all, and for those experiments $\Delta\chi^2$ is very small. The actual $68\%$ quantile of $\Delta\chi^2$ at such a point is therefore far below $2.30$. Wilks' prescription nevertheless accepts the point whenever $\Delta\chi^2 < 2.30$, and so retains regions of parameter space that the single observed event actually disfavors. Conversely, at points predicting $N(\vec\theta) \sim 1$ the distribution of $\Delta\chi^2$ can be broader than the asymptotic one, and there Wilks' prescription is slightly too restrictive. The quoted confidence level of the Wilks regions is therefore not reliable for this measurement.

\subsection*{Feldman-Cousins Construction}
The FC approach~\cite{Feldman:1997qc} avoids any asymptotic assumption by determining the distribution of $\Delta\chi^2$ explicitly at every parameter point. For each $\vec\theta$ on the scan grid, we simulate the experiment many times assuming $\vec\theta$ is true and record $\Delta\chi^2$ for each pseudo-experiment, maximising the likelihood over the full grid as for the real data. The critical value $\Delta\chi^2_c(\vec\theta)$ is the $68\%$ ($95\%$) quantile of this distribution, and $\vec\theta$ is included in the $\sim 1\sigma$ ($\sim 2\sigma$) region if the observed $\Delta\chi^2(\vec\theta) < \Delta\chi^2_c(\vec\theta)$. Since the critical value is adapted to each point, the region contains the true parameters with the stated probability, irrespective of the number of events. Moreover, ordering the outcomes by the likelihood ratio ensures a smooth transition between 2-sided regions and 1-sided limits, and it never yields empty regions. Practically, pseudo-experiments with zero or single event are treated exactly, since their $\Delta\chi^2$ depends only on $N(\vec\theta)$ and on the recoil energy of the event, respectively. Pseudo-experiments with two or more events are generated by Monte Carlo, with $5000$ pseudo-experiments per parameter point near the boundaries of the regions.

\begin{figure}[htb!]
  \centering
  \includegraphics[width=0.475\textwidth]{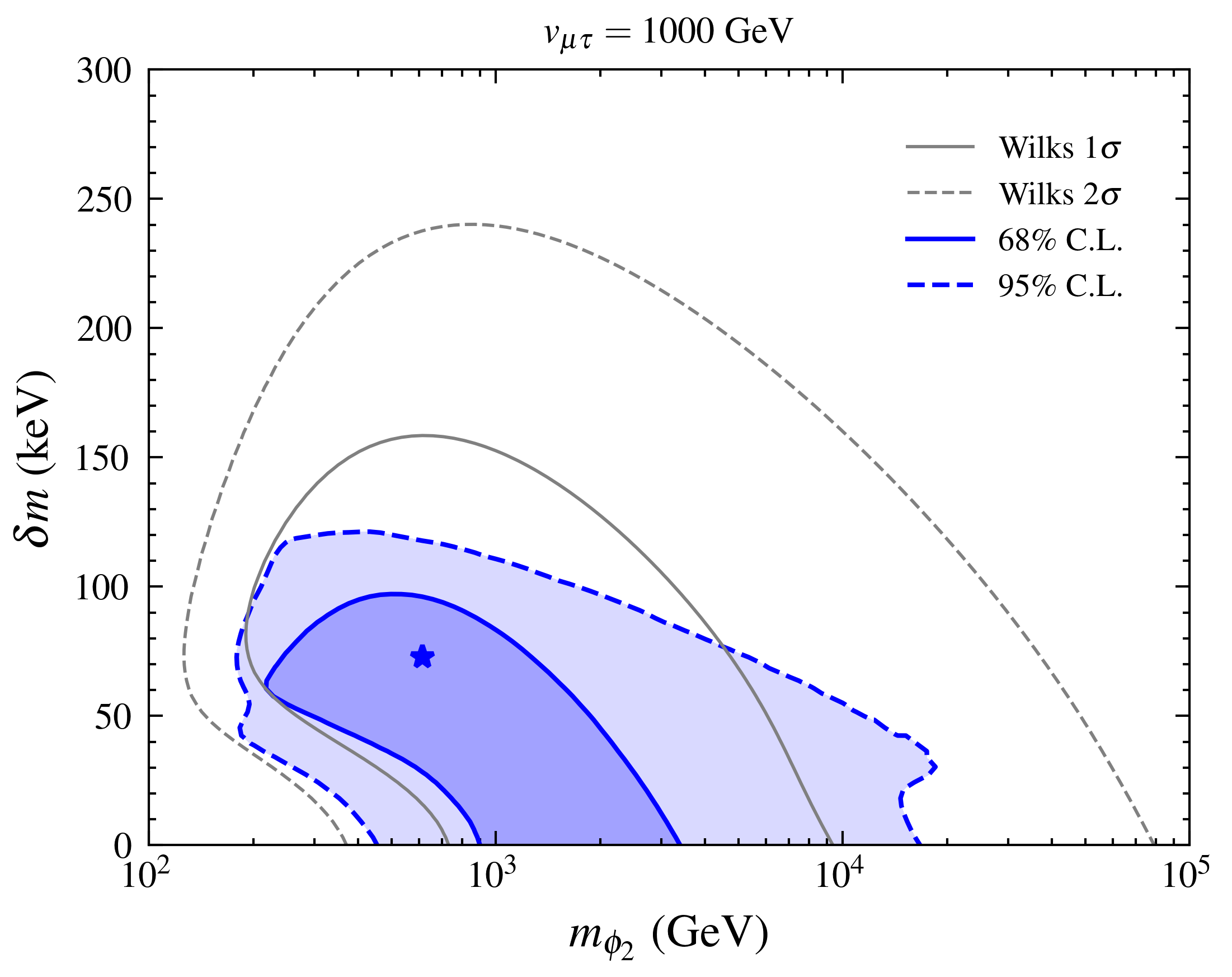}
  \includegraphics[width=0.475\textwidth]{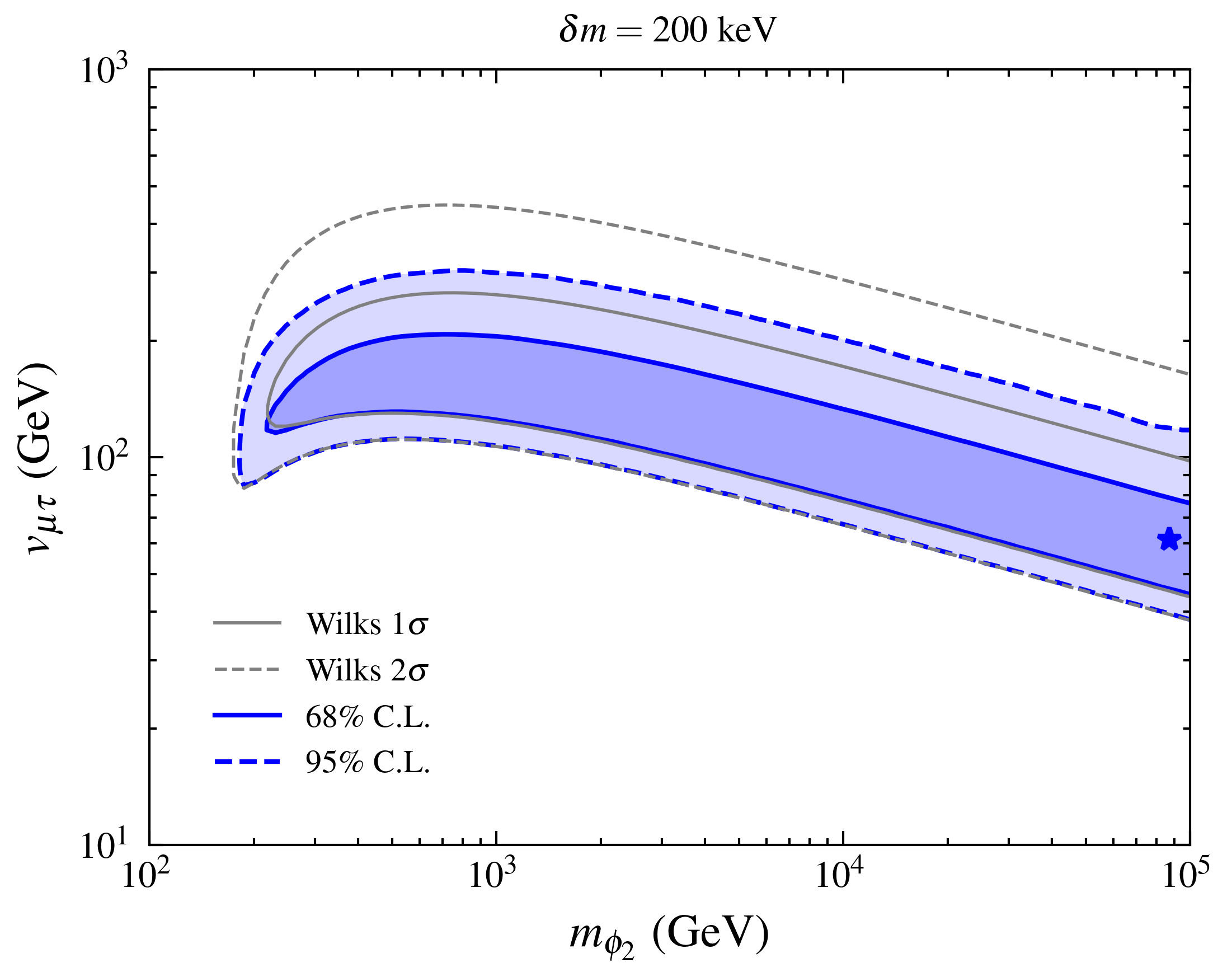}  
  \caption{\textit{Left}: Comparison of the confidence regions in the $m_{\phi_2}-\delta m$ plane for $v_{\mu\tau} = 1000$ GeV. Here, shaded regions correspond to FC $68\%$ (dark) and $95\%$ (light) C.L. regions; while black lines correspond to Wilks $1\sigma$ (solid) and $2\sigma$ (dashed) contours. Stars mark the best-fit points. \textit{Right}: Same as \textit{left}, but on the $m_{\phi_2}-v_{\mu\tau}$ plane for $\delta m = 200$ keV.}
  \label{fig:fc_wilks}
\end{figure}

\autoref{fig:fc_wilks} compare the two prescriptions in the $m_{\phi_2}-\delta m$ (\textit{left}) and $m_{\phi_2}-v_{\mu\tau}$ (\textit{right}) planes, respectively. In both cases the boundary on the high-rate side, where the model predicts too many events to be compatible with a single event, is almost unchanged. The differences arise on the low-rate side. In both the plane, the FC regions are smaller than the Wilks regions by roughly a factor of $2-3$, depending on the confidence intervals. Overall, the Wilks prescription overestimates the size of the favoured regions for the single event LZ excess, and thus, we use the FC regions throughout this work.

\bibliographystyle{JHEP}
\bibliography{biblio.bib}
\end{document}